\documentclass[aip,jcp,reprint,amsmath,amssymb,floatfix,citeautoscript,nofootinbib]{revtex4-2}
\usepackage{times}
\usepackage{mathptmx}
\DeclareMathAlphabet{\mathcal}{OMS}{cmsy}{m}{n} 
\usepackage[T1]{fontenc}
\usepackage{algorithm}
\usepackage{algpseudocode}
\usepackage{bm}
\usepackage{amsmath}
\usepackage{amssymb}
\usepackage{xcolor}
\usepackage{graphicx}
\usepackage{hyperref}
\hypersetup{
    colorlinks=true,
    linkcolor=blue,
    urlcolor=blue,
    citecolor=blue,
}
\usepackage{cleveref}
\usepackage{braket}
\usepackage[version=4]{mhchem}
\usepackage{multirow}
\usepackage{booktabs}

\usepackage{lipsum}

\usepackage{array}[=2016-10-06]
\usepackage{tabularx}

\newcommand{\me}{\mathrm{e}}

\definecolor{myred}{rgb}{0.8,0,0}

\DeclareUnicodeCharacter{2009}{\,}

\begin{document}

    \title{Correlation-Consistent Gaussian Basis Sets for Periodic Solids.~I.~Hydrogen and $p$-block Elements from Boron to Chlorine}

    \author{Jincheng Yu}
    \affiliation{Department of Chemistry and Biochemistry, University of Maryland, College Park, Maryland, 20742, United States}

    \author{Hong-Zhou Ye}
    \email{hzye@umd.edu}
    \affiliation{Department of Chemistry and Biochemistry, University of Maryland, College Park, Maryland, 20742, United States}
    \affiliation{Institute for Physical Science and Technology, University of Maryland, College Park, Maryland, 20742, United States}

    \date{\today}

    \begin{abstract}
        We present the cc-pV$X$Z-s series ($X =$ D, T, and Q) of correlation-consistent Gaussian basis sets for solids, covering hydrogen and the $p$-block elements from boron to chlorine.
        The series includes basis sets of consistent quality for the all-electron potential and for various pseudopotentials and effective core potentials commonly used in periodic calculations.
        Our design combines Dunning's construction of the molecular cc-pV$X$Z series with a revised material-constrained atomic optimization (MCAO) algorithm for improved numerical performance.
        MCAO regularizes standard atomic optimization to suppress overly diffuse basis functions and ensure numerical stability in a set of reference solids.
        Benchmarks against standard cc-pV$X$Z calculations for molecules and plane-wave calculations for periodic solids establish the accuracy, transferability, and numerical stability of cc-pV$X$Z-s in both mean-field and correlated wavefunction calculations and confirm reliable extrapolation to the complete-basis-set (CBS) limit.
        As an application, we use all-electron cc-pV$X$Z-s to compute basis-set-converged lattice constants, bulk moduli, and cohesive energies for $7$ bulk semiconductors.
        Using a hierarchy of correlated wavefunction methods up to local coupled cluster with singles, doubles, and perturbative triples [CCSD(T)], we find systematic convergence toward experimental bulk moduli and cohesive energies, while a small systematic overestimation of lattice constants suggests potential limitations in the approximate finite-size treatment.
        This work provides a foundation for the further development of systematically convergent Gaussian basis sets for periodic calculations.
    \end{abstract}

    \maketitle

    \section{Introduction}
    \label{sec:intro}

    Atom-centered Gaussian basis sets are a fundamental ingredient of modern quantum chemistry.
    Standard libraries of well-benchmarked Gaussian basis sets, such as the correlation-consistent cc-pV$X$Z series~\cite{Dunning89JCP}, the def2 series~\cite{Weigend03JCP,Weigend05PCCP}, and the polarization-consistent series~\cite{Jensen01JCP,Jensen02JCP}, have made molecular quantum chemical simulations routine and systematically improvable at both mean-field and post-mean-field levels of theory.
    Augmentation with diffuse functions~\cite{Kendall92JCP} or core-correlating functions~\cite{Woon95JCP} and adaptation to relativistic treatments~\cite{Jong01JCP,Pollak17JCTC} have further expanded the scope of molecular quantum chemistry.
    These libraries and their extensions provide a reliable foundation for the development and application of new methods and algorithms.~\cite{Hill13IJQC,Peterson07ARCC}

    The compactness and systematic improvability of Gaussian basis sets also make them attractive for correlated wavefunction calculations of periodic solids.~\cite{Ye26arXiv}
    Most Gaussian basis sets used in molecular calculations are optimized on isolated atoms~\cite{Dunning89JCP,Weigend03JCP,Weigend05PCCP} or small molecules~\cite{Jensen01JCP,Jensen02JCP}.
    Such optimization gives standard Gaussian basis sets broad transferability across chemical environments~\cite{Hill13IJQC}, but can also produce diffuse basis functions that cause linear dependence in periodic solids~\cite{VandeVondele07JCP,Peintinger13JCC,Ye22JCTC,Yu26arXiv}.
    This problem has been recognized since the early development of periodic Hartree-Fock~\cite{Szabo96Book} (HF) and density functional theory~\cite{Hohenberg64PR,Kohn65PR} (DFT) codes using Gaussian basis functions~\cite{Pisani88Book,Pisani96Inbook}.
    Basis-set linear dependence can destabilize self-consistent-field (SCF) calculations~\cite{Pisani88Book} and make it difficult to use the large basis sets needed for reliable convergence to the complete-basis-set (CBS) limit~\cite{Peterson07ARCC,Ye22JCTC}.
    Common remedies include removing diffuse primitives~\cite{Geipel97CPL,Kadek19PRB,Goldzak22JCP,Li26npjCM} (e.g.,~those with exponents below $0.05$ or $0.1$~a.u.) and discarding nearly redundant linear combinations through canonical orthogonalization~\cite{Kudin00PRB,Lee21JCP}.

    An alternative is to reoptimize basis-set parameters to improve numerical stability.
    Methods developed for this purpose fall into three broad classes.
    The first class combines molecular or solid-state energy optimization with an explicit linear-dependence penalty.
    In the molecular optimization (MOLOPT) approach of VandeVondele and Hutter~\cite{VandeVondele07JCP}, the objective combines the energy and the overlap-matrix condition number $\kappa$ evaluated on a set of reference molecules.
    Because $\kappa$ measures basis-set linear dependence in a given system, it serves as an explicit target for improving numerical stability.
    The resulting MOLOPT basis sets and their successors~\cite{Li21JPCL,Mirhosseini26JCP} generally show high numerical stability in periodic systems.
    A recent large-scale DFT benchmark study demonstrated this stability even for highly compact solids in unconventional phases~\cite{Bosoni24NRP}.
    Several groups independently adapted the MOLOPT strategy to optimize basis sets on selected reference solids~\cite{Daga20JCTC,Zhou21JCTC,Neufeld22JPCL,Dong24PRB}, minimizing a combination of the energy and $\kappa$ for those solids.
    The resulting basis sets are often system-specific~\cite{Daga20JCTC,Zhou21JCTC}, and their transferability beyond the reference solids remains less well characterized than that of MOLOPT basis sets.

    The second class uses direct solid-state energy optimization without an explicit linear-dependence penalty.
    The pob-TZVP basis sets developed by Peintinger and co-workers~\cite{Peintinger13JCC} are a prominent example.
    The authors selectively removed diffuse primitives from standard atom-optimized def2-TZVP basis sets~\cite{Weigend05PCCP} and reoptimized the remaining basis parameters on reference solids.
    Benchmarks using the CRYSTAL code~\cite{Erba23JCTC} showed that the pob-TZVP basis sets and their successors, the pob-rev2 series~\cite{Oliveira19JCC}, are accurate and numerically stable in periodic DFT calculations of compact solids.
    For challenging systems, however, the pob-TZVP basis sets may require augmentation with diffuse functions~\cite{Peintinger13JCC,Oliveira19JCC}, as recently demonstrated for Cu solids~\cite{Yu26arXiv}.
    Another example in this class is the work of Morales and Malone~\cite{Morales20JCP}, who optimized Gaussian basis sets for individual solids to accelerate basis-set convergence in auxiliary-field quantum Monte Carlo calculations.

    The third class regularizes standard atomic optimization to control linear dependence in periodic solids.
    One example is the GTH-cc-pV$X$Z series of Ye and Berkelbach~\cite{Ye22JCTC}, which comprises atom-optimized, correlation-consistent Gaussian basis sets for periodic calculations with Goedecker-Teter-Hutter (GTH) pseudopotentials~\cite{Goedecker96PRB,Hartwigsen98PRB}.
    The key design principle is to limit the number of valence primitives, thereby constraining the smallest Gaussian exponent in each valence angular momentum channel to improve the numerical stability of the full basis set.
    The GTH-cc-pV$X$Z basis sets have been used in several recent studies employing correlated wavefunction methods for both ground-state properties~\cite{Ye24JCTC,Haritan25JCTC,Zhang26PRX,Li26npjCM} and electronic excitations~\cite{Vo24JCP,Banerjee22JCTC,Laughon22JPCL,Li24PRL,Li26JCP}, demonstrating robust transferability across materials and methods, much like standard atom-optimized basis sets in molecules.
    We recently extended the Ye--Berkelbach approach through the material-constrained atomic optimization (MCAO) framework~\cite{Yu26arXiv}, which combines atomic energy optimization with a system-specific linear-dependence penalty to enable continuous control of linear dependence in periodic solids.
    We used MCAO to generate correlation-consistent Gaussian basis sets for Cu with various nuclear potentials and relativistic treatments and validated their accuracy, transferability, and numerical stability in both mean-field and correlated wavefunction calculations of Cu-containing solids, including systems beyond the reference solids used in the optimization.

    Building on these developments, we present the cc-pV$X$Z-s series ($X =$ D, T, and Q) of correlation-consistent Gaussian basis sets for solids, covering hydrogen and the $p$-block elements from boron to chlorine with various nuclear potentials.
    We extend the original MCAO algorithm to accommodate an arbitrary number of reference solids and improve its numerical performance.
    This allows us to make cc-pV$X$Z-s numerically stable across a wide range of solids, including highly compact unconventional phases that cause strong linear dependence in atom-optimized basis sets.
    Benchmarks against standard cc-pV$X$Z calculations for molecules and plane-wave (PW) calculations for periodic solids establish the accuracy, transferability, and numerical stability of cc-pV$X$Z-s in both mean-field and correlated wavefunction calculations, including the reliable extrapolation to the CBS limit.
    As an application, we use all-electron cc-pV$X$Z-s to compute basis-set-converged lattice constants, bulk moduli, and cohesive energies for $7$ bulk semiconductors.
    Using a hierarchy of correlated wavefunction methods up to local coupled cluster with singles, doubles, and perturbative triples~\cite{Raghavachari89CPL} [CCSD(T)], we find systematic convergence toward experimental bulk moduli and cohesive energies, while a small systematic overestimation of lattice constants suggests potential limitations in the approximate finite-size treatment.

    The rest of the paper is organized as follows.
    \Cref{sec:theory} presents general theoretical considerations for addressing basis-set linear dependence in solids and describes a revised MCAO algorithm with improved numerical performance.
    \Cref{sec:basis_design} presents the design of the cc-pV$X$Z-s basis sets for various nuclear-potential treatments.
    \Cref{sec:computational_details} gives the computational details.
    \Cref{sec:results} presents numerical calculations that validate various aspects of the cc-pV$X$Z-s basis sets and demonstrate their use in correlated wavefunction studies of solids.
    \Cref{sec:conclusion} concludes with an outlook on future work.

    \section{Theory}
    \label{sec:theory}

    \subsection{Periodic Gaussian basis sets}
    \label{subsec:backgrounds}

    Throughout this work, we consider a periodic solid described by $N_{\text{AO}}$ translationally adapted atom-centered Gaussian-type orbitals (GTOs) per unit cell,
    \begin{equation}    \label{eq:TAGTO}
        \chi_{\bm{k} \mu}(\bm{r})
            = \sum_{\bm{R}} \mathrm{e}^{\mathrm{i} \bm{k} \cdot \bm{R}}
            \mu(\bm{r} - \bm{R}),
    \end{equation}
    where $\mu(\bm{r} - \bm{R})$ denotes an isolated GTO translated to the unit cell at lattice vector $\bm{R}$, and
    $\bm{k} \in \mathcal{K}$ denotes a point in a uniform $k$-point mesh $\mathcal{K}$ of size $N_k$ sampling the first Brillouin zone.
    The isolated GTOs can be either primitive or contracted.
    A primitive GTO (pGTO) $\zeta_{i l m}$ is a single Gaussian function characterized by its exponent $\alpha_i$ and angular-momentum quantum numbers $l$ and $m$,
    \begin{equation}
        \zeta_{i l m}(\bm{r})
            = N_{i l} r^l \me^{-\alpha_i r^2} Y_{lm}(\hat{\bm{r}}),
    \end{equation}
    where $N_{il}$ is the normalization constant and $Y_{lm}$ is a real-valued spherical harmonic.
    A contracted GTO (cGTO) $\xi_{\mu l m}$ is formed by combining pGTOs with the same angular momentum using fixed contraction coefficients,
    \begin{equation}
        \xi_{\mu l m}(\bm{r})
            = \sum_{i} d_{i \mu} \zeta_{i l m}(\bm{r}).
    \end{equation}
    A Gaussian basis set is defined by its Gaussian exponents and contraction coefficients,
    \begin{equation}
        \bm{p}
            = \{\bm{\alpha}, \bm{d}\}.
    \end{equation}
    For given $\alpha$ and $l$, the $2l+1$ pGTOs with different values of $m$, i.e.,~$\{\zeta_{ilm}\}_{m=1}^{2l+1}$, form a primitive shell.
    Similarly, $\{\xi_{\mu lm}\}_{m=1}^{2l+1}$ forms a contracted shell.
    Basis-set optimization is performed at the shell level.

    \subsection{Characterization of linear dependence}
    \label{subsec:characterization_lindep}

    Linear dependence among GTOs is characterized by the eigenvalues and eigenvectors of the basis-set overlap matrix,
    \begin{equation}
        S_{\bm{k}\mu\nu}
            \equiv \braket{\chi_{\bm{k}\mu} | \chi_{\bm{k}\nu}}
            = \sum_{p=1}^{N_{\text{AO}}} \lambda_{\bm{k}p}
            U_{\bm{k}\mu p} U_{\bm{k}\nu p}^*.
    \end{equation}
    In particular, eigenvectors associated with small eigenvalues, empirically $\lambda < 10^{-6} \sim 10^{-8}$, represent nearly redundant linear combinations of basis functions and thus indicate linear dependence~\cite{Pisani88Book,Kudin00PRB}.
    In this work, we quantify the linear dependence of a basis set using the condition number $\kappa$ and the linear-dependence weight $w$.

    The condition number $\kappa$, defined as the ratio of the largest to the smallest eigenvalue, is the most commonly used metric for quantifying linear dependence~\cite{VandeVondele07JCP,Daga20JCTC,Zhou21JCTC,Neufeld22JPCL,Ye22JCTC}.
    Most previous studies use a $k$-point definition of $\kappa$,
    \begin{equation}
        \kappa_{k\text{-point}}
            = \max_{\bm{k} \in \mathcal{K}}
            \frac{
                \max_{p} \lambda_{\bm{k}p}
            }{
                \min_{p} \lambda_{\bm{k}p}
            },
    \end{equation}
    whereas we use the supercell definition,
    \begin{equation}    \label{eq:kappa_def}
        \kappa_{\text{supercell}}
            = \frac{
                \max_{\bm{k} \in \mathcal{K}} \max_{p} \lambda_{\bm{k}p}
            }{
                \min_{\bm{k} \in \mathcal{K}} \min_{p} \lambda_{\bm{k}p}
            }.
    \end{equation}
    Because $\kappa_{\text{supercell}} \geq \kappa_{k\text{-point}}$, the supercell definition provides a more stringent measure of linear dependence.
    In addition, $\kappa_{\text{supercell}}$ remains a meaningful metric in the special case of a single basis function per unit cell, whereas $\kappa_{k\text{-point}} = 1$ regardless of how diffuse the basis function is.
    We henceforth omit the subscript ``supercell''.

    Despite its simplicity and widespread use, $\kappa$ reflects only the conditioning of the most singular mode, whereas basis-set linear dependence involves \textit{all} singular modes.
    To account for these modes, we introduce the linear-dependence weight (LDW) as an alternative metric,
    \begin{equation}    \label{eq:w_def}
        w(\lambda_0)
            = \frac{1}{N_k} \sum_{\bm{k} \in \mathcal{K}}
            \sum_{p=1}^{N_{\text{AO}}} f(\lambda_{\bm{k}p}; \lambda_0),
    \end{equation}
    where $f(\lambda; \lambda_0)$ is a filter function that takes appreciable values when $\lambda \ll \lambda_0$ and vanishes when $\lambda \gg \lambda_0$, with $\lambda_0$ a user-selected threshold for identifying linear dependence.
    In practice, $\lambda_0$ can be related to a chosen condition-number threshold $\kappa_0$ by
    \begin{equation}    \label{eq:lambda_0_choice}
        \lambda_0
            = \frac{
                \max_{\bm{k}\in\mathcal{K}} \max_{p} \lambda_{\bm{k}p}
            }{\kappa_0}.
    \end{equation}
    In what follows, we use the thresholds $\lambda_0$ and $\kappa_0$ interchangeably.
    The filter function $f$ is not unique.
    Here, we choose $f$ to be a squared-softplus function,
    \begin{equation}    \label{eq:filter_function}
        f_{\text{softplus}^2}(\lambda; \lambda_0, \beta)
            = \left\{
                \beta^{-1} \ln \left[
                    1 + \left(
                        \frac{\lambda_0}{\lambda}
                    \right)^{\beta}
                \right]
            \right\}^2,
    \end{equation}
    with $\beta = 2$ providing a reasonably sharp transition near $\lambda = \lambda_0$.
    For $\lambda \ll \lambda_0$, $f_{\text{softplus}^2}(\lambda; \lambda_0)$ grows asymptotically as $[\ln (\lambda_0/\lambda)]^2$, thereby assigning larger weights to eigenvectors associated with more severe linear dependence.

    In addition to the total LDW in \cref{eq:w_def}, we also define the shell-resolved LDW,
    \begin{equation}    \label{eq:wmul_def}
        w_{\mu l}(\lambda_0)
            = \frac{1}{N_k} \sum_{\bm{k} \in \mathcal{K}}
            \sum_{p=1}^{N_{\text{AO}}} f(\lambda_{\bm{k}p}; \lambda_0)
            \sum_{m=1}^{2l+1} |U_{\bm{k}(\mu lm) p}|^2,
    \end{equation}
    which identifies the primitive or contracted shells that contribute most to the overall linear dependence of a basis set.
    Normalization of the overlap eigenvectors implies that the total LDW is the sum of all shell-resolved weights,
    \begin{equation}
        w(\lambda_0)
            = \sum_{\mu l} w_{\mu l}(\lambda_0).
    \end{equation}

    Both $\kappa$ and $w$ exhibit finite-size effects, meaning that their values depend on the size of the $k$-point mesh $\mathcal{K}$ or, equivalently, the supercell.
    Unless otherwise specified, all $\kappa$ and $w$ values reported in this work were evaluated using a sufficiently large mesh $\mathcal{K}$, with a corresponding supercell containing at least $300$ atoms.

    \subsection{Factors affecting a basis set's linear dependence}
    \label{subsec:example_studies}

    \begin{figure}[!t]
        \centering
        \includegraphics[width=3in]{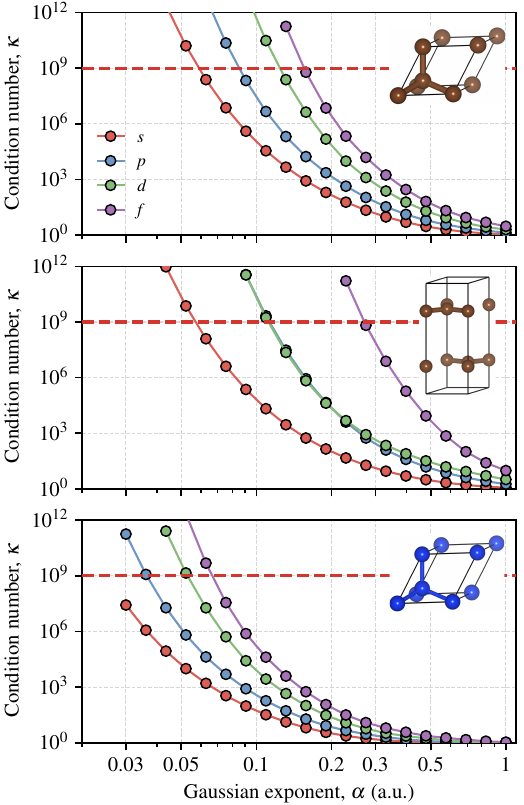}
        \caption{Condition number of a basis set consisting of a single primitive Gaussian shell per atom with exponent $\alpha$ and angular momentum $l$ evaluated for three elemental solids at their respective equilibrium geometries, namely diamond (top), graphite (middle), and silicon (bottom).
        The red dashed horizontal line indicates $\kappa = 10^{9}$, an empirical threshold for strong basis-set linear dependence~\cite{Ye22JCTC,Yu26arXiv}.
        }
        \label{fig:model_cond}
    \end{figure}

    Before presenting our approach to controlling linear dependence, we use a few examples to illustrate the factors that affect it.
    Consider an elemental solid described by a minimal basis set consisting of a single primitive shell per atom with Gaussian exponent $\alpha$ and angular momentum $l$.
    The linear dependence of this basis set---characterized by the metrics defined in \cref{subsec:characterization_lindep}---is then fully determined by $\alpha$ and $l$ for a given solid.
    \Cref{fig:model_cond} shows the condition number $\kappa$ as a function of $\alpha$ for different values of $l$ in diamond, graphite, and silicon at their respective equilibrium geometries.
    As expected, $\kappa$ increases monotonically as $\alpha$ decreases in all cases, illustrating that increasingly diffuse basis functions eventually lead to strong linear dependence (e.g.,~$\kappa \gtrsim 10^{9}$).
    The smallest $\alpha$ that can be used before encountering strong linear dependence depends strongly on angular momentum, crystal structure, and elemental species.
    \begin{enumerate}
        \item In general, $\kappa$ is more sensitive to $\alpha$ for primitives with higher $l$.
        For example, in diamond, an $s$ primitive with $\alpha \approx 0.07$ gives a value of $\kappa$ similar to that of an $f$ primitive with $\alpha \approx 0.2$.
        \item Structural anisotropy can make certain $l$ channels more sensitive than others, as illustrated by graphite.
        Compared with diamond, the layered structure of graphite leads to a markedly different ordering of the $\kappa$--$\alpha$ profiles.
        The $p$ and $d$ profiles nearly overlap, while the $f$ channel exhibits substantially faster growth in $\kappa$.
        \item Finally, different elements can exhibit very different $\kappa$--$\alpha$ profiles even when they share the same crystal structure.
        Silicon, for example, has the same crystal structure as diamond but can tolerate more diffuse primitives than diamond in all $l$ channels.
        This difference can be attributed to the larger equilibrium lattice constant ($5.431~\text{\AA}$ \textit{versus} $3.567~\text{\AA}$) and hence the greater interatomic separation in silicon, which reduces the overlap between neighboring basis functions.
    \end{enumerate}
    An important lesson from this model study is that the notion of a ``diffuse primitive'' is not well defined by the Gaussian exponent alone.
    Angular momentum, crystal structure, and elemental species all affect whether a particular primitive should be considered diffuse in the context of linear dependence.
    For this reason, discarding all primitives whose exponents fall below a fixed threshold (e.g.,~$\alpha < 0.1$~a.u.)~is an imperfect remedy for basis-set linear dependence.
    This criterion may remove important primitives with small exponents that cause little linear dependence while retaining problematic primitives with larger exponents.
    Consistent with this observation, recent periodic wavefunction-based calculations of solids have used system-specific truncation thresholds to address strong linear dependence in standard cc-pV$X$Z basis sets.~\cite{Goldzak22JCP,Li26npjCM}

    \begin{table}[!h]
        \centering
        \caption{Condition number $\kappa$ of different Gaussian basis sets for three solids at their respective equilibrium geometries.}
        \label{tab:lindep_cond}
        \begin{tabular}{lccc}
            \toprule
            Solid & cc-pVQZ & def2-QZVP & cc-pVQZ-s \\
            \midrule
            Diamond & $3.4 \times 10^{12}$ & $9.9 \times 10^{15}$ & $3.5 \times 10^{8}$  \\
            Graphite & $7.9 \times 10^{12}$ & $3.7 \times 10^{16}$ & $8.4 \times 10^{7}$  \\
            Silicon & $1.1 \times 10^{10}$ & $1.0 \times 10^{11}$ & $1.4 \times 10^{7}$  \\
            \bottomrule
        \end{tabular}
    \end{table}

\begin{figure}[!t]
        \centering
        \includegraphics[width=3in]{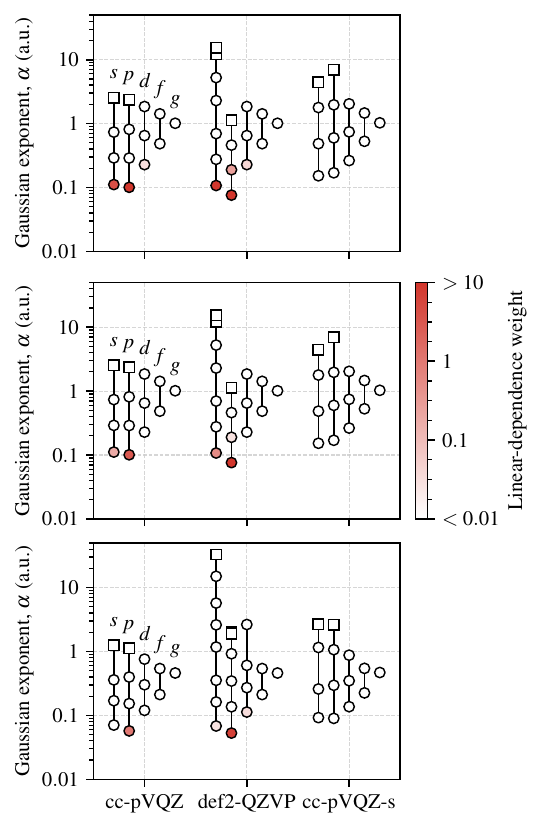}
        \caption{Shell-resolved linear-dependence weight [\cref{eq:wmul_def}] for the molecular cc-pVQZ and def2-QZVP basis sets and the solid-state cc-pVQZ-s basis sets developed in this work.
        Results are shown for diamond (top), graphite (middle), and silicon (bottom).
        Circles and boxes denote primitive and contracted shells, respectively, with each contracted shell placed at the exponent of its most diffuse primitive.
        The magnitude of the linear-dependence weight is indicated by the color scale shown on the right.
        The threshold $\lambda_0$ is determined using $\kappa_0 = 10^{9}$.
        }
        \label{fig:lindep_weight}
    \end{figure}

    The model study also helps understand the linear dependence of realistic multi-shell basis sets.
    \Cref{tab:lindep_cond} shows the condition numbers of two standard atom-optimized basis sets, cc-pVQZ~\cite{Dunning89JCP} and def2-QZVP~\cite{Weigend03JCP}, evaluated for the three solids studied above.
    All $\kappa$ values are significantly higher than $10^{9}$, indicating strong linear dependence in both basis sets for these solids.
    To identify the primitive or contracted shells responsible for this linear dependence, \cref{fig:lindep_weight} shows the shell-resolved LDWs $w_{\mu l}$ [\cref{eq:wmul_def}] alongside the basis-set structures.
    For diamond, the most diffuse primitives in both the $s$ and $p$ channels have $w_{\mu l} > 10$, indicating strong contributions to near-singular overlap eigenvectors.
    The most diffuse $d$ primitive also makes a noticeable contribution despite its relatively large exponent of ca.~$0.2$.
    For graphite, structural anisotropy causes the most diffuse $p$ primitive to acquire a noticeably larger weight than its $s$ counterpart.
    For silicon, the $s$ channel contributes little to linear dependence despite having a smaller minimum exponent than that in the corresponding carbon basis, whereas the most diffuse $p$ primitive has a moderate $w_{\mu l} \approx 1$ for cc-pVQZ and a more problematic $w_{\mu l} \approx 7$ for def2-QZVP.
    These trends are fully consistent with those observed in \cref{fig:model_cond}.

    To preview the basis sets developed in this work, \cref{tab:lindep_cond} and \cref{fig:lindep_weight} also show $\kappa$ and $w_{\mu l}$ for our cc-pVQZ-s basis sets, which are derived from the standard cc-pVQZ sets through controlled reoptimization with MCAO (\cref{sec:basis_design}).
    The cc-pVQZ-s basis sets are substantially more numerically stable than cc-pVQZ, with $\kappa < 10^{9}$ for all three solids and $w_{\mu l} < 0.01$ for all shells.
    This improved stability arises primarily from increasing the Gaussian exponents of the most diffuse $s$ and $p$ shells relative to those in cc-pVQZ, consistent with the LDW analysis.
    Importantly, the higher-$l$ polarization functions remain largely unchanged from cc-pVQZ, preserving the correlation-consistent character of the parent basis sets, as we will demonstrate numerically in \cref{subsec:validation}.

    \subsection{Material-constrained atomic optimization with improved numerical performance}
    \label{subsec:mcao}

    The numerical experiments above motivate us to ask whether we can largely preserve the structure of an atom-optimized basis set, and thus its accuracy, transferability, and systematic improvability, while regularizing its linear dependence in solids.
    If so, across what range of solids can these regularized atom-optimized basis sets maintain both numerical stability and accuracy?
    We developed material-constrained atomic optimization~\cite{Yu26arXiv} (MCAO) to address the first question.
    We address the second through systematic numerical validation and benchmarking in \cref{sec:results}.

    MCAO determines the Gaussian exponents $\bm{\alpha}$ of a basis set by minimizing the objective function
    \begin{equation}    \label{eq:mcao_loss}
        L(\bm{\alpha}; \{M\}, \varepsilon_0, \kappa_0)
            = E_{\text{atom}}(\bm{\alpha}) + \varepsilon_0 P(\bm{\alpha}; \{M\}, \kappa_0),
    \end{equation}
    where $E_{\text{atom}}$ is the atomic energy and $P$ is a dimensionless linear-dependence penalty evaluated over a set of reference solids $\{M\}$.
    The strength and onset of the penalty are controlled by two parameters, $\varepsilon_0$ and $\kappa_0$.
    $\varepsilon_0$ has units of energy and sets the energy scale of the penalty, whereas $\kappa_0$ acts approximately as a target condition number and serves as the primary parameter for tuning the conditioning of the optimized basis sets.
    The key hypothesis behind MCAO is that optimizing $E_{\text{atom}}$, as in conventional atomic optimization, helps retain the overall structure of atom-optimized basis sets while reducing sensitivity to the choice of reference solids.
    MCAO thus provides a middle ground between conventional atomic optimization~\cite{Dunning89JCP,Weigend03JCP,Weigend05PCCP}, which minimizes only $E_{\text{atom}}$, and solid-specific optimizations~\cite{Daga20JCTC,Zhou21JCTC,Neufeld22JPCL,Dong24PRB}, which minimize both $E$ and $\kappa$ for solids.

    In our previous application of MCAO to Cu basis-set optimization~\cite{Yu26arXiv}, we considered the simple case of a single reference solid with a $\kappa$-based penalty,
    \begin{equation}    \label{eq:LDP_old}
        P_{\kappa}(\bm{\alpha}; M, \kappa_0)
            = \frac{\kappa(\bm{\alpha}; M)}{\kappa_0}.
    \end{equation}
    In this work, we introduce two extensions to the original MCAO algorithm to improve its numerical performance.
    First, we define a revised penalty based on the LDW [\cref{eq:w_def}] that naturally accommodates multiple reference solids,
    \begin{equation}    \label{eq:LDP_new}
        P_{w}(\bm{\alpha}; \{M\}, \kappa_0)
            = \sum_{M} w(\bm{\alpha}; M, \kappa_0),
    \end{equation}
    where $w$ is evaluated using the squared-softplus filter function defined in \cref{eq:filter_function}.
    We find that $P_{w}$ substantially improves the stability and convergence rate of the optimization by smoothly incorporating all ill-conditioned overlap eigenvalues.
    By contrast, $P_{\kappa}$ is determined by the extremal eigenvalues, whose identities may change between optimization cycles, introducing nonsmoothness into the objective function.

    Second, if the initial basis set is extremely ill-conditioned for the chosen reference solids, evaluating the penalty in standard double-precision floating-point arithmetic can introduce substantial numerical errors and destabilize the optimization.
    In our previous work~\cite{Yu26arXiv}, we mitigated this issue by manually restricting the valence space of the initial basis set so that its condition number was reasonably close to the target value $\kappa_0$.
    Here, we introduce a lattice-scaling procedure that allows MCAO to start from basis sets with essentially arbitrary conditioning.
    Specifically, we determine an initial scaling factor $\gamma \geq 1$ that uniformly expands the lattice constants of all reference solids so that the largest $\kappa$ of the initial basis set across the reference solids is slightly below $\min(\kappa_0, 10^{10})$.
    Starting from these expanded geometries, we decrease $\gamma$ toward $1$ in steps of $0.03$ until the original reference solids are recovered.
    At each value of $\gamma$, we perform a complete MCAO optimization using the corresponding $\gamma$-scaled reference solids, with the optimized basis set from the preceding value of $\gamma$ serving as the initial guess.
    This procedure keeps the basis set within a numerically well-conditioned regime throughout the optimization and avoids instabilities associated with an excessively ill-conditioned starting point.

    \section{The cc-pV$X$Z-s basis sets for H and $p$-block elements from B to Cl}
    \label{sec:basis_design}

    \subsection{Overall design principle}
    \label{subsec:overall_strategy}

    Our design of the cc-pV$X$Z-s basis sets for $X = $ D, T, and Q closely follows Dunning's construction of the cc-pV$X$Z series~\cite{Dunning89JCP}.
    We make slight modifications to the structure of the lower-zeta basis sets, as described below, to improve their accuracy while preserving their correlation consistency.
    The selected elements---H and $p$-block elements from B to Cl---provide an excellent test set for investigating the trade-off between basis-set accuracy and numerical stability, while avoiding complications from (semi)core electrons~\cite{Woon95JCP}, multireference character~\cite{Balabanov05JCP}, and relativistic effects~\cite{Jong01JCP} associated with other elements.
    For each element, we first generate atom-optimized basis sets through the QZ level (\cref{subsec:atomic_optimization}), evaluate their linear dependence in the reference solids (\cref{subsec:reference_solids}), and apply MCAO to restore numerical stability when strong linear dependence is found (\cref{subsec:mcao_ccpvxzs}).
    The same protocol is applied to generate cc-pV$X$Z-s basis sets suitable for calculations using different nuclear-potential treatments (\cref{subsec:nuc_potential}).
    Compact auxiliary basis sets optimized for density fitting~\cite{Whitten73JCP,Dunlap00PCCP} are also developed (\cref{subsec:aux_basis}).

    \subsection{Atomic optimization}
    \label{subsec:atomic_optimization}

    For each element, we first construct atom-optimized cc-pV$X$Z-s basis sets using the following steps.
    \begin{enumerate}
        \item A valence set consisting of low-$l$ primitives ($s$ for H and $s$ and $p$ for $p$-block elements) is optimized by minimizing the atomic restricted open-shell HF (ROHF) energy,
        \begin{equation}    \label{eq:val_opt}
            \bm{\alpha}_{\text{val}}^*
                \gets \min_{\bm{\alpha}_{\text{val}}}
                E_{\text{HF}}(\bm{\alpha}_{\text{val}}).
        \end{equation}

        \item With the optimized valence primitives held fixed, a polarization set consisting of high-$l$ primitives ($l \geq 1$ for H and $l \geq 2$ for $p$-block elements) is optimized by minimizing the frozen-core atomic coupled-cluster singles and doubles~\cite{Bartlett07RMP} (CCSD) energy,
        \begin{equation}    \label{eq:pol_opt}
            \bm{\alpha}_{\text{pol}}^*
                \gets \min_{\bm{\alpha}_{\text{pol}}}
                E_{\text{CCSD}}(\bm{\alpha}_{\text{val}}^*, \bm{\alpha}_{\text{pol}}).
        \end{equation}
        For H, the atomic CCSD correlation energy is zero, so we instead minimize the CCSD correlation energy of \ce{H2} at a bond length of $0.7414~\text{\AA}$~\cite{Ye22JCTC}.

        \item The optimized valence primitives from step 1 are contracted using ROHF atomic natural orbital~\cite{Widmark90TCA} (ANO) coefficients and then combined with the optimized polarization functions from step 2.
        Optionally, the polarization primitives are contracted using frozen-core CCSD ANO coefficients (\textit{vide infra}).

        \item Additional valence-space polarization primitives are optimized in the presence of the contracted valence and polarization shells to further increase basis-set flexibility.
    \end{enumerate}

    This general protocol requires several choices.
    The first is the size of the valence set, i.e.,~the number of primitives in each valence $l$ channel.
    In conventional cc-pV$X$Z, the valence set grows with zeta level, while its size remains fixed across each period at a given zeta level.
    This leads to two issues in periodic solids.
    First, valence sets for elements early in a period (e.g.,~groups 13 and 14) contain very diffuse primitives at TZ and QZ levels, resulting in strong linear dependence in solids containing these elements.
    Second, the HF energy error increases with atomic number in each period, making basis sets for later elements (e.g.,~groups 15--17) systematically less accurate than those for earlier elements.
    To address these issues, we use a sufficiently large valence set at all zeta levels, subject to the linear-dependence constraints described in \cref{subsec:mcao_ccpvxzs}.
    Specifically, we use valence sets of QZ quality for group 15--17 elements, TZ quality for group 14 elements and H, and DZ quality for group 13 elements.
    The precise choice is less critical because MCAO automatically adjusts the most diffuse valence exponents in each $l$ channel during optimization.

    The second choice is the size of the polarization set.
    Conventional cc-pV$X$Z uses $1d$, $2d1f$, and $3d2f1g$ primitives for $p$-block elements and $1p$, $2p1d$, and $3p2d1f$ for H at the DZ, TZ, and QZ levels, respectively, without contraction.
    This hierarchy defines the correlation-consistent structure~\cite{Dunning89JCP} and provides the foundation for the systematic improvability of conventional cc-pV$X$Z.~\cite{Peterson07ARCC}
    However, large basis-set errors at lower zeta levels have been reported for molecules containing third-period elements, which require additional compact $d$ primitives, as in the cc-pV($X+d$)Z basis sets~\cite{Dunning01JCP}.
    We find a similar issue in our periodic benchmarks.
    With the conventional polarization structures, solids containing third-period elements can exhibit larger errors than their second-period counterparts at DZ and sometimes TZ.
    We therefore adopt a $(3d) \to [1d]$ structure at DZ and $(3d1f) \to [2d1f]$ at TZ for all $p$-block elements, while retaining the QZ structure.
    Here, $(nl) \to [ml]$ denotes contraction of $n$ $l$-type primitive shells into $m$ $l$-type contracted shells.
    For H, the conventional structures, $1p$ for DZ and $2p1d$ for TZ, are retained because there are no CCSD ANOs for a one-electron system.
    The QZ polarization structure is retained throughout, and the DZ and TZ contraction patterns preserve the overall basis-set size relative to cc-pV$X$Z.
    As we demonstrate numerically in \cref{subsec:validation}, the revised polarization structure improves the accuracy of cc-pV$X$Z-s at DZ and TZ while preserving the reliable CBS extrapolation of conventional cc-pV$X$Z.

    Finally, for the additional valence-space polarization functions, we follow Dunning's original recipe and add one and two primitives per valence $l$ channel at DZ and TZ levels, respectively.
    To avoid additional linear dependence, we decontract an existing valence primitive instead of adding a new function when its exponent is close to that of an optimized valence-space polarization function.
    We define closeness by an exponent ratio in $[\gamma^{-1}, \gamma]$ with $\gamma = 1.3$.
    In addition, if a valence-space polarization function is more diffuse than the most diffuse primitive in a given $l$ channel, the latter is decontracted to avoid introducing a more diffuse primitive.
    At QZ, we directly decontract the three most diffuse valence primitives from the contracted valence shells.

    \subsection{Reference solids}
    \label{subsec:reference_solids}

    \begin{figure}[!t]
        \centering
        \includegraphics[width=3.in]{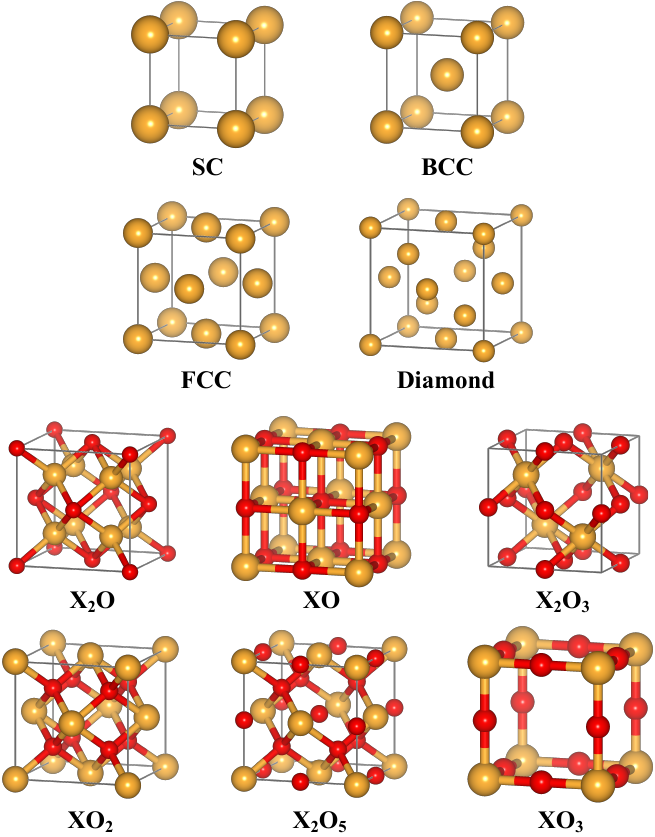}
        \caption{Crystal structures of the Bosoni reference solids~\cite{Bosoni24NRP} used for MCAO, comprising 4 elemental solids with simple cubic (SC), body-centered cubic (BCC), face-centered cubic (FCC), and diamond structures, and 6 oxides with increasing oxygen-to-element ratios, \ce{X2O}, \ce{XO}, \ce{X2O3}, \ce{XO2}, \ce{X2O5}, and \ce{XO3}.
        Conventional cubic cells are shown for BCC, FCC, diamond, \ce{X2O}, XO, and \ce{XO2}, while the calculations use the corresponding primitive cells.
        Yellow and red denote X and O, respectively.
        }
        \label{fig:bosoni_structure}
    \end{figure}

    The choice of reference solids for evaluating linear dependence and performing MCAO directly controls the range of systems to which the resulting basis sets can be applied.
    In this work, we adopt the set of solids constructed by Bosoni and co-workers~\cite{Bosoni24NRP} for a recent large-scale DFT benchmark study (henceforth referred to as the Bosoni set).
    For each element, the Bosoni set contains $10$ solids spanning a range of coordination environments and oxidation states, as shown in \cref{fig:bosoni_structure}.
    These comprise $4$ elemental solids in simple cubic (SC), body-centered cubic (BCC), face-centered cubic (FCC), and diamond phases, and $6$ oxides with different stoichiometries, \ce{X2O}, \ce{XO}, \ce{X2O3}, \ce{XO2}, \ce{X2O5}, and \ce{XO3}.
    For many elements, the Bosoni set covers thermodynamically stable phases at ambient conditions (e.g.,~diamond for C and Si, FCC for Al), and also includes highly compact unconventional phases that cause strong linear dependence in atom-optimized basis sets.
    The latter are particularly useful for developing broadly applicable basis sets.
    Other thermodynamically stable solids lie outside the Bosoni set (e.g.,~binary solids formed by group 13 and 15 elements, such as BN and AlN).
    For the elements considered here, we will show that the cc-pV$X$Z-s basis sets optimized using the Bosoni set are also well-conditioned for these more conventional phases.
    Nonetheless, the Bosoni set can also be augmented with additional solids when broader coverage is needed.

    \begin{figure}[!t]
        \centering
        \includegraphics[width=3.5in]{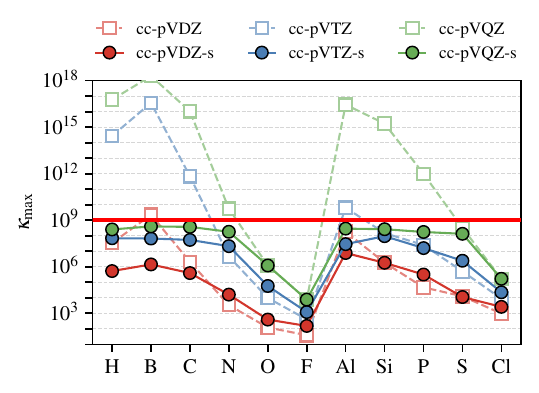}
        \caption{Maximum condition number ($\kappa_{\text{max}}$) over all $10$ reference solids of each element evaluated with the conventional cc-pV$X$Z basis sets and the cc-pV$X$Z-s sets developed in this work.
        The cc-pV$X$Z-s sets have $\kappa_{\text{max}} < 10^{9}$ for all elements at all zeta levels, as indicated by the red horizontal line.
        }
        \label{fig:bosoni_cond}
    \end{figure}

    To assess the linear dependence of atom-optimized basis sets, \cref{fig:bosoni_cond} shows the largest condition number $\kappa_{\text{max}}$ over the $10$ reference solids for each element using standard cc-pV$X$Z basis sets.
    Strong linear dependence is observed at TZ and QZ for H and group 13 and 14 elements, and at QZ for group 15 elements.
    These basis sets are natural targets for MCAO.
    Basis sets for elements in group 16 and higher remain largely well-conditioned even at QZ, consistent with their less diffuse primitives arising from larger effective nuclear charges.
    In particular, atom-optimized O basis sets remain numerically stable through QZ.
    This simplifies MCAO with the Bosoni set because the penalty for an oxide can be evaluated while keeping the atom-optimized O basis set fixed.

    Figs.~S1 and S2 show $\kappa_{\text{max}}$ over the Bosoni set for two other atom-optimized basis sets, the def2-SVP/TZVP/QZVP series~\cite{Weigend03JCP,Weigend05PCCP} and the ccECP-cc-pV$X$Z series optimized for the ccECP~\cite{Bennett17JCP}.
    Both series show patterns similar to those in \cref{fig:bosoni_cond}, demonstrating that linear dependence is a general issue for atom-optimized basis sets across basis structures and nuclear potentials.
    Fig.~S3 also shows strong linear dependence for the GTH-cc-pV$X$Z sets~\cite{Ye22JCTC}---which were designed to be numerically stable for a manually selected set of thermodynamically stable solids---for H and group 13--15 elements at TZ and QZ.
    The comparison therefore highlights the broader range of environments covered by the Bosoni reference solids.

    \subsection{MCAO}
    \label{subsec:mcao_ccpvxzs}

    We perform MCAO to improve the numerical stability of all atom-optimized basis sets from \cref{subsec:atomic_optimization} with $\kappa_{\text{max}} > 10^{7}$ on the Bosoni reference solids.
    In particular, MCAO is used to regularize the valence and polarization primitives, i.e.,~steps 1 and 2 in \cref{subsec:atomic_optimization}.
    The subsequent contraction of primitives (step 3) and determination of valence-space polarization primitives (step 4) use only atomic calculations, and neither introduces additional linear dependence.
    In conventional atomic optimization, valence and polarization primitives are determined separately [\cref{eq:val_opt,eq:pol_opt}], whereas MCAO couples these optimizations through the penalty term.
    We therefore alternate between valence and polarization optimization until convergence,
    \begin{subequations}    \label{eq:mcao_coupled_opt}
    \begin{align}
        \bm{\alpha}^{(n+1)}_{\text{val}}
            &\gets \min_{\bm{\alpha}_{\text{val}}}
            E_{\text{HF}}(\bm{\alpha}_{\text{val}})
            + \varepsilon_{0}^{\text{val}}
            P_{w}(
                \bm{\alpha}_{\text{val}},
                \bm{\alpha}_{\text{pol}}^{(n)}
            )   \\
        \bm{\alpha}^{(n+1)}_{\text{pol}}
            &\gets \min_{\bm{\alpha}_{\text{pol}}}
            E_{\text{CCSD}}(
                \bm{\alpha}^{(n+1)}_{\text{val}},
                \bm{\alpha}_{\text{pol}}
            )
            + \varepsilon_{0}^{\text{pol}}
            P_{w}(
                \bm{\alpha}_{\text{val}}^{(n+1)},
                \bm{\alpha}_{\text{pol}}
            )
    \end{align}
    \end{subequations}
    where $n$ labels the MCAO macrocycle, and the dependence of $P_w$ on $\kappa_0$ and $\{M\}$ is omitted for brevity.
    The basis parameters typically converge within $3$ macrocycles~\cite{Yu26arXiv}, and we use a conservative choice of $5$ cycles for all basis sets reported here.

    The energy scaling parameter $\varepsilon_0$ is set to $10~\text{m}E_{\text{h}}$ and $1~\text{m}E_{\text{h}}$ for the valence and polarization primitives, respectively.
    The weaker penalty on the polarization set helps retain the basis set's correlation-consistent structure~\cite{Dunning89JCP,Yu26arXiv}.
    The linear-dependence penalty $P_w$ [\cref{eq:LDP_new}] is evaluated only for the subset of Bosoni reference solids satisfying $\kappa \geq 0.01 \kappa_{\text{max}}$, with condition numbers computed using the initial atom-optimized basis sets.
    This subset remains fixed throughout the optimization.
    For most elements, this criterion selects one (typically the simple cubic phase) or a few elemental solids, reducing the cost of penalty evaluation.
    For selected cases, we verified that the optimized basis sets are identical to those obtained from the more expensive MCAO with the full Bosoni set.

\begin{figure}[!t]
        \centering
        \includegraphics[width=3.3in]{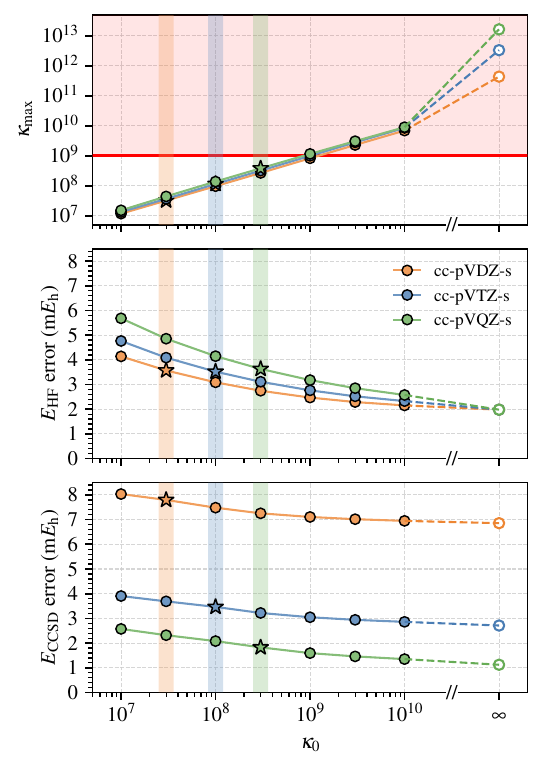}
        \caption{Variation of the highest condition number $\kappa_{\text{max}}$ over the Bosoni solids (top), the atomic ROHF energy error (middle), and the atomic CCSD correlation energy error (bottom) for carbon cc-pV$X$Z-s basis sets optimized using MCAO with different $\kappa_0$.
        The energy errors are relative to reference values obtained with the standard cc-pV5Z basis set.
        Results for the initial atom-optimized basis sets are denoted by $\kappa_0 = \infty$.
        In the top panel, the red shaded area indicates strong linear dependence with $\kappa_{\text{max}} \geq 10^{9}$.
        The final cc-pVD/T/QZ-s basis sets selected by $\kappa_0 = 3\times10^{7}$, $10^{8}$, and $3\times10^{8}$, respectively, are highlighted by stars and vertical shaded bands in all panels.
        }
        \label{fig:C_mcao}
    \end{figure}

    We use the optimization of all-electron carbon cc-pV$X$Z-s basis sets to illustrate practical aspects of MCAO and the choice of $\kappa_0$.
    This parameter controls the strength and onset of the linear-dependence penalty and balances energy accuracy against numerical conditioning.
    We vary $\kappa_0$ from $10^{10}$ for a weak penalty to $10^{7}$ for a strong penalty.
    We monitor linear dependence through $\kappa_{\text{max}}$ over the Bosoni reference solids and energy accuracy through atomic HF energy and CCSD correlation-energy errors relative to cc-pV5Z reference calculations.
    In all cases, MCAO starts from atom-optimized cc-pV$X$Z-s basis sets generated using the protocol in \cref{subsec:atomic_optimization}.
    The results are summarized in \cref{fig:C_mcao}.

    The top panel shows that MCAO effectively tunes numerical conditioning.
    As $\kappa_0$ is tightened from $10^{10}$ to $10^{7}$, the optimized $\kappa_{\text{max}}$ decreases smoothly and closely follows $\kappa_0$.
    All three zeta levels exhibit nearly identical $\kappa_{\text{max}}$--$\kappa_0$ profiles despite their very different initial conditioning, indicating that MCAO simultaneously adjusts both the valence and polarization exponents.

    Improved numerical conditioning comes at the cost of larger HF energy and CCSD correlation-energy errors, as shown in the lower two panels of \cref{fig:C_mcao}.
    At a given $\kappa_0$, the increase in both errors relative to unconstrained atomic optimization (denoted by $\kappa_0 = \infty$ in the figure) grows with zeta level, because the larger polarization sets at higher zeta levels require more substantial exponent adjustments to achieve comparable numerical conditioning.
    To maintain consistent increases in energy error across zeta levels, we use $\kappa_0 = 3\times10^{7}$, $10^8$, and $3\times10^8$ to select the final cc-pV$X$Z-s basis sets at DZ, TZ, and QZ, respectively, as highlighted by stars in \cref{fig:C_mcao}.
    With this choice, the increase in error relative to atom-optimized basis sets is consistent across zeta levels, \textit{ca.}~$1.5~\text{m}E_{\text{h}}$ for HF energies and $1~\text{m}E_{\text{h}}$ for CCSD correlation energies.
    The latter preserves the hierarchy of correlation-energy accuracy found in atom-optimized cc-pV$X$Z-s.

    \begin{figure}[!h]
        \centering
        \includegraphics[width=3.3in]{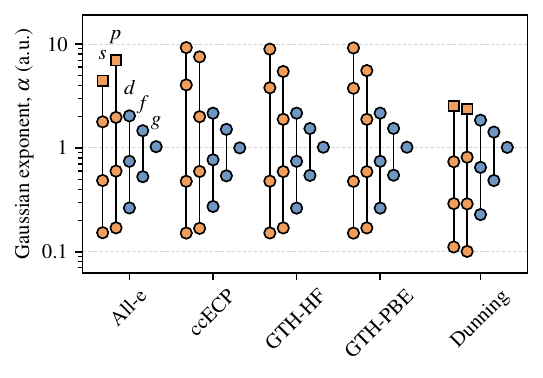}
        \caption{Gaussian exponents of carbon cc-pVQZ-s basis sets optimized for different nuclear potentials, including the all-electron potential, the GTH-HF and GTH-PBE pseudopotentials, and the ccECP.
        Dunning's atom-optimized all-electron cc-pVQZ basis set is also included for comparison.
        Circles denote primitive shells.
        Squares mark the smallest exponent in each contracted shell.
        Orange and blue denote valence and polarization channels, respectively.
        }
        \label{fig:C_mcao_basis}
    \end{figure}

    The Gaussian exponents of the final all-electron cc-pVQZ-s basis set of carbon are shown in \cref{fig:C_mcao_basis}.
    Relative to conventional cc-pVQZ, MCAO increases the valence primitive exponents and effectively reduces the valence set to between DZ and TZ quality.
    The smallest exponent in the $d$ polarization channel increases slightly, while the remaining polarization exponents are largely unchanged.
    These modifications are fully consistent with the ill-conditioned primitives identified by the LDW analysis in \cref{subsec:characterization_lindep}, validating the effectiveness of MCAO to identify and regularize these problematic primitives.

    The same choices of $\kappa_0$ are used to select the final cc-pV$X$Z-s basis sets for all other elements to which MCAO is applied.
    With these choices, the final all-electron cc-pV$X$Z-s basis sets have $\kappa_{\text{max}} < 10^{9}$ for all elements considered here at all zeta levels, as shown in \cref{fig:bosoni_cond}.
    This improved numerical stability across a range of solids enables reliable calculations with cc-pV$X$Z-s up to QZ and extrapolation to the CBS limit, as demonstrated numerically in \cref{sec:results}.

    \subsection{Pseudopotentials and effective core potentials}
    \label{subsec:nuc_potential}

    In addition to all-electron basis sets, we use the same protocol to generate cc-pV$X$Z-s basis sets for the GTH pseudopotentials optimized for HF calculations (GTH-HF~\cite{HutterPP}) and DFT calculations with the Perdew--Burke--Ernzerhof~\cite{Perdew96PRL} (PBE) exchange-correlation functional (GTH-PBE~\cite{Krack05TCA}) and the correlation-consistent effective core potential~\cite{Bennett17JCP} (ccECP) optimized for correlated wavefunction calculations.
    MCAO is largely insensitive to the choice of nuclear potential because differences between potentials primarily affect compact primitives in the core region.
    The cc-pV$X$Z-s basis sets optimized for all four nuclear potentials therefore have nearly identical valence exponents, as illustrated for carbon cc-pVQZ-s in \cref{fig:C_mcao_basis}.
    They also exhibit similar numerical stability across the Bosoni reference solids, as shown in figs.~S2 and S3.
    Consequently, calculations using cc-pV$X$Z-s with different nuclear potentials are expected to exhibit very similar basis-set convergence, as demonstrated numerically in \cref{subsubsec:mp2_eos}.

    This unique feature allows different aspects of cc-pV$X$Z-s to be benchmarked against suitable references.
    For example, the accuracy of cc-pV$X$Z-s in mean-field calculations can be assessed using the basis sets optimized for the GTH pseudopotentials, for which a sufficiently large PW basis provides a CBS reference.
    The accuracy of cc-pV$X$Z-s in correlated wavefunction calculations can be assessed using the all-electron basis sets and comparing the results with conventional cc-pV$X$Z calculations.
    We use these comparisons to validate the cc-pV$X$Z-s basis sets in \cref{subsec:validation}.

    \subsection{Auxiliary and MINAO basis sets}
    \label{subsec:aux_basis}

    To enable efficient periodic calculations using density fitting~\cite{Whitten73JCP,Dunlap00PCCP} (DF), we also developed optimized auxiliary basis sets for cc-pV$X$Z-s.
    The details of the optimization are provided in the Supporting Information.
    These auxiliary basis sets are termed cc-pV$X$Z-s-FIT and are designed for both mean-field and correlated wavefunction calculations.
    The average size ratio of cc-pV$X$Z-s-FIT to cc-pV$X$Z-s is approximately $8$, $5$, and $3$ for the all-electron DZ, TZ, and QZ sets and $4$, $3$, and $2$ for the GTH- and ccECP-based DZ, TZ, and QZ sets, respectively.
    For all-electron calculations, cc-pV$X$Z-s-FIT is therefore slightly larger than the commonly used cc-pV$X$Z-JKFIT~\cite{Weigend02PCCP} sets optimized for cc-pV$X$Z ($6$, $3$, and $2$ for DZ, TZ, and QZ, respectively), especially at lower zeta levels.
    This increase mainly reflects the enlarged valence and $d$ polarization primitive sets in cc-pV$X$Z-s.
    The cc-pV$X$Z-s-FIT sets are typically much smaller than automatically generated auxiliary basis sets, such as the even-tempered basis (ETB) used in PySCF~\cite{Sun18WIRCMS,Sun20JCP,Sun26JCP} and the AutoAux basis~\cite{Stoychev17JCTC} used by ORCA~\cite{Neese12WIRCMS}.

    Finally, we also generate the minimal atomic orbital (MINAO) basis sets corresponding to cc-pV$X$Z-s by fully contracting the atom-optimized valence primitives before MCAO.
    The MINAO basis sets can serve as projection bases for population analysis and the generation of intrinsic atomic orbitals~\cite{Knizia13JCTC} (IAOs).
    These IAOs serve as the localized-orbital basis for the periodic local coupled-cluster calculations in \cref{subsec:bulk_prop}.

    \section{Computational details}
    \label{sec:computational_details}

    \subsection{Basis-set optimization}

    Atomic optimization and MCAO are implemented in the open-source PyGTO code~\cite{pygto}, which uses PySCF~\cite{Sun18WIRCMS,Sun20JCP,Sun26JCP} for electronic structure calculations.
    The Gaussian exponents are optimized in logarithmic space using the Broyden--Fletcher--Goldfarb--Shanno (BFGS) algorithm~\cite{Nocedal80MC} with numerical gradients.
    An optimization is considered converged when changes in the objective function and logarithmic Gaussian exponents fall below $10^{-9}~E_{\text{h}}$ and $10^{-3}$,~respectively.
    For some elements treated with pseudopotentials or ECPs, HF energy minimization can cause two exponents to become nearly equal, leading to numerical instability.
    We avoid this variational collapse~\cite{Petersson03JCP} by penalizing the ratio of each pair of adjacent exponents $\alpha_1 < \alpha_2$,
    \begin{equation}
        P_{\text{exp-rat}}(\alpha_1, \alpha_2)
            = \varepsilon_{\text{exp-rat}}
            f_{\text{softplus}^2}(\alpha_2/\alpha_1; r_{\text{min}}, \beta),
    \end{equation}
    where $\varepsilon_{\text{exp-rat}} = 10~E_{\text{h}}$, $f_{\text{softplus}^2}$ is the squared-softplus function defined in \cref{eq:filter_function}, $r_{\text{min}} = 1.5$ is the minimum allowed ratio, and $\beta = 100$ tightly constrain the exponent ratio to $r \geq r_{\text{min}}$.
    The contribution of each reference solid to $P_{w}$ is evaluated from the overlap-matrix eigenvalues for the smallest supercell containing more than $300$ atoms.
    For cubic Bosoni reference solids, the supercells are constructed by repeating the unit cell equally along all three lattice directions.

    \subsection{Test sets}

    We benchmark different aspects of the cc-pV$X$Z-s basis sets using the following tests.
    \begin{enumerate}
        \item Atomization energies of $19$ molecular oxides, \ce{AlO} ($2$), \ce{BO} ($2$), \ce{ClO} ($2$), \ce{ClO2} ($2$), \ce{CO} ($1$), \ce{CO2} ($1$), \ce{FO} ($2$), \ce{H2O} ($1$), \ce{H2O2} ($1$), \ce{NO} ($2$), \ce{NO2} ($2$), \ce{O2} ($3$), \ce{O3} ($1$), \ce{PO} ($2$), \ce{PO2} ($2$), \ce{SiO} ($1$), \ce{SiO2} ($1$), \ce{SO2} ($1$), and \ce{SO3} ($1$), with spin multiplicities $2S+1$ shown in parentheses.

        \item Equations of state (EOS) for $110$ Bosoni solids, comprising $4$ elemental solids (simple cubic, FCC, BCC, and diamond) and $6$ oxides (\ce{X2O}, \ce{XO}, \ce{X2O3}, \ce{XO2}, \ce{X2O5}, and \ce{XO3}) for H, B, C, N, O, F, Al, Si, P, S, and Cl.

        \item Band structures and EOS for $7$ semiconductors, BN (B3), BP (B3), AlN (B3), AlP (B3), C (A4), Si (A4), and SiC (B3), with Strukturbericht designations shown in parentheses.
        The band-structure tests also include FCC Al as a representative bulk metal.
    \end{enumerate}
    Structures of all test systems are provided in the Supporting Information.

    \subsection{General computational details}

    We use the PySCF code for all Gaussian-basis calculations.
    Electron-repulsion integrals (ERIs) are evaluated using DF with auxiliary basis sets adapted to each calculation.
    For periodic PBE calculations, we use fast Fourier transform density fitting~\cite{Sun17JCP} (FFTDF) with a sufficiently large PW fitting basis to make DF errors negligible.
    This enables direct comparison of our Gaussian-basis PBE results with PW references calculated in Quantum ESPRESSO~\cite{Giannozzi09JPCM,Giannozzi17JPCM} using the same GTH-PBE pseudopotentials.

    For HF and correlated wavefunction calculations, we use the range-separated density fitting~\cite{Ye21JCPa,Ye21JCPb} (RSDF) with optimized Gaussian auxiliary basis sets generated by the algorithm developed in this work.
    The only exception is molecular atomization-energy calculations with the all-electron cc-pV$X$Z basis sets, for which we use the standard cc-pV$X$Z-JKFIT auxiliary basis sets~\cite{Weigend02PCCP}.
    The standard auxiliary basis sets cause numerical problems in periodic calculations because of strong linear dependence in the DF metric~\cite{Goldzak22JCP,Dinh26JCP}.
    Our reoptimized auxiliary basis sets resolve these problems.
    Canonical orthogonalization is not used in any periodic calculations.

    Results obtained with finite Gaussian basis sets are extrapolated to the CBS limit using standard protocols from the literature.
    We extrapolate HF energies using the exponential formula~\cite{Varandas19PCCP},
    \begin{equation}    \label{eq:hf_cbs}
        E_{\text{HF}}(X)
            = E_{\text{HF}}(\text{CBS}) + A \exp(-\beta X),
    \end{equation}
    where $\beta = 1.54$ for (D,T) and (T,Q) extrapolations and $1.95$ for (Q,5) extrapolations.
    Correlation energies are extrapolated using the $X^{-3}$ formula~\cite{Helgaker97JCP,Halkier98CPL},
    \begin{equation}    \label{eq:corr_cbs}
        E_{\text{corr}}(X)
            = E_{\text{corr}}(\text{CBS}) + A X^{-3}.
    \end{equation}
    In correlated calculations, we freeze the He and Ne cores for elements in the second and third periods, respectively.

    All EOS calculations use $7$ cell volumes uniformly spanning $[0.94, 1.06]$ times the reference volume $V_0$.
    For the Bosoni set, the reference volumes are taken from the PBE-optimized structures in ref~\citenum{Bosoni24NRP}, while experimental volumes compiled by Schimka and co-workers~\cite{Schimka11JCP} are used for the $7$ semiconductors.
    The resulting energy curves for wavefunction calculations and free-energy curves for PBE calculations (\textit{vide infra}) are fitted to the Birch--Murnaghan EOS to extract equilibrium lattice constants $a_0$ and bulk moduli $B_0$.
    The cohesive energy $E_{\text{coh}}$ is calculated at the reference volume $V_0$, with atomic reference energies corrected for basis-set superposition error~\cite{Boys70MP} (BSSE) using two shells of ghost atoms.

    We use box plots to visualize error distributions.
    Each box spans the middle $50\%$ of the data (i.e.,~the interquartile range), and the horizontal line inside the box marks the median.
    The whiskers extend to the most extreme points within $1.5$ times the interquartile range of the corresponding quartile.
    Data points beyond the whiskers are shown individually as outliers.

    \subsection{PBE calculations}

    For the Bosoni set, EOS calculations use spin-restricted PBE with the GTH-PBE pseudopotentials.
    Following ref~\citenum{Bosoni24NRP}, we use Fermi smearing with a width of $0.045~\text{Ry} = 0.0225~E_{\text{h}}$ and fit the EOS using the free energy $F = E - TS$, where $S$ is the electronic entropy.
    The Brillouin zone is sampled using uniform $k$-point meshes of size $4 \times 4 \times 4$ for BCC, FCC, and simple cubic lattices, $3 \times 3 \times 3$ for diamond and \ce{XO} lattices, and $2 \times 2 \times 2$ for \ce{X2O}, \ce{XO2}, \ce{XO3}, \ce{X2O3}, and \ce{X2O5} lattices, respectively.
    We reduce finite-size errors by shifting the $k$-point meshes by the Baldereschi mean-value point~\cite{Baldereschi73PRB,Drummond26JPCM}, $(0.1477, 0.3112, 0.4588)$ for FCC, diamond, \ce{X2O}, \ce{XO}, and \ce{XO2} lattices, $(0.25, 0.25, 0.5833)$ for the BCC lattice, and $(0.25, 0.25, 0.25)$ for SC, \ce{X2O3}, \ce{X2O5}, and \ce{XO3} lattices (all expressed in units of the supercell reciprocal lattice vectors).

    The band structures of the $7$ semiconductors and aluminum are also calculated using spin-restricted PBE with the GTH-PBE pseudopotentials.
    SCF calculations use an unshifted $4\times4\times4$ $k$-point mesh, followed by non-SCF calculations along the selected band path.
    Calculations for the $7$ semiconductors use zero temperature without smearing, while aluminum calculations use the same smearing setup as the Bosoni set.

    \subsection{Wavefunction-based calculations}
    \label{subsec:comp_details_wavefunc}

    We calculate the EOS of the $7$ semiconductors at zero temperature using HF, MP2, SCS-MP2~\cite{Grimme03JCP,Goldzak22JCP}, CCSD, and CCSD(T).
    The integrable divergence in the HF exchange energy is treated using the probe-charge Ewald method~\cite{Paier05JCP,Broqvist09PRB}.
    This shifts the occupied HF orbital energies by the supercell Madelung constant $v_{\text{M}}$ and the HF total energy by $N v_{\text{M}}/2$, where $N$ is the number of electrons per unit cell.
    The (SCS-)MP2 correlation energies and the (T) contribution to CCSD(T) use Madelung-corrected HF orbital energies, while CCSD calculations use uncorrected HF orbital energies.
    With these choices, all energy components have a leading-order finite-size scaling of $O(N_k^{-1})$~\cite{Xing24PRX,Ye26arXiv}.
    For the basis-set convergence tests in \cref{subsubsec:mp2_eos}, we use an unshifted $3\times3\times3$ $k$-point mesh for all $7$ semiconductors.

    For the comparison with experiment in \cref{subsec:bulk_prop}, we first extrapolate the calculated energy--volume curves to the combined basis-set and thermodynamic limits (TDL), then fit them to the EOS to extract $a_0$ and $B_0$.
    For HF and (SCS-)MP2, we use the composite correction
    \begin{equation}    \label{eq:HFMP2_CBS_TDL}
    \begin{split}
        E(\text{CBS},\text{TDL})
            &= E(\text{CBS}, 3 \times 3 \times 3)
            + E(\text{TZ}, \text{TDL})  \\
            &\quad{} - E(\text{TZ}, 3 \times 3 \times 3),
    \end{split}
    \end{equation}
    where CBS extrapolation on a $3\times3\times3$ $k$-point mesh uses TZ and QZ basis sets.
    The TZ results on $n \times n \times n$ $k$-point meshes with $n = 3, 4$ and $5$ are fitted to the asymptotic form~\cite{Goldzak22JCP}
    \begin{equation}
        E(N_k)
            = E(\text{TDL}) + A_1 N_k^{-1} + A_2 N_k^{-2}.
    \end{equation}
    For $E_{\text{coh}}$, we refine the TDL correction in \cref{eq:HFMP2_CBS_TDL} using $n = 4, 5$ and $6$.

    For CCSD and CCSD(T), we use the local natural orbital (LNO) approximation~\cite{Rolik11JCP,Rolik13JCP,Ye24JCTC} with systematically tightened truncation thresholds to obtain accurate estimates of canonical results with TZ basis sets and a $3\times3\times3$ $k$-point mesh.
    The atomic CCSD and CCSD(T) reference energies used to evaluate $E_{\text{coh}}$ are calculated within the frozen natural orbital (FNO) approximation~\cite{Sosa89CPL,Taube08JCP} using one shell of ghost atoms.
    We then apply an MP2-based composite correction for the change from one to two shells of ghost atoms.
    The FNO calculations use a tight percentage occupation threshold~\cite{Landau10JCP} of $99.995\%$ to select the natural orbitals.
    The estimated errors arising from the LNO approximation are below $0.003~\text{\AA}$ for $a_0$, $0.5$~GPa for $B_0$, and $0.02$~eV for $E_{\text{coh}}$, sufficient for the present benchmarks.
    Detailed convergence studies are provided in the Supporting Information.
    We treat the remaining basis-set and finite-size errors in the CC correlation energy approximately using a scaled MP2 correction,
    \begin{equation}    \label{eq:CC_CBS_TDL}
    \begin{split}
        E_{\text{CC}}(\text{CBS}, \text{TDL})
            &\approx E_{\text{CC}}(\text{TZ}, 3\times3\times3)
            + \alpha \big[ E_{\text{MP2}}(\text{CBS}, \text{TDL})   \\
            &\quad{} - E_{\text{MP2}}(\text{TZ}, 3\times3\times3) \big],
    \end{split}
    \end{equation}
    where $E_{\text{MP2}}(\text{CBS}, \text{TDL})$ is obtained using \cref{eq:HFMP2_CBS_TDL}.
    The scaling factor $\alpha$ accounts for differences in the basis-set and finite-size convergence of CC and MP2 correlation energies.
    We estimate $\alpha$ as $0.7$ using canonical MP2 and CCSD correlation energies reported in ref~\citenum{Li26npjCM} for four bulk semiconductors, BN, BP, Si, and C, at their equilibrium geometries.
    We apply this fitted $\alpha$ to all $7$ semiconductors and all cell volumes considered here.
    We denote these CC results as L-CC$+\Delta$MP2 to emphasize the approximate treatment of basis-set and finite-size errors.

    \section{Results and discussion}
    \label{sec:results}

    \subsection{Validation of cc-pV$X$Z-s basis sets}
    \label{subsec:validation}

    \subsubsection{Molecular atomization energies}
    \label{subsubsec:atomization_energy}

    \begin{figure}[!t]
        \centering
        \includegraphics[width=3.3in]{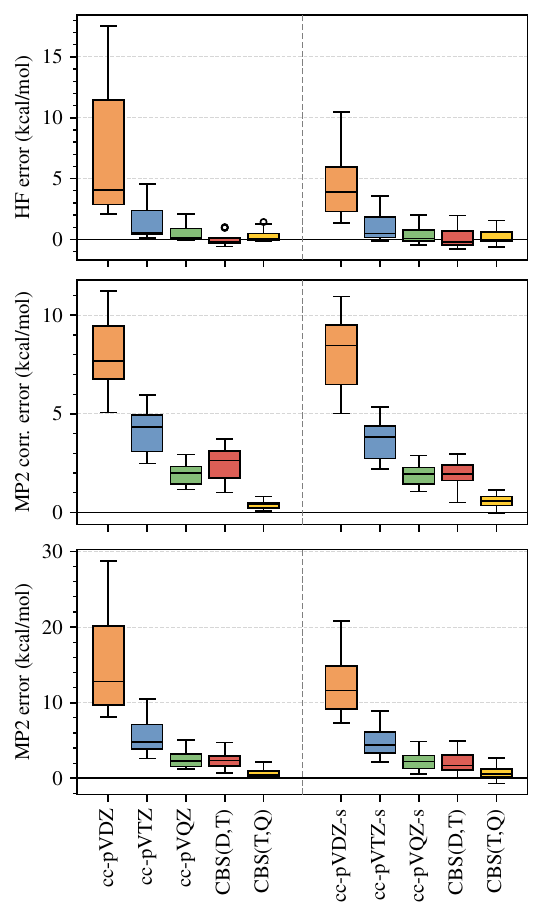}
        \caption{Box plots of basis-set errors in atomization energies of 19 molecular oxides calculated at the HF and MP2 levels using cc-pV$X$Z and cc-pV$X$Z-s for $X = $ D, T, and Q and CBS extrapolations using (D,T) and (T,Q).
        The panels show HF, MP2 correlation, and MP2 total atomization-energy errors from top to bottom.
        The errors are relative to CBS(Q,5) reference values obtained using cc-pVQZ and cc-pV5Z and are normalized by the number of atoms in each molecule.
        }
        \label{fig:mol_atomization}
    \end{figure}

    We start by benchmarking cc-pV$X$Z-s against standard cc-pV$X$Z in molecular calculations.
    Our aim is to assess whether cc-pV$X$Z-s retains the accuracy and reliable CBS extrapolation of standard cc-pV$X$Z in both mean-field and correlated wavefunction calculations.
    We select atomization energies because they directly probe the total energy recovered by a basis set, with less benefit from error cancellation.

    \Cref{fig:mol_atomization} shows the basis-set errors in HF and MP2 atomization energies for 19 molecular oxides calculated at DZ, TZ, and QZ, as well as CBS extrapolations using (D,T) and (T,Q) for both cc-pV$X$Z-s and standard cc-pV$X$Z.
    The reference values are obtained by CBS(Q,5) extrapolation with the standard basis sets.
    Overall, cc-pV$X$Z-s converges rapidly with zeta level and extrapolates to the CBS reference values at both the HF and MP2 levels.
    At the HF level, cc-pV$X$Z-s substantially outperforms standard cc-pV$X$Z at DZ and, to a lesser extent, at TZ.
    The improvement is greatest for third-period elements and can be attributed to the expanded $d$-polarization primitive set in cc-pV$X$Z-s, which serves a role similar to the additional $d$ primitives in cc-pV($X+d$)Z~\cite{Dunning01JCP}.
    At the MP2 level, cc-pV$X$Z-s closely follows the rapid convergence of standard cc-pV$X$Z, yielding reasonable accuracy with CBS(D,T) extrapolation and chemical accuracy with CBS(T,Q).
    Taken together, these results suggest that cc-pV$X$Z-s largely retains the systematic convergence of standard cc-pV$X$Z despite being regularized to improve numerical conditioning.

    \subsubsection{PBE equation of state}
    \label{subsubsec:pbe_eos}

    \begin{figure}[!t]
        \centering
        \includegraphics[width=2.7in]{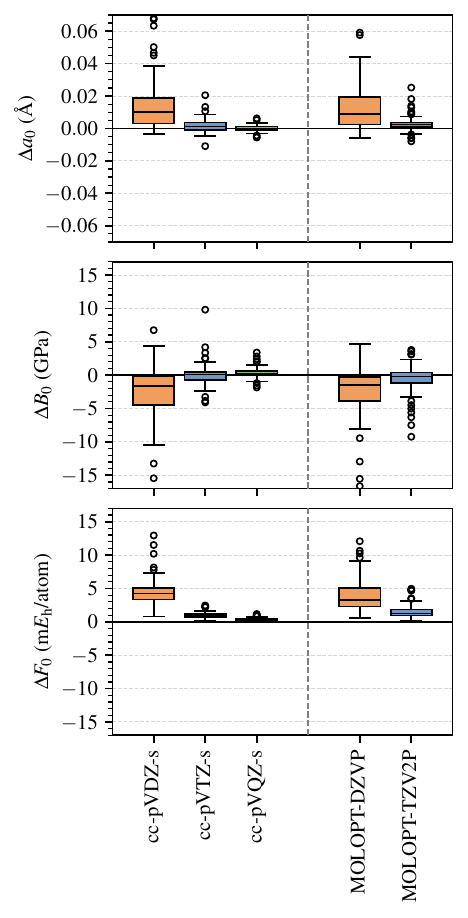}
        \caption{Box plots of basis-set errors in lattice constants, bulk moduli, and equilibrium free energies per atom calculated at the PBE level for the Bosoni test set using cc-pV$X$Z-s for $X =$ D, T, and Q.
        MOLOPT-DZVP and MOLOPT-TZV2P results are included for comparison.
        Each box plot includes results for $110$ solids corresponding to the $11$ elements considered in this work.
        All calculations employed the GTH-PBE pseudopotentials.
        The reference data were generated with a large plane-wave basis and the same pseudopotentials.
        }
        \label{fig:bosoni_pbe_cmp}
    \end{figure}

    Turning to periodic solids, we first use the Bosoni set to validate the mean-field accuracy of cc-pV$X$Z-s.
    The diverse lattice structures and chemical environments covered by the $110$ solids in the Bosoni set make this set an excellent testbed for probing the accuracy, transferability, and numerical stability of our basis sets.
    We perform PBE EOS calculations using cc-pV$X$Z-s with the GTH-PBE pseudopotentials.
    We compare the resulting lattice constants $a_0$, bulk moduli $B_0$, and free energies per atom $F_0$ at the equilibrium volume with reference results obtained using a converged PW basis.
    The basis-set errors are summarized in \cref{fig:bosoni_pbe_cmp}.
    A breakdown of these errors by element is shown in figs.~S4--S14.

    For all three properties, cc-pV$X$Z-s converges rapidly toward the PW reference as the zeta level increases, with improvements in both the median error and its spread.
    DZ already gives reasonable median errors of approximately $0.01~\text{\AA}$ for $a_0$, $-2$~GPa for $B_0$, and $4~\text{m}E_{\text{h}}$ for $F_0$.
    The error distributions are relatively wide, however, with several sizable outliers.
    Increasing the basis size to TZ and QZ further improves the medians and systematically narrows the error distributions.
    At QZ, the median errors for all three properties are close to zero.
    The absolute errors across all $110$ solids are below $0.01~\text{\AA}$ for $a_0$, $4$~GPa for $B_0$, and $2~\text{m}E_{\text{h}}$ for $F_0$, demonstrating consistent accuracy across this diverse test set.

    For comparison, \cref{fig:bosoni_pbe_cmp} includes results for the MOLOPT-DZVP and MOLOPT-TZV2P basis sets optimized using the UZH protocol for PBE calculations~\cite{Mirhosseini26JCP}.
    Both basis sets are molecularly optimized~\cite{VandeVondele07JCP} and have the same sizes as cc-pVDZ-s and cc-pVTZ-s, respectively, for the elements considered here.
    The cc-pVDZ-s and cc-pVTZ-s results closely match their MOLOPT counterparts for $a_0$ and $B_0$, while cc-pVTZ-s gives smaller errors in $F_0$.
    Both basis families show similar patterns of large outliers at DZ, which suggests that the errors primarily reflect limitations of the DZ basis structure, particularly the absence of $f$ functions, rather than differences in parameterization.
    The \ce{SO3} solid, for example, has the largest absolute errors in bulk modulus and free energy with both DZ basis sets.
    This can be attributed to the high formal oxidation state of S, whose accurate description requires $f$ functions and hence a TZ basis set.
    Overall, the results in \cref{fig:bosoni_pbe_cmp} suggest that the correlation-consistent cc-pV$X$Z-s basis sets, optimized using HF and correlated wavefunction methods, can be reliably applied to periodic DFT calculations without re-optimization and match the accuracy of state-of-the-art DFT basis sets.

    \subsubsection{PBE band structures}
    \label{subsubsec:pbe_band_structure}

    \begin{figure*}[!t]
        \centering
        \includegraphics[width=6in]{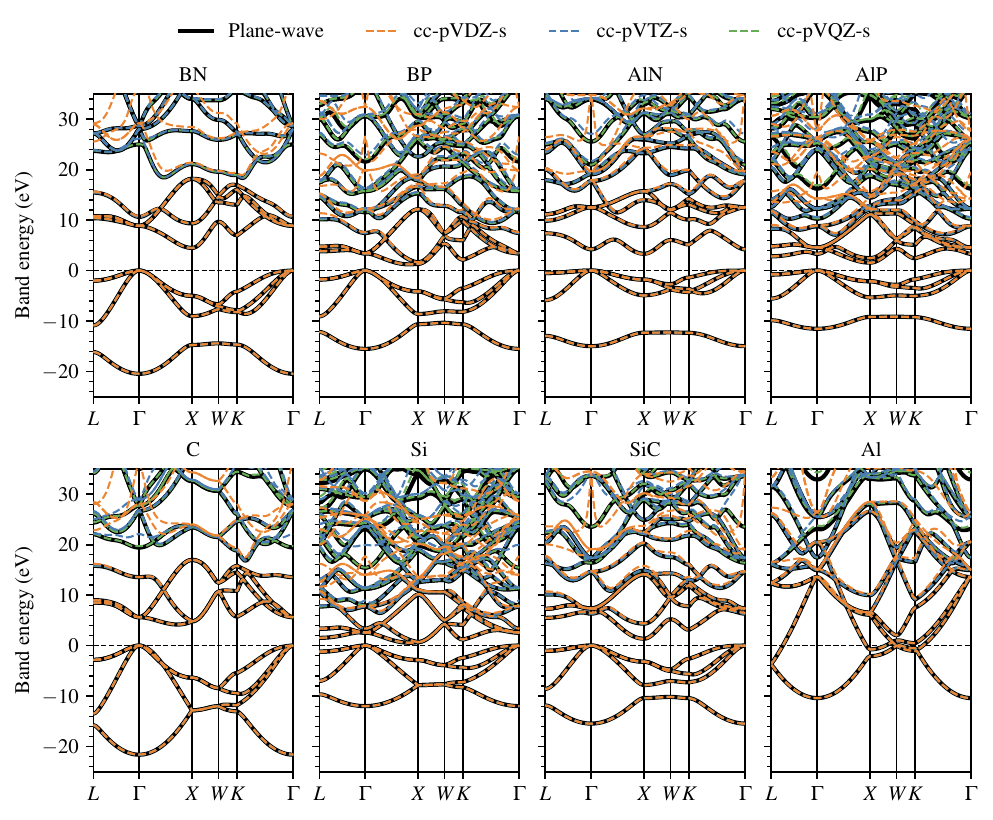}
        \caption{PBE band structures for the $7$ semiconductors and FCC Al calculated with the cc-pV$X$Z-s Gaussian basis sets and a plane-wave basis, using the GTH-PBE pseudopotentials.
        }
        \label{fig:bosoni_pbe_bands}
    \end{figure*}

    \Cref{fig:bosoni_pbe_bands} further illustrates the accuracy of cc-pV$X$Z-s through PBE band structures for the $7$ semiconductors and aluminum metal.
    Relative to the PW reference, cc-pVDZ-s already achieves quantitative accuracy for all occupied bands and the lowest few virtual bands.
    Higher-energy virtual bands are more challenging for cc-pVDZ-s, but their accuracy improves systematically with cc-pVTZ-s and cc-pVQZ-s.
    The latter closely reproduces all PW reference bands within $30$ eV above the Fermi level, including strongly entangled virtual bands in small-gap semiconductors (e.g.,~Si and AlP) and aluminum metal.
    These results suggest that accurate band structures emerge naturally from our regularized atom-optimized Gaussian basis sets.
    This differs from previous approaches that explicitly optimize basis sets against PW band structures of reference solids~\cite{Zhou21JCTC,Dong24PRB}.

    \subsubsection{MP2 equation of state}
    \label{subsubsec:mp2_eos}

    \begin{figure*}[!t]
        \centering
        \includegraphics[width=7.0in]{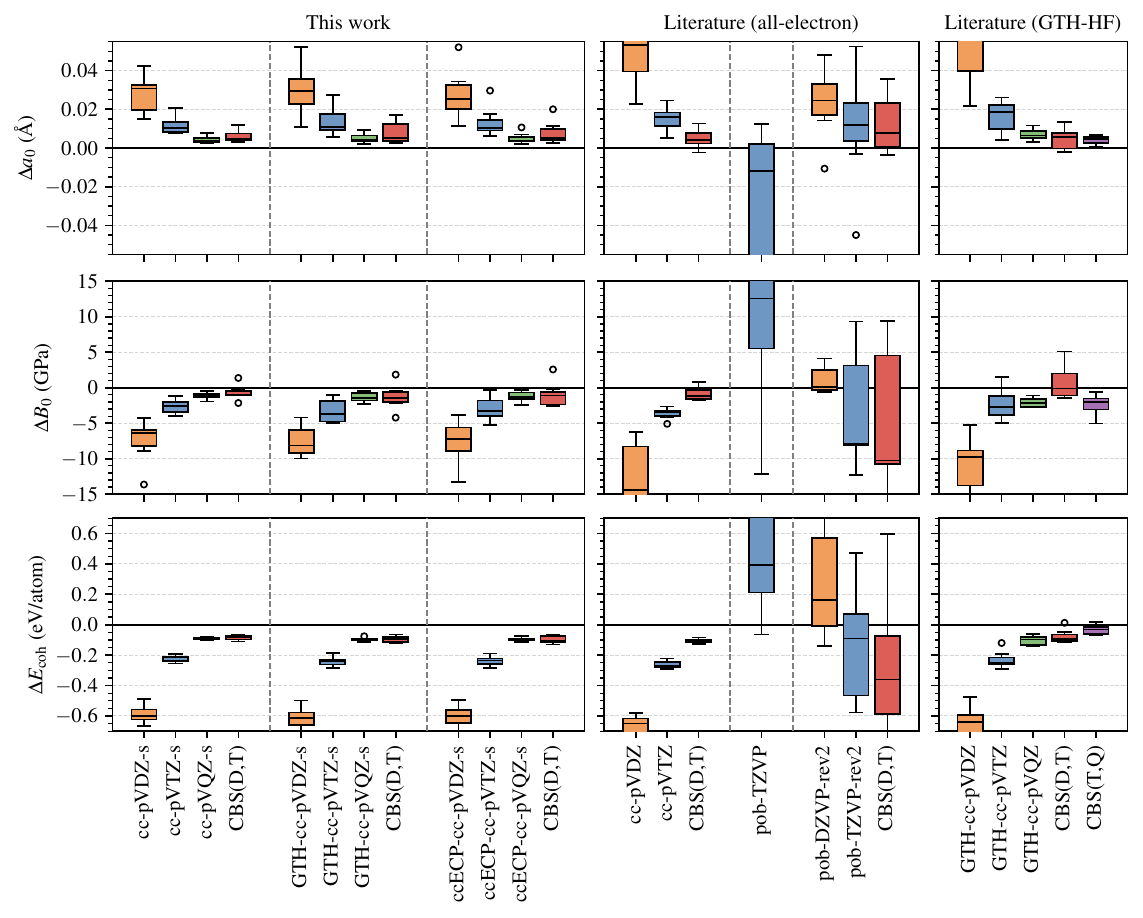}
        \caption{Box plots of basis-set errors in $a_0$, $B_0$, and $E_{\text{coh}}$ calculated at the MP2 level for the $7$ semiconductors using different Gaussian basis sets and nuclear potentials.
        The left panels show results obtained using the cc-pV$X$Z-s ($X = $ D, T, and Q) basis sets developed in this work for the all-electron potential, the GTH-HF pseudopotential, and the ccECP.
        Errors are relative to the respective CBS(T,Q) extrapolations.
        The two right columns show all-electron and GTH-HF results obtained with basis sets from the literature.
        The all-electron results use cc-pV$X$Z~\cite{Dunning89JCP} ($X = $ D and T), pob-TZVP~\cite{Peintinger13JCC}, and pob-$X$ZVP-rev2~\cite{Oliveira19JCC} ($X = $ D and T), while the GTH-HF results use GTH-cc-pV$X$Z~\cite{Ye22JCTC} ($X = $ D, T, and Q).
        Errors are relative to CBS(T,Q) extrapolations obtained with our cc-pV$X$Z-s basis sets for the corresponding nuclear potential.
        }
        \label{fig:schimka_mp2_cmp}
    \end{figure*}

    The molecular MP2 and periodic PBE benchmarks above establish the accuracy of cc-pV$X$Z-s in both mean-field and correlated calculations.
    As a final validation, we examine cc-pV$X$Z-s in periodic wavefunction calculations through MP2 EOS calculations for the $7$ semiconductors.
    The left column of \cref{fig:schimka_mp2_cmp} shows the basis-set errors in $a_0$, $B_0$, and $E_{\text{coh}}$ for cc-pV$X$Z-s optimized for the all-electron potential, the GTH-HF pseudopotential, and the ccECP.
    Here, errors are calculated against the CBS(T,Q) extrapolations using the respective cc-pV$X$Z-s basis sets.
    For all three properties, the cc-pV$X$Z-s basis sets developed for different nuclear potentials exhibit nearly identical rapid convergence with zeta level.
    The cc-pVDZ-s basis sets achieve reasonable accuracy for the two structural properties, with a median error of \textit{ca.}~$0.03~\text{\AA}$ for $a_0$ and $-7$~GPa for $B_0$, reduced to below $0.01~\text{\AA}$ and $-2$~GPa with cc-pVQZ-s.
    As with the molecular atomization energies in \cref{subsubsec:atomization_energy}, $E_{\text{coh}}$ shows stronger basis-set dependence.
    The smallest basis, cc-pVDZ-s, gives a sizable median error of \textit{ca.}~$-0.6$~eV/atom, which is reduced to \textit{ca.}~$-0.2$ and $-0.1$~eV/atom using cc-pVTZ-s and cc-pVQZ-s, respectively.
    For all three properties, CBS(D,T) extrapolation achieves accuracy comparable to cc-pVQZ-s and offers a computationally efficient alternative.
    Similar systematic convergence of cc-pV$X$Z-s in HF EOS calculations is presented in fig.~S15.
    In these calculations, both the median error and its spread converge even faster with zeta level, with cc-pVTZ-s essentially reaching CBS accuracy.

    The two right columns of \cref{fig:schimka_mp2_cmp} show all-electron and GTH-HF results obtained with basis sets from the literature.
    In all-electron calculations, standard cc-pV$X$Z is less accurate than cc-pV$X$Z-s at DZ for all three properties, while the two families achieve similar accuracy at TZ and with CBS(D,T) extrapolation.
    For GTH-HF, the GTH-cc-pV$X$Z basis sets~\cite{Ye22JCTC} show a similar trend, with larger errors than cc-pV$X$Z-s at DZ and comparable accuracy at TZ and QZ.
    The CBS(T,Q) extrapolations from the two families also agree closely, with median differences of \textit{ca.}~$0.005~\text{\AA}$ for $a_0$, $3$~GPa for $B_0$, and $0.05$~eV/atom for $E_{\text{coh}}$.
    Together with the molecular benchmarks in \cref{subsubsec:atomization_energy}, these results support the accuracy of cc-pV$X$Z-s in periodic wavefunction calculations.
    Regularized atomic optimization also enables numerically stable calculations with cc-pVQZ-s in solids where conventional cc-pVQZ suffers from strong linear dependence~\cite{Li26npjCM}.

    Comparison with the solid-optimized all-electron pob-family basis sets in \cref{fig:schimka_mp2_cmp} provides additional insight.
    For all three properties, pob-TZVP gives substantially larger errors than cc-pVTZ-s in both the median and spread.
    This can be attributed to the overly compact valence primitives in pob-TZVP, which cause sizable BSSE in molecular or periodic calculations~\cite{Oliveira19JCC}.
    The pob-rev2 series was designed to mitigate this limitation.~\cite{Oliveira19JCC}
    As shown in \cref{fig:schimka_mp2_cmp}, pob-DZVP-rev2 and pob-TZVP-rev2 improve upon pob-TZVP, especially for $a_0$.
    Both pob-rev2 basis sets, however, still have much wider error distributions than cc-pV$X$Z-s, and the improvement from DZVP to TZVP is less clear, making them less appealing for periodic wavefunction calculations.
    As suggested previously~\cite{Peintinger13JCC}, augmenting pob-family basis sets with suitable diffuse valence primitives may substantially improve their accuracy.
    In principle, MCAO provides a systematic way to determine these additional valence primitives as demonstrated in our previous work~\cite{Yu26arXiv}, although we do not explore this possibility here.


    \subsection{Bulk properties of semiconductors from wavefunction methods}
    \label{subsec:bulk_prop}

    \begin{figure}[!t]
        \centering
        \includegraphics[width=2.5in]{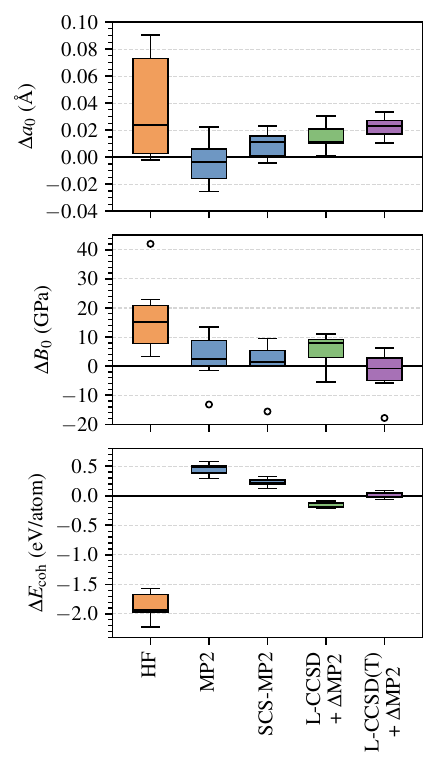}
        \caption{Box plots of errors in $a_0$, $B_0$, and $E_{\text{coh}}$ calculated using HF, MP2, SCS-MP2, L-CCSD$+\Delta$MP2, and L-CCSD(T)$+\Delta$MP2 for the $7$ semiconductors, relative to experimental references corrected for zero-point anharmonic effects~\cite{Schimka11JCP}.
        All calculations have been extrapolated to the combined basis-set and thermodynamic limits using the protocol described in \cref{subsec:comp_details_wavefunc}.
        }
        \label{fig:schimka_expt_cmp}
    \end{figure}

    To illustrate the practical use of cc-pV$X$Z-s, we calculate $a_0$, $B_0$, and $E_{\text{coh}}$ for the $7$ semiconductors using a hierarchy of wavefunction methods, including HF, MP2, SCS-MP2, L-CCSD$+\Delta$MP2, and L-CCSD(T)$+\Delta$MP2.
    All calculations are performed using the all-electron potential and extrapolated to the combined basis-set and thermodynamic limits using the protocol described in \cref{subsec:comp_details_wavefunc}.
    The final results are visualized in \cref{fig:schimka_expt_cmp} and summarized in \cref{app:bulk_prop}.

    For all three properties, the theoretical predictions converge systematically as the treatment of electron correlation is refined.
    For $B_0$ and $E_{\text{coh}}$, our highest level of theory, L-CCSD(T)$+\Delta$MP2, achieves excellent overall agreement with the experimental references.
    For $a_0$, however, the median error is smallest at MP2 and increases at higher levels of theory, with L-CCSD(T)$+\Delta$MP2 systematically overestimating the experimental values by \textit{ca.}~$0.02~\text{\AA}$.
    Given the reliable CBS extrapolations enabled by cc-pV$X$Z-s, the remaining errors in our CCSD(T) calculations likely arise from the approximate MP2-based finite-size treatment in \cref{eq:CC_CBS_TDL}.
    These results therefore demonstrate the predictive power of correlated wavefunction methods for solids while motivating the development of more scalable periodic local correlation methods~\cite{Ye24JCTC}.

    \begin{table}[!h]
        \centering
        \caption{Bulk modulus of cubic boron nitride from theoretical calculations and experimental measurements.
        All calculations are extrapolated to the combined basis-set and thermodynamic limits using the all-electron cc-pV$X$Z-s basis sets developed in this work.
        The experimental values include corrections for zero-point motion ($+10$~GPa) and thermal expansion ($+2$~GPa).
        The uncorrected room-temperature values are listed in parentheses.
        }
        \label{tab:BN_bulk_moduli}
        \begin{tabular}{lc}
            \hline
            \hline
            Method & $B_0$ (GPa)    \\
            \hline
            HF & $429$  \\
            MP2 & $397$  \\
            SCS-MP2 & $395$  \\
            L-CCSD$+\Delta$MP2 & $405$  \\
            L-CCSD(T)$+\Delta$MP2 & $392$  \\
            &   \\
            Brillouin~\cite{Grimsditch94JAP} & $412 \pm 20$ ($400 \pm 20$)   \\
            Ultrasonic~\cite{Manghnani1998} & $384 \pm 4$ ($372 \pm 4$)   \\
            X-ray~\cite{Goncharov07PRB} & $399 \pm 4$ ($387 \pm 4$)   \\
            Brillouin\cite{Goncharov07HPR} & $398 \pm 4$ ($386 \pm 4$)   \\
            \hline
        \end{tabular}
    \end{table}

    Despite the good overall agreement with experiment, an outlier is apparent in the bulk-modulus comparison in \cref{fig:schimka_expt_cmp}.
    Closer inspection identifies this outlier as BN.
    \Cref{tab:BN_bulk_moduli} compares its bulk modulus from several theoretical methods with experimental values from multiple sources.
    Except for L-CCSD$+\Delta$MP2, which predicts a relatively high value of $405$~GPa, all correlated wavefunction predictions cluster around $395$~GPa.
    The experimental reference used in \cref{fig:schimka_expt_cmp} is $410$~GPa, obtained by Schimka and co-workers~\cite{Schimka11JCP} by adding a zero-point correction of \textit{ca.}~$10$~GPa to the room-temperature experimental value of $400$~GPa reported by Grimsditch and co-workers~\cite{Grimsditch94JAP}.
    Goncharov and co-workers~\cite{Goncharov07HPR} assigned this measurement a large error bar of $20$~GPa, making it a less precise benchmark for comparison with theory.
    \Cref{tab:BN_bulk_moduli} also lists $B_0$ from three other independent measurements using various experimental methods with smaller uncertainties~\cite{Manghnani1998,Goncharov07PRB,Goncharov07HPR}.
    After applying the zero-point correction of $10$~GPa and an estimated temperature correction of $2$~GPa (based on the thermal expansion coefficient reported by Goncharov and co-workers~\cite{Goncharov07PRB}), these measurements give reference values of $384$--$399$~GPa, with a mean of \textit{ca.}~$394$~GPa, in excellent agreement with the theoretical consensus.
    This case study illustrates the difficulty of rigorous comparisons between theory and experiment, even for ``simple'' solids like BN.
    The cc-pV$X$Z-s basis sets developed here help remove basis-set incompleteness as a prominent source of error in correlated wavefunction calculations, facilitating such comparisons in future studies.

    \section{Conclusion}
    \label{sec:conclusion}

    In summary, we have developed and numerically validated the cc-pV$X$Z-s series ($X =$ D, T, and Q) of correlation-consistent Gaussian basis sets for solids, covering hydrogen and the $p$-block elements from boron to chlorine.
    We provide basis sets for the all-electron potential, the GTH pseudopotentials optimized for HF and PBE calculations, and the ccECPs.
    Using an improved MCAO algorithm, we obtain cc-pV$X$Z-s basis sets that are numerically stable across a diverse set of reference materials drawn from a recent large-scale DFT benchmark study.
    Benchmarks against standard cc-pV$X$Z calculations for molecules and plane-wave calculations for periodic solids establish the accuracy, transferability, and numerical stability of cc-pV$X$Z-s in both mean-field and correlated wavefunction calculations and confirm reliable extrapolation to the CBS limit.
    Periodic wavefunction calculations with the all-electron cc-pV$X$Z-s basis sets demonstrate systematic convergence toward experimental bulk moduli and cohesive energies for the $7$ semiconductors, while a small systematic overestimation of lattice constants suggests potential limitations in the approximate finite-size treatment.

    The present work offers two main contributions to future solid-state basis-set design.
    Technically, it establishes MCAO as an effective tool for fine-tuning basis-set numerical stability across an essentially arbitrary number of reference solids.
    Chemically, for the selected H and $p$-block elements, it demonstrates that regularization against judiciously chosen reference solids yields basis sets that balance numerical stability, energy accuracy, and transferability across diverse solids.
    Extending MCAO to other elements will require element-specific adaptations, such as adding semicore-correlating functions for $s$-block metals~\cite{Prascher11TCA}, treating multireference character in $d$-block metals~\cite{Balabanov05JCP}, and accounting for relativistic effects in heavy elements~\cite{Jong01JCP,Pollak17JCTC}.
    The advances reported here provide a foundation for these extensions, which we plan to present in subsequent papers in this series.

    \appendix
    \section{Predicted lattice constants, bulk moduli, and cohesive energies for the $7$ semiconductors}
    \label{app:bulk_prop}

    \Cref{tab:sc7_a0,tab:sc7_b0,tab:sc7_ecoh} report our predictions of $a_0$, $B_0$, and $E_{\text{coh}}$ for the $7$ semiconductors at the HF, MP2, SCS-MP2, L-CCSD$+\Delta$MP2, and L-CCSD(T)$+\Delta$MP2 levels of theory.
    The tables also report the mean absolute error (MAE) and mean absolute relative error (MARE) of each method relative to experiment.

    \begin{table}[!h]
        \centering
        \caption{Predicted lattice constants at the indicated levels of theory, including the mean absolute error (MAE) and mean absolute relative error (MARE) relative to experiment.
        Experimental values, which have been corrected for zero-point motion, are from ref~\citenum{Schimka11JCP}.}
        \label{tab:sc7_a0}
        \begin{tabular}{lcccccc}
            \toprule
             & \multicolumn{6}{c}{Lattice constant (\AA)} \\
            \cmidrule{2-7}
            Solid & HF & MP2 & SCS-MP2 & \shortstack{L-CCSD\\$+\Delta$MP2} & \shortstack{L-CCSD(T)\\$+\Delta$MP2} & Expt. \\
            \midrule
            BN & $3.599$ & $3.596$ & $3.603$ & $3.601$ & $3.610$ & $3.592$ \\
            BP & $4.585$ & $4.500$ & $4.521$ & $4.536$ & $4.541$ & $4.525$ \\
            AlN & $4.367$ & $4.390$ & $4.391$ & $4.379$ & $4.391$ & $4.368$ \\
            AlP & $5.542$ & $5.448$ & $5.467$ & $5.482$ & $5.485$ & $5.451$ \\
            C & $3.551$ & $3.542$ & $3.552$ & $3.554$ & $3.563$ & $3.553$ \\
            Si & $5.507$ & $5.400$ & $5.424$ & $5.446$ & $5.449$ & $5.421$ \\
            SiC & $4.370$ & $4.354$ & $4.361$ & $4.362$ & $4.372$ & $4.346$ \\
            \midrule
            MAE (\AA) & $0.039$ & $0.014$ & $0.011$ & $0.015$ & $0.022$ & $\cdots$ \\
            MARE (\%) & $0.8$ & $0.3$ & $0.2$ & $0.3$ & $0.5$ & $\cdots$ \\
            \bottomrule
        \end{tabular}
    \end{table}

    \begin{table}[!h]
        \centering
        \caption{Predicted bulk moduli at the indicated levels of theory, including the mean absolute error (MAE) and mean absolute relative error (MARE) relative to experiment.
        Experimental values, corrected for zero-point motion, are from ref~\citenum{Schimka11JCP}, except for BN, for which we use the revised experimental average discussed in \cref{subsec:bulk_prop}.}
        \label{tab:sc7_b0}
        \begin{tabular}{lcccccc}
            \toprule
             & \multicolumn{6}{c}{Bulk modulus (GPa)} \\
            \cmidrule{2-7}
            Solid & HF & MP2 & SCS-MP2 & \shortstack{L-CCSD\\$+\Delta$MP2} & \shortstack{L-CCSD(T)\\$+\Delta$MP2} & Expt. \\
            \midrule
            BN & $429$ & $397$ & $395$ & $405$ & $392$ & $394$ \\
            BP & $176$ & $182$ & $178$ & $178$ & $173$ & $168$ \\
            AlN & $229$ & $205$ & $206$ & $214$ & $207$ & $206$ \\
            AlP & $95$ & $96$ & $95$ & $96$ & $94$ & $87$ \\
            C & $497$ & $464$ & $458$ & $466$ & $449$ & $455$ \\
            Si & $104$ & $103$ & $102$ & $103$ & $100$ & $101$ \\
            SiC & $244$ & $231$ & $230$ & $233$ & $225$ & $229$ \\
            \midrule
            MAE (GPa) & $19$ & $6$ & $3$ & $8$ & $3$ & $\cdots$ \\
            MARE (\%) & $7.5$ & $3.6$ & $2.4$ & $4.1$ & $2.1$ & $\cdots$ \\
            \bottomrule
        \end{tabular}
    \end{table}

    \begin{table}[!h]
        \centering
        \caption{Predicted cohesive energies at the indicated levels of theory, including the mean absolute error (MAE) and mean absolute relative error (MARE) relative to experiment.
        Experimental values, which have been corrected for zero-point motion, are from ref~\citenum{Schimka11JCP}.}
        \label{tab:sc7_ecoh}
        \begin{tabular}{lcccccc}
            \toprule
             & \multicolumn{6}{c}{Cohesive energy (eV/atom)} \\
            \cmidrule{2-7}
            Solid & HF & MP2 & SCS-MP2 & \shortstack{L-CCSD\\$+\Delta$MP2} & \shortstack{L-CCSD(T)\\$+\Delta$MP2} & Expt. \\
            \midrule
            BN & $4.76$ & $7.23$ & $7.05$ & $6.62$ & $6.77$ & $6.76$ \\
            BP & $3.43$ & $5.72$ & $5.47$ & $5.05$ & $5.23$ & $5.14$ \\
            AlN & $3.90$ & $6.16$ & $6.03$ & $5.63$ & $5.78$ & $5.85$ \\
            AlP & $2.75$ & $4.61$ & $4.45$ & $4.14$ & $4.30$ & $4.32$ \\
            C & $5.32$ & $8.08$ & $7.78$ & $7.35$ & $7.53$ & $7.55$ \\
            Si & $3.05$ & $5.14$ & $4.90$ & $4.57$ & $4.76$ & $4.68$ \\
            SiC & $4.55$ & $6.96$ & $6.69$ & $6.29$ & $6.46$ & $6.48$ \\
            \midrule
            MAE (eV/atom) & $1.86$ & $0.45$ & $0.23$ & $0.16$ & $0.04$ & $\cdots$ \\
            MARE (\%) & $32.4$ & $7.8$ & $3.9$ & $2.8$ & $0.8$ & $\cdots$ \\
            \bottomrule
        \end{tabular}
    \end{table}

    \section*{Supporting Information}

    See the Supporting Information for (i) geometry and basis-set data files; (ii) raw data for the molecular atomization energy and periodic EOS calculations; (iii) condition numbers of literature basis sets for solids in the Bosoni set; (iv) a breakdown of Bosoni-set PBE EOS errors by element; (v) HF EOS errors for the $7$ semiconductors; (vi) computational details of periodic local coupled cluster calculations; and (vii) the algorithm for optimizing auxiliary basis sets for density fitting.

    \section*{Conflict of interest}
    The authors declare no competing financial interest.

    \section*{Data availability}

    The data that support the findings of this study are available in the GitHub repository:
    \href{https://github.com/hongzhouye/supporting_data/tree/main/2026/ccpvxz_solids}{https://github.com/hongzhouye/supporting\_data/tree/main/\\2026/ccpvxz\_solids}

    \section*{Acknowledgments}

    This work was supported by the National Science Foundation under Grant No.~CHE-2543461.
    We acknowledge computing resources provided by the Division of Information Technology at the University of Maryland, College Park.

    \bibliography{refs}

\begin{thebibliography}{95}%
\makeatletter
\providecommand \@ifxundefined [1]{%
 \@ifx{#1\undefined}
}%
\providecommand \@ifnum [1]{%
 \ifnum #1\expandafter \@firstoftwo
 \else \expandafter \@secondoftwo
 \fi
}%
\providecommand \@ifx [1]{%
 \ifx #1\expandafter \@firstoftwo
 \else \expandafter \@secondoftwo
 \fi
}%
\providecommand \natexlab [1]{#1}%
\providecommand \enquote  [1]{``#1''}%
\providecommand \bibnamefont  [1]{#1}%
\providecommand \bibfnamefont [1]{#1}%
\providecommand \citenamefont [1]{#1}%
\providecommand \href@noop [0]{\@secondoftwo}%
\providecommand \href [0]{\begingroup \@sanitize@url \@href}%
\providecommand \@href[1]{\@@startlink{#1}\@@href}%
\providecommand \@@href[1]{\endgroup#1\@@endlink}%
\providecommand \@sanitize@url [0]{\catcode `\\12\catcode `\$12\catcode
  `\&12\catcode `\#12\catcode `\^12\catcode `\_12\catcode `\%12\relax}%
\providecommand \@@startlink[1]{}%
\providecommand \@@endlink[0]{}%
\providecommand \url  [0]{\begingroup\@sanitize@url \@url }%
\providecommand \@url [1]{\endgroup\@href {#1}{\urlprefix }}%
\providecommand \urlprefix  [0]{URL }%
\providecommand \Eprint [0]{\href }%
\providecommand \doibase [0]{https://doi.org/}%
\providecommand \selectlanguage [0]{\@gobble}%
\providecommand \bibinfo  [0]{\@secondoftwo}%
\providecommand \bibfield  [0]{\@secondoftwo}%
\providecommand \translation [1]{[#1]}%
\providecommand \BibitemOpen [0]{}%
\providecommand \bibitemStop [0]{}%
\providecommand \bibitemNoStop [0]{.\EOS\space}%
\providecommand \EOS [0]{\spacefactor3000\relax}%
\providecommand \BibitemShut  [1]{\csname bibitem#1\endcsname}%
\let\auto@bib@innerbib\@empty
\bibitem [{\citenamefont {Dunning}(1989)}]{Dunning89JCP}%
  \BibitemOpen
  \bibfield  {author} {\bibinfo {author} {\bibfnamefont {T.~H.}\ \bibnamefont
  {Dunning}},\ }\bibfield  {title} {\enquote {\bibinfo {title} {Gaussian basis
  sets for use in correlated molecular calculations. i. the atoms boron through
  neon and hydrogen},}\ }\href {https://doi.org/10.1063/1.456153} {\bibfield
  {journal} {\bibinfo  {journal} {J. Chem. Phys.}\ }\textbf {\bibinfo {volume}
  {90}},\ \bibinfo {pages} {1007--1023} (\bibinfo {year} {1989})}\BibitemShut
  {NoStop}%
\bibitem [{\citenamefont {Weigend}, \citenamefont {Furche},\ and\ \citenamefont
  {Ahlrichs}(2003)}]{Weigend03JCP}%
  \BibitemOpen
  \bibfield  {author} {\bibinfo {author} {\bibfnamefont {F.}~\bibnamefont
  {Weigend}}, \bibinfo {author} {\bibfnamefont {F.}~\bibnamefont {Furche}},\
  and\ \bibinfo {author} {\bibfnamefont {R.}~\bibnamefont {Ahlrichs}},\
  }\bibfield  {title} {\enquote {\bibinfo {title} {Gaussian basis sets of
  quadruple zeta valence quality for atoms h-kr},}\ }\href
  {https://doi.org/10.1063/1.1627293} {\bibfield  {journal} {\bibinfo
  {journal} {J. Chem. Phys.}\ }\textbf {\bibinfo {volume} {119}},\ \bibinfo
  {pages} {12753--12762} (\bibinfo {year} {2003})}\BibitemShut {NoStop}%
\bibitem [{\citenamefont {Weigend}\ and\ \citenamefont
  {Ahlrichs}(2005)}]{Weigend05PCCP}%
  \BibitemOpen
  \bibfield  {author} {\bibinfo {author} {\bibfnamefont {F.}~\bibnamefont
  {Weigend}}\ and\ \bibinfo {author} {\bibfnamefont {R.}~\bibnamefont
  {Ahlrichs}},\ }\bibfield  {title} {\enquote {\bibinfo {title} {Balanced basis
  sets of split valence{,} triple zeta valence and quadruple zeta valence
  quality for h to rn: Design and assessment of accuracy},}\ }\href
  {https://doi.org/10.1039/B508541A} {\bibfield  {journal} {\bibinfo  {journal}
  {Phys. Chem. Chem. Phys.}\ }\textbf {\bibinfo {volume} {7}},\ \bibinfo
  {pages} {3297--3305} (\bibinfo {year} {2005})}\BibitemShut {NoStop}%
\bibitem [{\citenamefont {Jensen}(2001)}]{Jensen01JCP}%
  \BibitemOpen
  \bibfield  {author} {\bibinfo {author} {\bibfnamefont {F.}~\bibnamefont
  {Jensen}},\ }\bibfield  {title} {\enquote {\bibinfo {title} {Polarization
  consistent basis sets: Principles},}\ }\href
  {https://doi.org/10.1063/1.1413524} {\bibfield  {journal} {\bibinfo
  {journal} {J. Chem. Phys.}\ }\textbf {\bibinfo {volume} {115}},\ \bibinfo
  {pages} {9113--9125} (\bibinfo {year} {2001})}\BibitemShut {NoStop}%
\bibitem [{\citenamefont {Jensen}(2002)}]{Jensen02JCP}%
  \BibitemOpen
  \bibfield  {author} {\bibinfo {author} {\bibfnamefont {F.}~\bibnamefont
  {Jensen}},\ }\bibfield  {title} {\enquote {\bibinfo {title} {Polarization
  consistent basis sets. ii. estimating the kohn–sham basis set limit},}\
  }\href {https://doi.org/10.1063/1.1465405} {\bibfield  {journal} {\bibinfo
  {journal} {J. Chem. Phys.}\ }\textbf {\bibinfo {volume} {116}},\ \bibinfo
  {pages} {7372--7379} (\bibinfo {year} {2002})}\BibitemShut {NoStop}%
\bibitem [{\citenamefont {Kendall}, \citenamefont {Dunning},\ and\
  \citenamefont {Harrison}(1992)}]{Kendall92JCP}%
  \BibitemOpen
  \bibfield  {author} {\bibinfo {author} {\bibfnamefont {R.~A.}\ \bibnamefont
  {Kendall}}, \bibinfo {author} {\bibfnamefont {T.~H.}\ \bibnamefont
  {Dunning}},\ and\ \bibinfo {author} {\bibfnamefont {R.~J.}\ \bibnamefont
  {Harrison}},\ }\bibfield  {title} {\enquote {\bibinfo {title} {Electron
  affinities of the first-row atoms revisited. systematic basis sets and wave
  functions},}\ }\href {https://doi.org/10.1063/1.462569} {\bibfield  {journal}
  {\bibinfo  {journal} {J. Chem. Phys.}\ }\textbf {\bibinfo {volume} {96}},\
  \bibinfo {pages} {6796--6806} (\bibinfo {year} {1992})}\BibitemShut {NoStop}%
\bibitem [{\citenamefont {Woon}\ and\ \citenamefont
  {Dunning}(1995)}]{Woon95JCP}%
  \BibitemOpen
  \bibfield  {author} {\bibinfo {author} {\bibfnamefont {D.~E.}\ \bibnamefont
  {Woon}}\ and\ \bibinfo {author} {\bibfnamefont {T.~H.}\ \bibnamefont
  {Dunning}},\ }\bibfield  {title} {\enquote {\bibinfo {title} {Gaussian basis
  sets for use in correlated molecular calculations. v. core-valence basis sets
  for boron through neon},}\ }\href {https://doi.org/10.1063/1.470645}
  {\bibfield  {journal} {\bibinfo  {journal} {J. Chem. Phys.}\ }\textbf
  {\bibinfo {volume} {103}},\ \bibinfo {pages} {4572--4585} (\bibinfo {year}
  {1995})}\BibitemShut {NoStop}%
\bibitem [{\citenamefont {de~Jong}, \citenamefont {Harrison},\ and\
  \citenamefont {Dixon}(2001)}]{Jong01JCP}%
  \BibitemOpen
  \bibfield  {author} {\bibinfo {author} {\bibfnamefont {W.~A.}\ \bibnamefont
  {de~Jong}}, \bibinfo {author} {\bibfnamefont {R.~J.}\ \bibnamefont
  {Harrison}},\ and\ \bibinfo {author} {\bibfnamefont {D.~A.}\ \bibnamefont
  {Dixon}},\ }\bibfield  {title} {\enquote {\bibinfo {title} {Parallel
  douglas-kroll energy and gradients in nwchem: Estimating scalar relativistic
  effects using douglas-kroll contracted basis sets},}\ }\href
  {https://doi.org/10.1063/1.1329891} {\bibfield  {journal} {\bibinfo
  {journal} {J. Chem. Phys.}\ }\textbf {\bibinfo {volume} {114}},\ \bibinfo
  {pages} {48} (\bibinfo {year} {2001})}\BibitemShut {NoStop}%
\bibitem [{\citenamefont {Pollak}\ and\ \citenamefont
  {Weigend}(2017)}]{Pollak17JCTC}%
  \BibitemOpen
  \bibfield  {author} {\bibinfo {author} {\bibfnamefont {P.}~\bibnamefont
  {Pollak}}\ and\ \bibinfo {author} {\bibfnamefont {F.}~\bibnamefont
  {Weigend}},\ }\bibfield  {title} {\enquote {\bibinfo {title} {Segmented
  contracted error-consistent basis sets of double- and triple-$\zeta$ valence
  quality for one- and two-component relativistic all-electron calculations},}\
  }\href {https://doi.org/10.1021/acs.jctc.7b00593} {\bibfield  {journal}
  {\bibinfo  {journal} {J. Chem. Theory Comput.}\ }\textbf {\bibinfo {volume}
  {13}},\ \bibinfo {pages} {3696--3705} (\bibinfo {year} {2017})}\BibitemShut
  {NoStop}%
\bibitem [{\citenamefont {Hill}(2013)}]{Hill13IJQC}%
  \BibitemOpen
  \bibfield  {author} {\bibinfo {author} {\bibfnamefont {J.~G.}\ \bibnamefont
  {Hill}},\ }\bibfield  {title} {\enquote {\bibinfo {title} {Gaussian basis
  sets for molecular applications},}\ }\href
  {https://doi.org/10.1002/qua.24355} {\bibfield  {journal} {\bibinfo
  {journal} {Int. J. Quantum Chem.}\ }\textbf {\bibinfo {volume} {113}},\
  \bibinfo {pages} {21--34} (\bibinfo {year} {2013})}\BibitemShut {NoStop}%
\bibitem [{\citenamefont {Peterson}(2007)}]{Peterson07ARCC}%
  \BibitemOpen
  \bibfield  {author} {\bibinfo {author} {\bibfnamefont {K.~A.}\ \bibnamefont
  {Peterson}},\ }\bibfield  {title} {\enquote {\bibinfo {title} {Gaussian basis
  sets exhibiting systematic convergence to the complete basis set limit},}\
  }in\ \href {https://doi.org/10.1016/S1574-1400(07)03011-3} {\emph {\bibinfo
  {booktitle} {Ann. Rep. Comput. Chem.}}},\ Vol.~\bibinfo {volume} {3}\
  (\bibinfo  {publisher} {Elsevier},\ \bibinfo {year} {2007})\ pp.\ \bibinfo
  {pages} {195--206}\BibitemShut {NoStop}%
\bibitem [{\citenamefont {Ye}\ and\ \citenamefont
  {Berkelbach}(2026)}]{Ye26arXiv}%
  \BibitemOpen
  \bibfield  {author} {\bibinfo {author} {\bibfnamefont {H.-Z.}\ \bibnamefont
  {Ye}}\ and\ \bibinfo {author} {\bibfnamefont {T.~C.}\ \bibnamefont
  {Berkelbach}},\ }\href {https://doi.org/10.48550/arXiv.2607.28554} {\enquote
  {\bibinfo {title} {Wavefunction-based periodic quantum chemistry},}\ }
  (\bibinfo {year} {2026}),\ \Eprint {https://arxiv.org/abs/2607.28554}
  {arXiv:2607.28554} \BibitemShut {NoStop}%
\bibitem [{\citenamefont {VandeVondele}\ and\ \citenamefont
  {Hutter}(2007)}]{VandeVondele07JCP}%
  \BibitemOpen
  \bibfield  {author} {\bibinfo {author} {\bibfnamefont {J.}~\bibnamefont
  {VandeVondele}}\ and\ \bibinfo {author} {\bibfnamefont {J.}~\bibnamefont
  {Hutter}},\ }\bibfield  {title} {\enquote {\bibinfo {title} {Gaussian basis
  sets for accurate calculations on molecular systems in gas and condensed
  phases},}\ }\href {https://doi.org/10.1063/1.2770708} {\bibfield  {journal}
  {\bibinfo  {journal} {J. Chem. Phys.}\ }\textbf {\bibinfo {volume} {127}},\
  \bibinfo {pages} {114105} (\bibinfo {year} {2007})}\BibitemShut {NoStop}%
\bibitem [{\citenamefont {Peintinger}, \citenamefont {Oliveira},\ and\
  \citenamefont {Bredow}(2013)}]{Peintinger13JCC}%
  \BibitemOpen
  \bibfield  {author} {\bibinfo {author} {\bibfnamefont {M.~F.}\ \bibnamefont
  {Peintinger}}, \bibinfo {author} {\bibfnamefont {D.~V.}\ \bibnamefont
  {Oliveira}},\ and\ \bibinfo {author} {\bibfnamefont {T.}~\bibnamefont
  {Bredow}},\ }\bibfield  {title} {\enquote {\bibinfo {title} {Consistent
  gaussian basis sets of triple-zeta valence with polarization quality for
  solid-state calculations},}\ }\href {https://doi.org/10.1002/jcc.23153}
  {\bibfield  {journal} {\bibinfo  {journal} {J. Comput. Chem.}\ }\textbf
  {\bibinfo {volume} {34}},\ \bibinfo {pages} {451--459} (\bibinfo {year}
  {2013})}\BibitemShut {NoStop}%
\bibitem [{\citenamefont {Ye}\ and\ \citenamefont
  {Berkelbach}(2022)}]{Ye22JCTC}%
  \BibitemOpen
  \bibfield  {author} {\bibinfo {author} {\bibfnamefont {H.-Z.}\ \bibnamefont
  {Ye}}\ and\ \bibinfo {author} {\bibfnamefont {T.~C.}\ \bibnamefont
  {Berkelbach}},\ }\bibfield  {title} {\enquote {\bibinfo {title}
  {Correlation-consistent gaussian basis sets for solids made simple},}\ }\href
  {https://doi.org/10.1021/acs.jctc.1c01245} {\bibfield  {journal} {\bibinfo
  {journal} {J. Chem. Theory Comput.}\ }\textbf {\bibinfo {volume} {18}},\
  \bibinfo {pages} {1595--1606} (\bibinfo {year} {2022})}\BibitemShut {NoStop}%
\bibitem [{\citenamefont {Yu}\ \emph {et~al.}(2026)\citenamefont {Yu},
  \citenamefont {Zhang}, \citenamefont {Zhang}, \citenamefont {Cao},
  \citenamefont {Sun},\ and\ \citenamefont {Ye}}]{Yu26arXiv}%
  \BibitemOpen
  \bibfield  {author} {\bibinfo {author} {\bibfnamefont {J.}~\bibnamefont
  {Yu}}, \bibinfo {author} {\bibfnamefont {X.}~\bibnamefont {Zhang}}, \bibinfo
  {author} {\bibfnamefont {M.-Y.}\ \bibnamefont {Zhang}}, \bibinfo {author}
  {\bibfnamefont {Y.}~\bibnamefont {Cao}}, \bibinfo {author} {\bibfnamefont
  {Q.}~\bibnamefont {Sun}},\ and\ \bibinfo {author} {\bibfnamefont {H.-Z.}\
  \bibnamefont {Ye}},\ }\href {https://doi.org/10.48550/arXiv.2607.11711}
  {\enquote {\bibinfo {title} {Correlation-consistent gaussian basis sets for
  copper solids from material-constrained atomic optimization},}\ } (\bibinfo
  {year} {2026}),\ \Eprint {https://arxiv.org/abs/2607.11711}
  {arXiv:2607.11711} \BibitemShut {NoStop}%
\bibitem [{\citenamefont {Szabo}\ and\ \citenamefont
  {Ostlund}(1996)}]{Szabo96Book}%
  \BibitemOpen
  \bibfield  {author} {\bibinfo {author} {\bibfnamefont {A.}~\bibnamefont
  {Szabo}}\ and\ \bibinfo {author} {\bibfnamefont {N.~S.}\ \bibnamefont
  {Ostlund}},\ }\href@noop {} {\emph {\bibinfo {title} {Modern Quantum
  Chemistry: Introduction to Advanced Electronic Structure Theory}}}\ (\bibinfo
   {publisher} {Dover Publications},\ \bibinfo {address} {Mineola, New York},\
  \bibinfo {year} {1996})\BibitemShut {NoStop}%
\bibitem [{\citenamefont {Hohenberg}\ and\ \citenamefont
  {Kohn}(1964)}]{Hohenberg64PR}%
  \BibitemOpen
  \bibfield  {author} {\bibinfo {author} {\bibfnamefont {P.}~\bibnamefont
  {Hohenberg}}\ and\ \bibinfo {author} {\bibfnamefont {W.}~\bibnamefont
  {Kohn}},\ }\bibfield  {title} {\enquote {\bibinfo {title} {Inhomogeneous
  electron gas},}\ }\href {https://doi.org/10.1103/PhysRev.136.B864} {\bibfield
   {journal} {\bibinfo  {journal} {Phys. Rev.}\ }\textbf {\bibinfo {volume}
  {136}},\ \bibinfo {pages} {B864--B871} (\bibinfo {year} {1964})}\BibitemShut
  {NoStop}%
\bibitem [{\citenamefont {Kohn}\ and\ \citenamefont {Sham}(1965)}]{Kohn65PR}%
  \BibitemOpen
  \bibfield  {author} {\bibinfo {author} {\bibfnamefont {W.}~\bibnamefont
  {Kohn}}\ and\ \bibinfo {author} {\bibfnamefont {L.~J.}\ \bibnamefont
  {Sham}},\ }\bibfield  {title} {\enquote {\bibinfo {title} {Self-consistent
  equations including exchange and correlation effects},}\ }\href
  {https://doi.org/10.1103/PhysRev.140.A1133} {\bibfield  {journal} {\bibinfo
  {journal} {Phys. Rev.}\ }\textbf {\bibinfo {volume} {140}},\ \bibinfo {pages}
  {A1133--A1138} (\bibinfo {year} {1965})}\BibitemShut {NoStop}%
\bibitem [{\citenamefont {Pisani}, \citenamefont {Dovesi},\ and\ \citenamefont
  {Roetti}(1988)}]{Pisani88Book}%
  \BibitemOpen
  \bibfield  {author} {\bibinfo {author} {\bibfnamefont {C.}~\bibnamefont
  {Pisani}}, \bibinfo {author} {\bibfnamefont {R.}~\bibnamefont {Dovesi}},\
  and\ \bibinfo {author} {\bibfnamefont {C.}~\bibnamefont {Roetti}},\ }\href
  {https://doi.org/10.1007/978-3-642-93385-1} {\emph {\bibinfo {title}
  {{Hartree--Fock} Ab Initio Treatment of Crystalline Systems}}},\ \bibinfo
  {series} {Lecture Notes in Chemistry}, Vol.~\bibinfo {volume} {48}\ (\bibinfo
   {publisher} {Springer},\ \bibinfo {address} {Berlin, Heidelberg},\ \bibinfo
  {year} {1988})\BibitemShut {NoStop}%
\bibitem [{\citenamefont {Pisani}(1996)}]{Pisani96Inbook}%
  \BibitemOpen
  \bibfield  {author} {\bibinfo {author} {\bibfnamefont {C.}~\bibnamefont
  {Pisani}},\ }\enquote {\bibinfo {title} {Ab-initio approaches to the
  quantum-mechanical treatment of periodic systems},}\ in\ \href
  {https://doi.org/10.1007/978-3-642-61478-1_3} {\emph {\bibinfo {booktitle}
  {Quantum-Mechanical Ab-initio Calculation of the Properties of Crystalline
  Materials}}},\ \bibinfo {editor} {edited by\ \bibinfo {editor} {\bibfnamefont
  {C.}~\bibnamefont {Pisani}}}\ (\bibinfo  {publisher} {Springer Berlin
  Heidelberg},\ \bibinfo {address} {Berlin, Heidelberg},\ \bibinfo {year}
  {1996})\ pp.\ \bibinfo {pages} {47--75}\BibitemShut {NoStop}%
\bibitem [{\citenamefont {Geipel}\ and\ \citenamefont
  {He{\ss}}(1997)}]{Geipel97CPL}%
  \BibitemOpen
  \bibfield  {author} {\bibinfo {author} {\bibfnamefont {N.~J.~M.}\
  \bibnamefont {Geipel}}\ and\ \bibinfo {author} {\bibfnamefont {B.~A.}\
  \bibnamefont {He{\ss}}},\ }\bibfield  {title} {\enquote {\bibinfo {title}
  {Scalar-relativistic effects in solids in the framework of a douglas-kroll
  transformed dirac-coulomb hamiltonian},}\ }\href
  {https://doi.org/10.1016/S0009-2614(97)00553-8} {\bibfield  {journal}
  {\bibinfo  {journal} {Chem. Phys. Lett.}\ }\textbf {\bibinfo {volume}
  {273}},\ \bibinfo {pages} {62--70} (\bibinfo {year} {1997})}\BibitemShut
  {NoStop}%
\bibitem [{\citenamefont {Kadek}, \citenamefont {Repisky},\ and\ \citenamefont
  {Ruud}(2019)}]{Kadek19PRB}%
  \BibitemOpen
  \bibfield  {author} {\bibinfo {author} {\bibfnamefont {M.}~\bibnamefont
  {Kadek}}, \bibinfo {author} {\bibfnamefont {M.}~\bibnamefont {Repisky}},\
  and\ \bibinfo {author} {\bibfnamefont {K.}~\bibnamefont {Ruud}},\ }\bibfield
  {title} {\enquote {\bibinfo {title} {All-electron fully relativistic
  kohn-sham theory for solids based on the dirac-coulomb hamiltonian and
  gaussian-type functions},}\ }\href
  {https://doi.org/10.1103/PhysRevB.99.205103} {\bibfield  {journal} {\bibinfo
  {journal} {Phys. Rev. B}\ }\textbf {\bibinfo {volume} {99}},\ \bibinfo
  {pages} {205103} (\bibinfo {year} {2019})}\BibitemShut {NoStop}%
\bibitem [{\citenamefont {Goldzak}\ \emph {et~al.}(2022)\citenamefont
  {Goldzak}, \citenamefont {Wang}, \citenamefont {Ye},\ and\ \citenamefont
  {Berkelbach}}]{Goldzak22JCP}%
  \BibitemOpen
  \bibfield  {author} {\bibinfo {author} {\bibfnamefont {T.}~\bibnamefont
  {Goldzak}}, \bibinfo {author} {\bibfnamefont {X.}~\bibnamefont {Wang}},
  \bibinfo {author} {\bibfnamefont {H.-Z.}\ \bibnamefont {Ye}},\ and\ \bibinfo
  {author} {\bibfnamefont {T.~C.}\ \bibnamefont {Berkelbach}},\ }\bibfield
  {title} {\enquote {\bibinfo {title} {Accurate thermochemistry of covalent and
  ionic solids from spin-component-scaled mp2},}\ }\href
  {https://doi.org/10.1063/5.0119633} {\bibfield  {journal} {\bibinfo
  {journal} {J. Chem. Phys.}\ }\textbf {\bibinfo {volume} {157}},\ \bibinfo
  {pages} {174112} (\bibinfo {year} {2022})}\BibitemShut {NoStop}%
\bibitem [{\citenamefont {Li}\ \emph {et~al.}(2026{\natexlab{a}})\citenamefont
  {Li}, \citenamefont {Zhai}, \citenamefont {Evangelista},\ and\ \citenamefont
  {Berkelbach}}]{Li26npjCM}%
  \BibitemOpen
  \bibfield  {author} {\bibinfo {author} {\bibfnamefont {S.}~\bibnamefont
  {Li}}, \bibinfo {author} {\bibfnamefont {H.}~\bibnamefont {Zhai}}, \bibinfo
  {author} {\bibfnamefont {F.~A.}\ \bibnamefont {Evangelista}},\ and\ \bibinfo
  {author} {\bibfnamefont {T.~C.}\ \bibnamefont {Berkelbach}},\ }\bibfield
  {title} {\enquote {\bibinfo {title} {Reaching the thermodynamic limit of
  periodic ccsd cohesive energies and band gaps},}\ }\href
  {https://doi.org/10.1038/s41524-026-02372-8} {\bibfield  {journal} {\bibinfo
  {journal} {npj Comput. Mater.}\ } (\bibinfo {year} {2026}{\natexlab{a}}),\
  10.1038/s41524-026-02372-8}\BibitemShut {NoStop}%
\bibitem [{\citenamefont {Kudin}\ and\ \citenamefont
  {Scuseria}(2000)}]{Kudin00PRB}%
  \BibitemOpen
  \bibfield  {author} {\bibinfo {author} {\bibfnamefont {K.~N.}\ \bibnamefont
  {Kudin}}\ and\ \bibinfo {author} {\bibfnamefont {G.~E.}\ \bibnamefont
  {Scuseria}},\ }\bibfield  {title} {\enquote {\bibinfo {title} {Linear-scaling
  density-functional theory with gaussian orbitals and periodic boundary
  conditions: Efficient evaluation of energy and forces via the fast multipole
  method},}\ }\href {https://doi.org/10.1103/PhysRevB.61.16440} {\bibfield
  {journal} {\bibinfo  {journal} {Phys. Rev. B}\ }\textbf {\bibinfo {volume}
  {61}},\ \bibinfo {pages} {16440--16453} (\bibinfo {year} {2000})}\BibitemShut
  {NoStop}%
\bibitem [{\citenamefont {Lee}\ \emph {et~al.}(2021)\citenamefont {Lee},
  \citenamefont {Feng}, \citenamefont {Cunha}, \citenamefont {Gonthier},
  \citenamefont {Epifanovsky},\ and\ \citenamefont {Head-Gordon}}]{Lee21JCP}%
  \BibitemOpen
  \bibfield  {author} {\bibinfo {author} {\bibfnamefont {J.}~\bibnamefont
  {Lee}}, \bibinfo {author} {\bibfnamefont {X.}~\bibnamefont {Feng}}, \bibinfo
  {author} {\bibfnamefont {L.~A.}\ \bibnamefont {Cunha}}, \bibinfo {author}
  {\bibfnamefont {J.~F.}\ \bibnamefont {Gonthier}}, \bibinfo {author}
  {\bibfnamefont {E.}~\bibnamefont {Epifanovsky}},\ and\ \bibinfo {author}
  {\bibfnamefont {M.}~\bibnamefont {Head-Gordon}},\ }\bibfield  {title}
  {\enquote {\bibinfo {title} {Approaching the basis set limit in
  gaussian-orbital-based periodic calculations with transferability:
  Performance of pure density functionals for simple semiconductors},}\ }\href
  {https://doi.org/10.1063/5.0069177} {\bibfield  {journal} {\bibinfo
  {journal} {J. Chem. Phys.}\ }\textbf {\bibinfo {volume} {155}},\ \bibinfo
  {pages} {164102} (\bibinfo {year} {2021})}\BibitemShut {NoStop}%
\bibitem [{\citenamefont {Li}\ \emph {et~al.}(2021)\citenamefont {Li},
  \citenamefont {Chen}, \citenamefont {Rossomme}, \citenamefont {Head-Gordon},\
  and\ \citenamefont {Head-Gordon}}]{Li21JPCL}%
  \BibitemOpen
  \bibfield  {author} {\bibinfo {author} {\bibfnamefont {W.-L.}\ \bibnamefont
  {Li}}, \bibinfo {author} {\bibfnamefont {K.}~\bibnamefont {Chen}}, \bibinfo
  {author} {\bibfnamefont {E.}~\bibnamefont {Rossomme}}, \bibinfo {author}
  {\bibfnamefont {M.}~\bibnamefont {Head-Gordon}},\ and\ \bibinfo {author}
  {\bibfnamefont {T.}~\bibnamefont {Head-Gordon}},\ }\bibfield  {title}
  {\enquote {\bibinfo {title} {Optimized pseudopotentials and basis sets for
  semiempirical density functional theory for electrocatalysis applications},}\
  }\href {https://doi.org/10.1021/acs.jpclett.1c02918} {\bibfield  {journal}
  {\bibinfo  {journal} {J. Phys. Chem. Lett.}\ }\textbf {\bibinfo {volume}
  {12}},\ \bibinfo {pages} {10304--10309} (\bibinfo {year} {2021})}\BibitemShut
  {NoStop}%
\bibitem [{\citenamefont {Mirhosseini}\ \emph {et~al.}(2026)\citenamefont
  {Mirhosseini}, \citenamefont {M\"{u}ller}, \citenamefont {Krack},
  \citenamefont {Kühne},\ and\ \citenamefont {Hutter}}]{Mirhosseini26JCP}%
  \BibitemOpen
  \bibfield  {author} {\bibinfo {author} {\bibfnamefont {H.}~\bibnamefont
  {Mirhosseini}}, \bibinfo {author} {\bibfnamefont {T.~M.~A.}\ \bibnamefont
  {M\"{u}ller}}, \bibinfo {author} {\bibfnamefont {M.}~\bibnamefont {Krack}},
  \bibinfo {author} {\bibfnamefont {T.~D.}\ \bibnamefont {Kühne}},\ and\
  \bibinfo {author} {\bibfnamefont {J.}~\bibnamefont {Hutter}},\ }\bibfield
  {title} {\enquote {\bibinfo {title} {The uzh protocol: Separating errors and
  constructing improved cp2k basis sets and pseudopotentials},}\ }\href
  {https://doi.org/10.1063/5.0347392} {\bibfield  {journal} {\bibinfo
  {journal} {J. Chem. Phys.}\ }\textbf {\bibinfo {volume} {165}},\ \bibinfo
  {pages} {104103} (\bibinfo {year} {2026})}\BibitemShut {NoStop}%
\bibitem [{\citenamefont {Bosoni}\ \emph {et~al.}(2024)\citenamefont {Bosoni},
  \citenamefont {Beal}, \citenamefont {Bercx}, \citenamefont {Blaha},
  \citenamefont {Bl{\"u}gel}, \citenamefont {Br{\"o}der}, \citenamefont
  {Callsen}, \citenamefont {Cottenier}, \citenamefont {Degomme}, \citenamefont
  {Dikan}, \citenamefont {Eimre}, \citenamefont {Flage-Larsen}, \citenamefont
  {Fornari}, \citenamefont {Garcia}, \citenamefont {Genovese}, \citenamefont
  {Giantomassi}, \citenamefont {Huber}, \citenamefont {Janssen}, \citenamefont
  {Kastlunger}, \citenamefont {Krack}, \citenamefont {Kresse}, \citenamefont
  {K{\"u}hne}, \citenamefont {Lejaeghere}, \citenamefont {Madsen},
  \citenamefont {Marsman}, \citenamefont {Marzari}, \citenamefont {Michalicek},
  \citenamefont {Mirhosseini}, \citenamefont {M{\"u}ller}, \citenamefont
  {Petretto}, \citenamefont {Pickard}, \citenamefont {Ponc{\'e}}, \citenamefont
  {Rignanese}, \citenamefont {Rubel}, \citenamefont {Ruh}, \citenamefont
  {Sluydts}, \citenamefont {Vanpoucke}, \citenamefont {Vijay}, \citenamefont
  {Wolloch}, \citenamefont {Wortmann}, \citenamefont {Yakutovich},
  \citenamefont {Yu}, \citenamefont {Zadoks}, \citenamefont {Zhu},\ and\
  \citenamefont {Pizzi}}]{Bosoni24NRP}%
  \BibitemOpen
  \bibfield  {author} {\bibinfo {author} {\bibfnamefont {E.}~\bibnamefont
  {Bosoni}}, \bibinfo {author} {\bibfnamefont {L.}~\bibnamefont {Beal}},
  \bibinfo {author} {\bibfnamefont {M.}~\bibnamefont {Bercx}}, \bibinfo
  {author} {\bibfnamefont {P.}~\bibnamefont {Blaha}}, \bibinfo {author}
  {\bibfnamefont {S.}~\bibnamefont {Bl{\"u}gel}}, \bibinfo {author}
  {\bibfnamefont {J.}~\bibnamefont {Br{\"o}der}}, \bibinfo {author}
  {\bibfnamefont {M.}~\bibnamefont {Callsen}}, \bibinfo {author} {\bibfnamefont
  {S.}~\bibnamefont {Cottenier}}, \bibinfo {author} {\bibfnamefont
  {A.}~\bibnamefont {Degomme}}, \bibinfo {author} {\bibfnamefont
  {V.}~\bibnamefont {Dikan}}, \bibinfo {author} {\bibfnamefont
  {K.}~\bibnamefont {Eimre}}, \bibinfo {author} {\bibfnamefont
  {E.}~\bibnamefont {Flage-Larsen}}, \bibinfo {author} {\bibfnamefont
  {M.}~\bibnamefont {Fornari}}, \bibinfo {author} {\bibfnamefont
  {A.}~\bibnamefont {Garcia}}, \bibinfo {author} {\bibfnamefont
  {L.}~\bibnamefont {Genovese}}, \bibinfo {author} {\bibfnamefont
  {M.}~\bibnamefont {Giantomassi}}, \bibinfo {author} {\bibfnamefont {S.~P.}\
  \bibnamefont {Huber}}, \bibinfo {author} {\bibfnamefont {H.}~\bibnamefont
  {Janssen}}, \bibinfo {author} {\bibfnamefont {G.}~\bibnamefont {Kastlunger}},
  \bibinfo {author} {\bibfnamefont {M.}~\bibnamefont {Krack}}, \bibinfo
  {author} {\bibfnamefont {G.}~\bibnamefont {Kresse}}, \bibinfo {author}
  {\bibfnamefont {T.~D.}\ \bibnamefont {K{\"u}hne}}, \bibinfo {author}
  {\bibfnamefont {K.}~\bibnamefont {Lejaeghere}}, \bibinfo {author}
  {\bibfnamefont {G.~K.~H.}\ \bibnamefont {Madsen}}, \bibinfo {author}
  {\bibfnamefont {M.}~\bibnamefont {Marsman}}, \bibinfo {author} {\bibfnamefont
  {N.}~\bibnamefont {Marzari}}, \bibinfo {author} {\bibfnamefont
  {G.}~\bibnamefont {Michalicek}}, \bibinfo {author} {\bibfnamefont
  {H.}~\bibnamefont {Mirhosseini}}, \bibinfo {author} {\bibfnamefont
  {T.~M.~A.}\ \bibnamefont {M{\"u}ller}}, \bibinfo {author} {\bibfnamefont
  {G.}~\bibnamefont {Petretto}}, \bibinfo {author} {\bibfnamefont {C.~J.}\
  \bibnamefont {Pickard}}, \bibinfo {author} {\bibfnamefont {S.}~\bibnamefont
  {Ponc{\'e}}}, \bibinfo {author} {\bibfnamefont {G.-M.}\ \bibnamefont
  {Rignanese}}, \bibinfo {author} {\bibfnamefont {O.}~\bibnamefont {Rubel}},
  \bibinfo {author} {\bibfnamefont {T.}~\bibnamefont {Ruh}}, \bibinfo {author}
  {\bibfnamefont {M.}~\bibnamefont {Sluydts}}, \bibinfo {author} {\bibfnamefont
  {D.~E.~P.}\ \bibnamefont {Vanpoucke}}, \bibinfo {author} {\bibfnamefont
  {S.}~\bibnamefont {Vijay}}, \bibinfo {author} {\bibfnamefont
  {M.}~\bibnamefont {Wolloch}}, \bibinfo {author} {\bibfnamefont
  {D.}~\bibnamefont {Wortmann}}, \bibinfo {author} {\bibfnamefont {A.~V.}\
  \bibnamefont {Yakutovich}}, \bibinfo {author} {\bibfnamefont
  {J.}~\bibnamefont {Yu}}, \bibinfo {author} {\bibfnamefont {A.}~\bibnamefont
  {Zadoks}}, \bibinfo {author} {\bibfnamefont {B.}~\bibnamefont {Zhu}},\ and\
  \bibinfo {author} {\bibfnamefont {G.}~\bibnamefont {Pizzi}},\ }\bibfield
  {title} {\enquote {\bibinfo {title} {How to verify the precision of
  density-functional-theory implementations via reproducible and universal
  workflows},}\ }\href {https://doi.org/10.1038/s42254-023-00655-3} {\bibfield
  {journal} {\bibinfo  {journal} {Nat. Rev. Phys.}\ }\textbf {\bibinfo {volume}
  {6}},\ \bibinfo {pages} {45--58} (\bibinfo {year} {2024})}\BibitemShut
  {NoStop}%
\bibitem [{\citenamefont {Daga}, \citenamefont {Civalleri},\ and\ \citenamefont
  {Maschio}(2020)}]{Daga20JCTC}%
  \BibitemOpen
  \bibfield  {author} {\bibinfo {author} {\bibfnamefont {L.~E.}\ \bibnamefont
  {Daga}}, \bibinfo {author} {\bibfnamefont {B.}~\bibnamefont {Civalleri}},\
  and\ \bibinfo {author} {\bibfnamefont {L.}~\bibnamefont {Maschio}},\
  }\bibfield  {title} {\enquote {\bibinfo {title} {Gaussian basis sets for
  crystalline solids: All-purpose basis set libraries vs system-specific
  optimizations},}\ }\href {https://doi.org/10.1021/acs.jctc.9b01004}
  {\bibfield  {journal} {\bibinfo  {journal} {J. Chem. Theory Comput.}\
  }\textbf {\bibinfo {volume} {16}},\ \bibinfo {pages} {2192--2201} (\bibinfo
  {year} {2020})}\BibitemShut {NoStop}%
\bibitem [{\citenamefont {Zhou}, \citenamefont {Gull},\ and\ \citenamefont
  {Zgid}(2021)}]{Zhou21JCTC}%
  \BibitemOpen
  \bibfield  {author} {\bibinfo {author} {\bibfnamefont {Y.}~\bibnamefont
  {Zhou}}, \bibinfo {author} {\bibfnamefont {E.}~\bibnamefont {Gull}},\ and\
  \bibinfo {author} {\bibfnamefont {D.}~\bibnamefont {Zgid}},\ }\bibfield
  {title} {\enquote {\bibinfo {title} {Material-specific optimization of
  gaussian basis sets against plane wave data},}\ }\href
  {https://doi.org/10.1021/acs.jctc.1c00491} {\bibfield  {journal} {\bibinfo
  {journal} {J. Chem. Theory Comput.}\ }\textbf {\bibinfo {volume} {17}},\
  \bibinfo {pages} {5611--5622} (\bibinfo {year} {2021})}\BibitemShut {NoStop}%
\bibitem [{\citenamefont {Neufeld}, \citenamefont {Ye},\ and\ \citenamefont
  {Berkelbach}(2022)}]{Neufeld22JPCL}%
  \BibitemOpen
  \bibfield  {author} {\bibinfo {author} {\bibfnamefont {V.~A.}\ \bibnamefont
  {Neufeld}}, \bibinfo {author} {\bibfnamefont {H.-Z.}\ \bibnamefont {Ye}},\
  and\ \bibinfo {author} {\bibfnamefont {T.~C.}\ \bibnamefont {Berkelbach}},\
  }\bibfield  {title} {\enquote {\bibinfo {title} {Ground-state properties of
  metallic solids from ab initio coupled-cluster theory},}\ }\href
  {https://doi.org/10.1021/acs.jpclett.2c01828} {\bibfield  {journal} {\bibinfo
   {journal} {J. Phys. Chem. Lett.}\ }\textbf {\bibinfo {volume} {13}},\
  \bibinfo {pages} {7497--7503} (\bibinfo {year} {2022})}\BibitemShut {NoStop}%
\bibitem [{\citenamefont {Dong}\ \emph {et~al.}(2024)\citenamefont {Dong},
  \citenamefont {Jiang}, \citenamefont {Li}, \citenamefont {Pei}, \citenamefont
  {Han}, \citenamefont {Xu}, \citenamefont {Garoufalis}, \citenamefont
  {Baskoutas}, \citenamefont {Zhou}, \citenamefont {Zhao},\ and\ \citenamefont
  {Zeng}}]{Dong24PRB}%
  \BibitemOpen
  \bibfield  {author} {\bibinfo {author} {\bibfnamefont {T.}~\bibnamefont
  {Dong}}, \bibinfo {author} {\bibfnamefont {Y.}~\bibnamefont {Jiang}},
  \bibinfo {author} {\bibfnamefont {Y.}~\bibnamefont {Li}}, \bibinfo {author}
  {\bibfnamefont {M.}~\bibnamefont {Pei}}, \bibinfo {author} {\bibfnamefont
  {P.}~\bibnamefont {Han}}, \bibinfo {author} {\bibfnamefont {D.}~\bibnamefont
  {Xu}}, \bibinfo {author} {\bibfnamefont {C.~S.}\ \bibnamefont {Garoufalis}},
  \bibinfo {author} {\bibfnamefont {S.}~\bibnamefont {Baskoutas}}, \bibinfo
  {author} {\bibfnamefont {S.}~\bibnamefont {Zhou}}, \bibinfo {author}
  {\bibfnamefont {J.}~\bibnamefont {Zhao}},\ and\ \bibinfo {author}
  {\bibfnamefont {Z.}~\bibnamefont {Zeng}},\ }\bibfield  {title} {\enquote
  {\bibinfo {title} {Dimensionally consistent gaussian basis set bridging bulk
  solids and nanomaterials},}\ }\href
  {https://doi.org/10.1103/PhysRevB.110.245415} {\bibfield  {journal} {\bibinfo
   {journal} {Phys. Rev. B}\ }\textbf {\bibinfo {volume} {110}},\ \bibinfo
  {pages} {245415} (\bibinfo {year} {2024})}\BibitemShut {NoStop}%
\bibitem [{\citenamefont {Erba}\ \emph {et~al.}(2023)\citenamefont {Erba},
  \citenamefont {Desmarais}, \citenamefont {Casassa}, \citenamefont
  {Civalleri}, \citenamefont {Donà}, \citenamefont {Bush}, \citenamefont
  {Searle}, \citenamefont {Maschio}, \citenamefont {Edith-Daga}, \citenamefont
  {Cossard}, \citenamefont {Ribaldone}, \citenamefont {Ascrizzi}, \citenamefont
  {Marana}, \citenamefont {Flament},\ and\ \citenamefont
  {Kirtman}}]{Erba23JCTC}%
  \BibitemOpen
  \bibfield  {author} {\bibinfo {author} {\bibfnamefont {A.}~\bibnamefont
  {Erba}}, \bibinfo {author} {\bibfnamefont {J.~K.}\ \bibnamefont {Desmarais}},
  \bibinfo {author} {\bibfnamefont {S.}~\bibnamefont {Casassa}}, \bibinfo
  {author} {\bibfnamefont {B.}~\bibnamefont {Civalleri}}, \bibinfo {author}
  {\bibfnamefont {L.}~\bibnamefont {Donà}}, \bibinfo {author} {\bibfnamefont
  {I.~J.}\ \bibnamefont {Bush}}, \bibinfo {author} {\bibfnamefont
  {B.}~\bibnamefont {Searle}}, \bibinfo {author} {\bibfnamefont
  {L.}~\bibnamefont {Maschio}}, \bibinfo {author} {\bibfnamefont
  {L.}~\bibnamefont {Edith-Daga}}, \bibinfo {author} {\bibfnamefont
  {A.}~\bibnamefont {Cossard}}, \bibinfo {author} {\bibfnamefont
  {C.}~\bibnamefont {Ribaldone}}, \bibinfo {author} {\bibfnamefont
  {E.}~\bibnamefont {Ascrizzi}}, \bibinfo {author} {\bibfnamefont {N.~L.}\
  \bibnamefont {Marana}}, \bibinfo {author} {\bibfnamefont {J.-P.}\
  \bibnamefont {Flament}},\ and\ \bibinfo {author} {\bibfnamefont
  {B.}~\bibnamefont {Kirtman}},\ }\bibfield  {title} {\enquote {\bibinfo
  {title} {Crystal23: A program for computational solid state physics and
  chemistry},}\ }\href {https://doi.org/10.1021/acs.jctc.2c00958} {\bibfield
  {journal} {\bibinfo  {journal} {J. Chem. Theory Comput.}\ }\textbf {\bibinfo
  {volume} {19}},\ \bibinfo {pages} {6891--6932} (\bibinfo {year}
  {2023})}\BibitemShut {NoStop}%
\bibitem [{\citenamefont {Vilela~Oliveira}\ \emph {et~al.}(2019)\citenamefont
  {Vilela~Oliveira}, \citenamefont {Laun}, \citenamefont {Peintinger},\ and\
  \citenamefont {Bredow}}]{Oliveira19JCC}%
  \BibitemOpen
  \bibfield  {author} {\bibinfo {author} {\bibfnamefont {D.}~\bibnamefont
  {Vilela~Oliveira}}, \bibinfo {author} {\bibfnamefont {J.}~\bibnamefont
  {Laun}}, \bibinfo {author} {\bibfnamefont {M.~F.}\ \bibnamefont
  {Peintinger}},\ and\ \bibinfo {author} {\bibfnamefont {T.}~\bibnamefont
  {Bredow}},\ }\bibfield  {title} {\enquote {\bibinfo {title} {Bsse-correction
  scheme for consistent gaussian basis sets of double- and triple-zeta valence
  with polarization quality for solid-state calculations},}\ }\href
  {https://doi.org/10.1002/jcc.26013} {\bibfield  {journal} {\bibinfo
  {journal} {J. Comput. Chem.}\ }\textbf {\bibinfo {volume} {40}},\ \bibinfo
  {pages} {2364--2376} (\bibinfo {year} {2019})}\BibitemShut {NoStop}%
\bibitem [{\citenamefont {Morales}\ and\ \citenamefont
  {Malone}(2020)}]{Morales20JCP}%
  \BibitemOpen
  \bibfield  {author} {\bibinfo {author} {\bibfnamefont {M.~A.}\ \bibnamefont
  {Morales}}\ and\ \bibinfo {author} {\bibfnamefont {F.~D.}\ \bibnamefont
  {Malone}},\ }\bibfield  {title} {\enquote {\bibinfo {title} {Accelerating the
  convergence of auxiliary-field quantum monte carlo in solids with optimized
  gaussian basis sets},}\ }\href {https://doi.org/10.1063/5.0025390} {\bibfield
   {journal} {\bibinfo  {journal} {J. Chem. Phys.}\ }\textbf {\bibinfo {volume}
  {153}},\ \bibinfo {pages} {194111} (\bibinfo {year} {2020})}\BibitemShut
  {NoStop}%
\bibitem [{\citenamefont {Goedecker}, \citenamefont {Teter},\ and\
  \citenamefont {Hutter}(1996)}]{Goedecker96PRB}%
  \BibitemOpen
  \bibfield  {author} {\bibinfo {author} {\bibfnamefont {S.}~\bibnamefont
  {Goedecker}}, \bibinfo {author} {\bibfnamefont {M.}~\bibnamefont {Teter}},\
  and\ \bibinfo {author} {\bibfnamefont {J.}~\bibnamefont {Hutter}},\
  }\bibfield  {title} {\enquote {\bibinfo {title} {Separable dual-space
  gaussian pseudopotentials},}\ }\href
  {https://doi.org/10.1103/PhysRevB.54.1703} {\bibfield  {journal} {\bibinfo
  {journal} {Phys. Rev. B}\ }\textbf {\bibinfo {volume} {54}},\ \bibinfo
  {pages} {1703--1710} (\bibinfo {year} {1996})}\BibitemShut {NoStop}%
\bibitem [{\citenamefont {Hartwigsen}, \citenamefont {Goedecker},\ and\
  \citenamefont {Hutter}(1998)}]{Hartwigsen98PRB}%
  \BibitemOpen
  \bibfield  {author} {\bibinfo {author} {\bibfnamefont {C.}~\bibnamefont
  {Hartwigsen}}, \bibinfo {author} {\bibfnamefont {S.}~\bibnamefont
  {Goedecker}},\ and\ \bibinfo {author} {\bibfnamefont {J.}~\bibnamefont
  {Hutter}},\ }\bibfield  {title} {\enquote {\bibinfo {title} {Relativistic
  separable dual-space gaussian pseudopotentials from h to rn},}\ }\href
  {https://doi.org/10.1103/PhysRevB.58.3641} {\bibfield  {journal} {\bibinfo
  {journal} {Phys. Rev. B}\ }\textbf {\bibinfo {volume} {58}},\ \bibinfo
  {pages} {3641--3662} (\bibinfo {year} {1998})}\BibitemShut {NoStop}%
\bibitem [{\citenamefont {Ye}\ and\ \citenamefont
  {Berkelbach}(2024)}]{Ye24JCTC}%
  \BibitemOpen
  \bibfield  {author} {\bibinfo {author} {\bibfnamefont {H.-Z.}\ \bibnamefont
  {Ye}}\ and\ \bibinfo {author} {\bibfnamefont {T.~C.}\ \bibnamefont
  {Berkelbach}},\ }\bibfield  {title} {\enquote {\bibinfo {title} {Periodic
  local coupled-cluster theory for insulators and metals},}\ }\href
  {https://doi.org/10.1021/acs.jctc.4c00936} {\bibfield  {journal} {\bibinfo
  {journal} {J. Chem. Theory Comput.}\ }\textbf {\bibinfo {volume} {20}},\
  \bibinfo {pages} {8948--8959} (\bibinfo {year} {2024})}\BibitemShut {NoStop}%
\bibitem [{\citenamefont {Haritan}, \citenamefont {Wang},\ and\ \citenamefont
  {Goldzak}(2025)}]{Haritan25JCTC}%
  \BibitemOpen
  \bibfield  {author} {\bibinfo {author} {\bibfnamefont {I.}~\bibnamefont
  {Haritan}}, \bibinfo {author} {\bibfnamefont {X.}~\bibnamefont {Wang}},\ and\
  \bibinfo {author} {\bibfnamefont {T.}~\bibnamefont {Goldzak}},\ }\bibfield
  {title} {\enquote {\bibinfo {title} {An efficient scaled opposite-spin {MP2}
  method for periodic systems},}\ }\href
  {https://doi.org/10.1021/acs.jctc.5c00394} {\bibfield  {journal} {\bibinfo
  {journal} {J. Chem. Theory Comput.}\ }\textbf {\bibinfo {volume} {21}},\
  \bibinfo {pages} {6823--6833} (\bibinfo {year} {2025})}\BibitemShut {NoStop}%
\bibitem [{\citenamefont {Zhang}\ \emph {et~al.}(2026)\citenamefont {Zhang},
  \citenamefont {Chen}, \citenamefont {Rettig}, \citenamefont {Jiang},
  \citenamefont {Robinson}, \citenamefont {Dinh}, \citenamefont {Ni},\ and\
  \citenamefont {Lee}}]{Zhang26PRX}%
  \BibitemOpen
  \bibfield  {author} {\bibinfo {author} {\bibfnamefont {J.}~\bibnamefont
  {Zhang}}, \bibinfo {author} {\bibfnamefont {M.-F.}\ \bibnamefont {Chen}},
  \bibinfo {author} {\bibfnamefont {A.}~\bibnamefont {Rettig}}, \bibinfo
  {author} {\bibfnamefont {T.}~\bibnamefont {Jiang}}, \bibinfo {author}
  {\bibfnamefont {P.~J.}\ \bibnamefont {Robinson}}, \bibinfo {author}
  {\bibfnamefont {H.~Q.}\ \bibnamefont {Dinh}}, \bibinfo {author}
  {\bibfnamefont {A.~Z.}\ \bibnamefont {Ni}},\ and\ \bibinfo {author}
  {\bibfnamefont {J.}~\bibnamefont {Lee}},\ }\bibfield  {title} {\enquote
  {\bibinfo {title} {Ab initio auxiliary-field quantum {Monte Carlo} in the
  thermodynamic limit},}\ }\href {https://doi.org/10.1103/dtly-b2lp} {\bibfield
   {journal} {\bibinfo  {journal} {Phys. Rev. X}\ }\textbf {\bibinfo {volume}
  {16}},\ \bibinfo {pages} {031044} (\bibinfo {year} {2026})}\BibitemShut
  {NoStop}%
\bibitem [{\citenamefont {Vo}, \citenamefont {Wang},\ and\ \citenamefont
  {Berkelbach}(2024)}]{Vo24JCP}%
  \BibitemOpen
  \bibfield  {author} {\bibinfo {author} {\bibfnamefont {E.~A.}\ \bibnamefont
  {Vo}}, \bibinfo {author} {\bibfnamefont {X.}~\bibnamefont {Wang}},\ and\
  \bibinfo {author} {\bibfnamefont {T.~C.}\ \bibnamefont {Berkelbach}},\
  }\bibfield  {title} {\enquote {\bibinfo {title} {Performance of periodic
  {EOM-CCSD} for bandgaps of inorganic semiconductors and insulators},}\ }\href
  {https://doi.org/10.1063/5.0187856} {\bibfield  {journal} {\bibinfo
  {journal} {J. Chem. Phys.}\ }\textbf {\bibinfo {volume} {160}},\ \bibinfo
  {pages} {044106} (\bibinfo {year} {2024})}\BibitemShut {NoStop}%
\bibitem [{\citenamefont {Banerjee}\ and\ \citenamefont
  {Sokolov}(2022)}]{Banerjee22JCTC}%
  \BibitemOpen
  \bibfield  {author} {\bibinfo {author} {\bibfnamefont {S.}~\bibnamefont
  {Banerjee}}\ and\ \bibinfo {author} {\bibfnamefont {A.~Y.}\ \bibnamefont
  {Sokolov}},\ }\bibfield  {title} {\enquote {\bibinfo {title} {Non-dyson
  algebraic diagrammatic construction theory for charged excitations in
  solids},}\ }\href {https://doi.org/10.1021/acs.jctc.2c00565} {\bibfield
  {journal} {\bibinfo  {journal} {J. Chem. Theory Comput.}\ }\textbf {\bibinfo
  {volume} {18}},\ \bibinfo {pages} {5337--5348} (\bibinfo {year}
  {2022})}\BibitemShut {NoStop}%
\bibitem [{\citenamefont {Laughon}, \citenamefont {Yu},\ and\ \citenamefont
  {Zhu}(2022)}]{Laughon22JPCL}%
  \BibitemOpen
  \bibfield  {author} {\bibinfo {author} {\bibfnamefont {K.}~\bibnamefont
  {Laughon}}, \bibinfo {author} {\bibfnamefont {J.~M.}\ \bibnamefont {Yu}},\
  and\ \bibinfo {author} {\bibfnamefont {T.}~\bibnamefont {Zhu}},\ }\bibfield
  {title} {\enquote {\bibinfo {title} {Periodic coupled-cluster {Green's}
  function for photoemission spectra of realistic solids},}\ }\href
  {https://doi.org/10.1021/acs.jpclett.2c02534} {\bibfield  {journal} {\bibinfo
   {journal} {J. Phys. Chem. Lett.}\ }\textbf {\bibinfo {volume} {13}},\
  \bibinfo {pages} {9122--9128} (\bibinfo {year} {2022})}\BibitemShut {NoStop}%
\bibitem [{\citenamefont {Li}\ and\ \citenamefont {Zhu}(2024)}]{Li24PRL}%
  \BibitemOpen
  \bibfield  {author} {\bibinfo {author} {\bibfnamefont {J.}~\bibnamefont
  {Li}}\ and\ \bibinfo {author} {\bibfnamefont {T.}~\bibnamefont {Zhu}},\
  }\bibfield  {title} {\enquote {\bibinfo {title} {Interacting-bath dynamical
  embedding for capturing nonlocal electron correlation in solids},}\ }\href
  {https://doi.org/10.1103/PhysRevLett.133.216402} {\bibfield  {journal}
  {\bibinfo  {journal} {Phys. Rev. Lett.}\ }\textbf {\bibinfo {volume} {133}},\
  \bibinfo {pages} {216402} (\bibinfo {year} {2024})}\BibitemShut {NoStop}%
\bibitem [{\citenamefont {Li}\ \emph {et~al.}(2026{\natexlab{b}})\citenamefont
  {Li}, \citenamefont {Hillenbrand}, \citenamefont {Venturella}, \citenamefont
  {Chen},\ and\ \citenamefont {Zhu}}]{Li26JCP}%
  \BibitemOpen
  \bibfield  {author} {\bibinfo {author} {\bibfnamefont {J.}~\bibnamefont
  {Li}}, \bibinfo {author} {\bibfnamefont {C.}~\bibnamefont {Hillenbrand}},
  \bibinfo {author} {\bibfnamefont {C.}~\bibnamefont {Venturella}}, \bibinfo
  {author} {\bibfnamefont {E.}~\bibnamefont {Chen}},\ and\ \bibinfo {author}
  {\bibfnamefont {T.}~\bibnamefont {Zhu}},\ }\bibfield  {title} {\enquote
  {\bibinfo {title} {Resolving finite-size errors in {EOM-CCSD} band gaps of
  solids with interacting-bath dynamical embedding theory},}\ }\href
  {https://doi.org/10.1063/5.0347878} {\bibfield  {journal} {\bibinfo
  {journal} {J. Chem. Phys.}\ }\textbf {\bibinfo {volume} {165}},\ \bibinfo
  {pages} {094108} (\bibinfo {year} {2026}{\natexlab{b}})}\BibitemShut
  {NoStop}%
\bibitem [{\citenamefont {Raghavachari}\ \emph {et~al.}(1989)\citenamefont
  {Raghavachari}, \citenamefont {Trucks}, \citenamefont {Pople},\ and\
  \citenamefont {Head-Gordon}}]{Raghavachari89CPL}%
  \BibitemOpen
  \bibfield  {author} {\bibinfo {author} {\bibfnamefont {K.}~\bibnamefont
  {Raghavachari}}, \bibinfo {author} {\bibfnamefont {G.~W.}\ \bibnamefont
  {Trucks}}, \bibinfo {author} {\bibfnamefont {J.~A.}\ \bibnamefont {Pople}},\
  and\ \bibinfo {author} {\bibfnamefont {M.}~\bibnamefont {Head-Gordon}},\
  }\bibfield  {title} {\enquote {\bibinfo {title} {A fifth-order perturbation
  comparison of electron correlation theories},}\ }\href
  {https://doi.org/10.1016/S0009-2614(89)87395-6} {\bibfield  {journal}
  {\bibinfo  {journal} {Chem. Phys. Lett.}\ }\textbf {\bibinfo {volume}
  {157}},\ \bibinfo {pages} {479--483} (\bibinfo {year} {1989})}\BibitemShut
  {NoStop}%
\bibitem [{\citenamefont {Balabanov}\ and\ \citenamefont
  {Peterson}(2005)}]{Balabanov05JCP}%
  \BibitemOpen
  \bibfield  {author} {\bibinfo {author} {\bibfnamefont {N.~B.}\ \bibnamefont
  {Balabanov}}\ and\ \bibinfo {author} {\bibfnamefont {K.~A.}\ \bibnamefont
  {Peterson}},\ }\bibfield  {title} {\enquote {\bibinfo {title} {Systematically
  convergent basis sets for transition metals. i. all-electron correlation
  consistent basis sets for the 3d elements sc-zn},}\ }\href
  {https://doi.org/10.1063/1.1998907} {\bibfield  {journal} {\bibinfo
  {journal} {J. Chem. Phys.}\ }\textbf {\bibinfo {volume} {123}},\ \bibinfo
  {pages} {064107} (\bibinfo {year} {2005})}\BibitemShut {NoStop}%
\bibitem [{\citenamefont {Whitten}(1973)}]{Whitten73JCP}%
  \BibitemOpen
  \bibfield  {author} {\bibinfo {author} {\bibfnamefont {J.~L.}\ \bibnamefont
  {Whitten}},\ }\bibfield  {title} {\enquote {\bibinfo {title} {Coulombic
  potential energy integrals and approximations},}\ }\href
  {https://doi.org/10.1063/1.1679012} {\bibfield  {journal} {\bibinfo
  {journal} {J. Chem. Phys.}\ }\textbf {\bibinfo {volume} {58}},\ \bibinfo
  {pages} {4496--4501} (\bibinfo {year} {1973})}\BibitemShut {NoStop}%
\bibitem [{\citenamefont {Dunlap}(2000)}]{Dunlap00PCCP}%
  \BibitemOpen
  \bibfield  {author} {\bibinfo {author} {\bibfnamefont {B.~I.}\ \bibnamefont
  {Dunlap}},\ }\bibfield  {title} {\enquote {\bibinfo {title} {Robust and
  variational fitting},}\ }\href {https://doi.org/10.1039/B000027M} {\bibfield
  {journal} {\bibinfo  {journal} {Phys. Chem. Chem. Phys.}\ }\textbf {\bibinfo
  {volume} {2}},\ \bibinfo {pages} {2113--2116} (\bibinfo {year}
  {2000})}\BibitemShut {NoStop}%
\bibitem [{\citenamefont {Bartlett}\ and\ \citenamefont
  {Musia\l{}}(2007)}]{Bartlett07RMP}%
  \BibitemOpen
  \bibfield  {author} {\bibinfo {author} {\bibfnamefont {R.~J.}\ \bibnamefont
  {Bartlett}}\ and\ \bibinfo {author} {\bibfnamefont {M.}~\bibnamefont
  {Musia\l{}}},\ }\bibfield  {title} {\enquote {\bibinfo {title}
  {Coupled-cluster theory in quantum chemistry},}\ }\href
  {https://doi.org/10.1103/RevModPhys.79.291} {\bibfield  {journal} {\bibinfo
  {journal} {Rev. Mod. Phys.}\ }\textbf {\bibinfo {volume} {79}},\ \bibinfo
  {pages} {291--352} (\bibinfo {year} {2007})}\BibitemShut {NoStop}%
\bibitem [{\citenamefont {Widmark}, \citenamefont {Malmqvist},\ and\
  \citenamefont {Roos}(1990)}]{Widmark90TCA}%
  \BibitemOpen
  \bibfield  {author} {\bibinfo {author} {\bibfnamefont {P.-O.}\ \bibnamefont
  {Widmark}}, \bibinfo {author} {\bibfnamefont {P.-{\AA}.}\ \bibnamefont
  {Malmqvist}},\ and\ \bibinfo {author} {\bibfnamefont {B.~O.}\ \bibnamefont
  {Roos}},\ }\bibfield  {title} {\enquote {\bibinfo {title} {Density matrix
  averaged atomic natural orbital (ano) basis sets for correlated molecular
  wave functions},}\ }\href {https://doi.org/10.1007/BF01120130} {\bibfield
  {journal} {\bibinfo  {journal} {Theor. Chim. Acta}\ }\textbf {\bibinfo
  {volume} {77}},\ \bibinfo {pages} {291--306} (\bibinfo {year}
  {1990})}\BibitemShut {NoStop}%
\bibitem [{\citenamefont {Dunning}, \citenamefont {Peterson},\ and\
  \citenamefont {Wilson}(2001)}]{Dunning01JCP}%
  \BibitemOpen
  \bibfield  {author} {\bibinfo {author} {\bibfnamefont {J.}~\bibnamefont
  {Dunning}, \bibfnamefont {Thom~H.}}, \bibinfo {author} {\bibfnamefont
  {K.~A.}\ \bibnamefont {Peterson}},\ and\ \bibinfo {author} {\bibfnamefont
  {A.~K.}\ \bibnamefont {Wilson}},\ }\bibfield  {title} {\enquote {\bibinfo
  {title} {Gaussian basis sets for use in correlated molecular calculations. x.
  the atoms aluminum through argon revisited},}\ }\href
  {https://doi.org/10.1063/1.1367373} {\bibfield  {journal} {\bibinfo
  {journal} {J. Chem. Phys.}\ }\textbf {\bibinfo {volume} {114}},\ \bibinfo
  {pages} {9244--9253} (\bibinfo {year} {2001})}\BibitemShut {NoStop}%
\bibitem [{\citenamefont {Bennett}\ \emph {et~al.}(2017)\citenamefont
  {Bennett}, \citenamefont {Melton}, \citenamefont {Annaberdiyev},
  \citenamefont {Wang}, \citenamefont {Shulenburger},\ and\ \citenamefont
  {Mitas}}]{Bennett17JCP}%
  \BibitemOpen
  \bibfield  {author} {\bibinfo {author} {\bibfnamefont {M.~C.}\ \bibnamefont
  {Bennett}}, \bibinfo {author} {\bibfnamefont {C.~A.}\ \bibnamefont {Melton}},
  \bibinfo {author} {\bibfnamefont {A.}~\bibnamefont {Annaberdiyev}}, \bibinfo
  {author} {\bibfnamefont {G.}~\bibnamefont {Wang}}, \bibinfo {author}
  {\bibfnamefont {L.}~\bibnamefont {Shulenburger}},\ and\ \bibinfo {author}
  {\bibfnamefont {L.}~\bibnamefont {Mitas}},\ }\bibfield  {title} {\enquote
  {\bibinfo {title} {A new generation of effective core potentials for
  correlated calculations},}\ }\href {https://doi.org/10.1063/1.4995643}
  {\bibfield  {journal} {\bibinfo  {journal} {J. Chem. Phys.}\ }\textbf
  {\bibinfo {volume} {147}},\ \bibinfo {pages} {224106} (\bibinfo {year}
  {2017})}\BibitemShut {NoStop}%
\bibitem [{\citenamefont {Hutter}(2019)}]{HutterPP}%
  \BibitemOpen
  \bibfield  {author} {\bibinfo {author} {\bibfnamefont {J.}~\bibnamefont
  {Hutter}},\ }\href@noop {} {\enquote {\bibinfo {title} {New optimization of
  {GTH} pseudopotentials for {PBE}, {SCAN}, {PBE0} functionals. {GTH}
  pseudopotentials for {H}artree-{F}ock. {NLCC} pseudopotentials for {PBE}.}}\
  }\bibinfo {howpublished} {\url{https://github.com/juerghutter/GTH}} (\bibinfo
  {year} {2019})\BibitemShut {NoStop}%
\bibitem [{\citenamefont {Perdew}, \citenamefont {Burke},\ and\ \citenamefont
  {Ernzerhof}(1996)}]{Perdew96PRL}%
  \BibitemOpen
  \bibfield  {author} {\bibinfo {author} {\bibfnamefont {J.~P.}\ \bibnamefont
  {Perdew}}, \bibinfo {author} {\bibfnamefont {K.}~\bibnamefont {Burke}},\ and\
  \bibinfo {author} {\bibfnamefont {M.}~\bibnamefont {Ernzerhof}},\ }\bibfield
  {title} {\enquote {\bibinfo {title} {Generalized gradient approximation made
  simple},}\ }\href {https://doi.org/10.1103/PhysRevLett.77.3865} {\bibfield
  {journal} {\bibinfo  {journal} {Phys. Rev. Lett.}\ }\textbf {\bibinfo
  {volume} {77}},\ \bibinfo {pages} {3865--3868} (\bibinfo {year}
  {1996})}\BibitemShut {NoStop}%
\bibitem [{\citenamefont {Krack}(2005)}]{Krack05TCA}%
  \BibitemOpen
  \bibfield  {author} {\bibinfo {author} {\bibfnamefont {M.}~\bibnamefont
  {Krack}},\ }\bibfield  {title} {\enquote {\bibinfo {title} {Pseudopotentials
  for h to kr optimized for gradient-corrected exchange-correlation
  functionals},}\ }\href {https://doi.org/10.1007/s00214-005-0655-y} {\bibfield
   {journal} {\bibinfo  {journal} {Theor. Chim. Acta}\ }\textbf {\bibinfo
  {volume} {114}},\ \bibinfo {pages} {145--152} (\bibinfo {year}
  {2005})}\BibitemShut {NoStop}%
\bibitem [{\citenamefont {Weigend}(2002)}]{Weigend02PCCP}%
  \BibitemOpen
  \bibfield  {author} {\bibinfo {author} {\bibfnamefont {F.}~\bibnamefont
  {Weigend}},\ }\bibfield  {title} {\enquote {\bibinfo {title} {A fully direct
  ri-hf algorithm: Implementation, optimised auxiliary basis sets,
  demonstration of accuracy and efficiency},}\ }\href
  {https://doi.org/10.1039/b204199p} {\bibfield  {journal} {\bibinfo  {journal}
  {Phys. Chem. Chem. Phys.}\ }\textbf {\bibinfo {volume} {4}},\ \bibinfo
  {pages} {4285--4291} (\bibinfo {year} {2002})}\BibitemShut {NoStop}%
\bibitem [{\citenamefont {Sun}\ \emph {et~al.}(2018)\citenamefont {Sun},
  \citenamefont {Berkelbach}, \citenamefont {Blunt}, \citenamefont {Booth},
  \citenamefont {Guo}, \citenamefont {Li}, \citenamefont {Liu}, \citenamefont
  {McClain}, \citenamefont {Sayfutyarova}, \citenamefont {Sharma},
  \citenamefont {Wouters},\ and\ \citenamefont {Chan}}]{Sun18WIRCMS}%
  \BibitemOpen
  \bibfield  {author} {\bibinfo {author} {\bibfnamefont {Q.}~\bibnamefont
  {Sun}}, \bibinfo {author} {\bibfnamefont {T.~C.}\ \bibnamefont {Berkelbach}},
  \bibinfo {author} {\bibfnamefont {N.~S.}\ \bibnamefont {Blunt}}, \bibinfo
  {author} {\bibfnamefont {G.~H.}\ \bibnamefont {Booth}}, \bibinfo {author}
  {\bibfnamefont {S.}~\bibnamefont {Guo}}, \bibinfo {author} {\bibfnamefont
  {Z.}~\bibnamefont {Li}}, \bibinfo {author} {\bibfnamefont {J.}~\bibnamefont
  {Liu}}, \bibinfo {author} {\bibfnamefont {J.~D.}\ \bibnamefont {McClain}},
  \bibinfo {author} {\bibfnamefont {E.~R.}\ \bibnamefont {Sayfutyarova}},
  \bibinfo {author} {\bibfnamefont {S.}~\bibnamefont {Sharma}}, \bibinfo
  {author} {\bibfnamefont {S.}~\bibnamefont {Wouters}},\ and\ \bibinfo {author}
  {\bibfnamefont {G.~K.-L.}\ \bibnamefont {Chan}},\ }\bibfield  {title}
  {\enquote {\bibinfo {title} {Pyscf: the python-based simulations of chemistry
  framework},}\ }\href {https://doi.org/10.1002/wcms.1340} {\bibfield
  {journal} {\bibinfo  {journal} {Wiley Interdiscip. Rev. Comput. Mol. Sci}\
  }\textbf {\bibinfo {volume} {8}},\ \bibinfo {pages} {e1340} (\bibinfo {year}
  {2018})}\BibitemShut {NoStop}%
\bibitem [{\citenamefont {Sun}\ \emph {et~al.}(2020)\citenamefont {Sun},
  \citenamefont {Zhang}, \citenamefont {Banerjee}, \citenamefont {Bao},
  \citenamefont {Barbry}, \citenamefont {Blunt}, \citenamefont {Bogdanov},
  \citenamefont {Booth}, \citenamefont {Chen}, \citenamefont {Cui},
  \citenamefont {Eriksen}, \citenamefont {Gao}, \citenamefont {Guo},
  \citenamefont {Hermann}, \citenamefont {Hermes}, \citenamefont {Koh},
  \citenamefont {Koval}, \citenamefont {Lehtola}, \citenamefont {Li},
  \citenamefont {Liu}, \citenamefont {Mardirossian}, \citenamefont {McClain},
  \citenamefont {Motta}, \citenamefont {Mussard}, \citenamefont {Pham},
  \citenamefont {Pulkin}, \citenamefont {Purwanto}, \citenamefont {Robinson},
  \citenamefont {Ronca}, \citenamefont {Sayfutyarova}, \citenamefont
  {Scheurer}, \citenamefont {Schurkus}, \citenamefont {Smith}, \citenamefont
  {Sun}, \citenamefont {Sun}, \citenamefont {Upadhyay}, \citenamefont {Wagner},
  \citenamefont {Wang}, \citenamefont {White}, \citenamefont {Whitfield},
  \citenamefont {Williamson}, \citenamefont {Wouters}, \citenamefont {Yang},
  \citenamefont {Yu}, \citenamefont {Zhu}, \citenamefont {Berkelbach},
  \citenamefont {Sharma}, \citenamefont {Sokolov},\ and\ \citenamefont
  {Chan}}]{Sun20JCP}%
  \BibitemOpen
  \bibfield  {author} {\bibinfo {author} {\bibfnamefont {Q.}~\bibnamefont
  {Sun}}, \bibinfo {author} {\bibfnamefont {X.}~\bibnamefont {Zhang}}, \bibinfo
  {author} {\bibfnamefont {S.}~\bibnamefont {Banerjee}}, \bibinfo {author}
  {\bibfnamefont {P.}~\bibnamefont {Bao}}, \bibinfo {author} {\bibfnamefont
  {M.}~\bibnamefont {Barbry}}, \bibinfo {author} {\bibfnamefont {N.~S.}\
  \bibnamefont {Blunt}}, \bibinfo {author} {\bibfnamefont {N.~A.}\ \bibnamefont
  {Bogdanov}}, \bibinfo {author} {\bibfnamefont {G.~H.}\ \bibnamefont {Booth}},
  \bibinfo {author} {\bibfnamefont {J.}~\bibnamefont {Chen}}, \bibinfo {author}
  {\bibfnamefont {Z.-H.}\ \bibnamefont {Cui}}, \bibinfo {author} {\bibfnamefont
  {J.~J.}\ \bibnamefont {Eriksen}}, \bibinfo {author} {\bibfnamefont
  {Y.}~\bibnamefont {Gao}}, \bibinfo {author} {\bibfnamefont {S.}~\bibnamefont
  {Guo}}, \bibinfo {author} {\bibfnamefont {J.}~\bibnamefont {Hermann}},
  \bibinfo {author} {\bibfnamefont {M.~R.}\ \bibnamefont {Hermes}}, \bibinfo
  {author} {\bibfnamefont {K.}~\bibnamefont {Koh}}, \bibinfo {author}
  {\bibfnamefont {P.}~\bibnamefont {Koval}}, \bibinfo {author} {\bibfnamefont
  {S.}~\bibnamefont {Lehtola}}, \bibinfo {author} {\bibfnamefont
  {Z.}~\bibnamefont {Li}}, \bibinfo {author} {\bibfnamefont {J.}~\bibnamefont
  {Liu}}, \bibinfo {author} {\bibfnamefont {N.}~\bibnamefont {Mardirossian}},
  \bibinfo {author} {\bibfnamefont {J.~D.}\ \bibnamefont {McClain}}, \bibinfo
  {author} {\bibfnamefont {M.}~\bibnamefont {Motta}}, \bibinfo {author}
  {\bibfnamefont {B.}~\bibnamefont {Mussard}}, \bibinfo {author} {\bibfnamefont
  {H.~Q.}\ \bibnamefont {Pham}}, \bibinfo {author} {\bibfnamefont
  {A.}~\bibnamefont {Pulkin}}, \bibinfo {author} {\bibfnamefont
  {W.}~\bibnamefont {Purwanto}}, \bibinfo {author} {\bibfnamefont {P.~J.}\
  \bibnamefont {Robinson}}, \bibinfo {author} {\bibfnamefont {E.}~\bibnamefont
  {Ronca}}, \bibinfo {author} {\bibfnamefont {E.~R.}\ \bibnamefont
  {Sayfutyarova}}, \bibinfo {author} {\bibfnamefont {M.}~\bibnamefont
  {Scheurer}}, \bibinfo {author} {\bibfnamefont {H.~F.}\ \bibnamefont
  {Schurkus}}, \bibinfo {author} {\bibfnamefont {J.~E.~T.}\ \bibnamefont
  {Smith}}, \bibinfo {author} {\bibfnamefont {C.}~\bibnamefont {Sun}}, \bibinfo
  {author} {\bibfnamefont {S.-N.}\ \bibnamefont {Sun}}, \bibinfo {author}
  {\bibfnamefont {S.}~\bibnamefont {Upadhyay}}, \bibinfo {author}
  {\bibfnamefont {L.~K.}\ \bibnamefont {Wagner}}, \bibinfo {author}
  {\bibfnamefont {X.}~\bibnamefont {Wang}}, \bibinfo {author} {\bibfnamefont
  {A.}~\bibnamefont {White}}, \bibinfo {author} {\bibfnamefont {J.~D.}\
  \bibnamefont {Whitfield}}, \bibinfo {author} {\bibfnamefont {M.~J.}\
  \bibnamefont {Williamson}}, \bibinfo {author} {\bibfnamefont
  {S.}~\bibnamefont {Wouters}}, \bibinfo {author} {\bibfnamefont
  {J.}~\bibnamefont {Yang}}, \bibinfo {author} {\bibfnamefont {J.~M.}\
  \bibnamefont {Yu}}, \bibinfo {author} {\bibfnamefont {T.}~\bibnamefont
  {Zhu}}, \bibinfo {author} {\bibfnamefont {T.~C.}\ \bibnamefont {Berkelbach}},
  \bibinfo {author} {\bibfnamefont {S.}~\bibnamefont {Sharma}}, \bibinfo
  {author} {\bibfnamefont {A.~Y.}\ \bibnamefont {Sokolov}},\ and\ \bibinfo
  {author} {\bibfnamefont {G.~K.-L.}\ \bibnamefont {Chan}},\ }\bibfield
  {title} {\enquote {\bibinfo {title} {{Recent developments in the PySCF
  program package}},}\ }\href {https://doi.org/10.1063/5.0006074} {\bibfield
  {journal} {\bibinfo  {journal} {J. Chem. Phys.}\ }\textbf {\bibinfo {volume}
  {153}},\ \bibinfo {pages} {024109} (\bibinfo {year} {2020})}\BibitemShut
  {NoStop}%
\bibitem [{\citenamefont {Sun}\ \emph {et~al.}(2026)\citenamefont {Sun},
  \citenamefont {Hermes}, \citenamefont {Wu}, \citenamefont {Zhai},
  \citenamefont {Zhang}, \citenamefont {Ahmed}, \citenamefont {Aucar},
  \citenamefont {Backhouse}, \citenamefont {Banerjee}, \citenamefont {Bao},
  \citenamefont {Bogdanov}, \citenamefont {Bystrom}, \citenamefont {Chapoton},
  \citenamefont {Chen}, \citenamefont {Chernyshov}, \citenamefont {Clifford},
  \citenamefont {Cohen-Janes}, \citenamefont {Cui}, \citenamefont {Damour},
  \citenamefont {Dattani}, \citenamefont {Dittmer}, \citenamefont {Ehlert},
  \citenamefont {Eriksen}, \citenamefont {Evangelista}, \citenamefont {Ewing},
  \citenamefont {Farahvash}, \citenamefont {Focke}, \citenamefont {Gao},
  \citenamefont {Gasperich}, \citenamefont {Gillispie}, \citenamefont
  {Greiner}, \citenamefont {Hennefarth}, \citenamefont {Hermann}, \citenamefont
  {Hillenbrand}, \citenamefont {Huhtasalo}, \citenamefont {Ibrahim},
  \citenamefont {Jangid}, \citenamefont {Javaremi}, \citenamefont {Jenkins},
  \citenamefont {Jin}, \citenamefont {King}, \citenamefont {Kooi},
  \citenamefont {Kurian}, \citenamefont {Larsson}, \citenamefont {Lau},
  \citenamefont {Lee}, \citenamefont {Lehtola}, \citenamefont {Li},
  \citenamefont {Li}, \citenamefont {Li}, \citenamefont {Li}, \citenamefont
  {Li}, \citenamefont {Lykhin}, \citenamefont {Mahajan}, \citenamefont
  {Mauger}, \citenamefont {Mazo-Sevillano}, \citenamefont {Moussa},
  \citenamefont {Nakano}, \citenamefont {Neufeld}, \citenamefont {Peng},
  \citenamefont {Pham}, \citenamefont {Pinski}, \citenamefont {Pokhilko},
  \citenamefont {Pu}, \citenamefont {Qian}, \citenamefont {Quiton},
  \citenamefont {Schulze}, \citenamefont {Scott}, \citenamefont {Seal},
  \citenamefont {Serna}, \citenamefont {Smith}, \citenamefont {Smyser},
  \citenamefont {Stahl}, \citenamefont {Sun}, \citenamefont {Sung},
  \citenamefont {Trushin}, \citenamefont {Upadhyay}, \citenamefont {Vo},
  \citenamefont {Vogels}, \citenamefont {Wang}, \citenamefont {Wang},
  \citenamefont {Wang}, \citenamefont {Wang}, \citenamefont {Wang},
  \citenamefont {Williamson}, \citenamefont {Yang}, \citenamefont {Ye},
  \citenamefont {Yeh}, \citenamefont {Yu}, \citenamefont {Yu}, \citenamefont
  {Yu}, \citenamefont {Zhang}, \citenamefont {Zhang}, \citenamefont {Zhang},
  \citenamefont {Zhao}, \citenamefont {Zhou}, \citenamefont {Zhu},
  \citenamefont {Zhu}, \citenamefont {Berkelbach}, \citenamefont {Gagliardi},
  \citenamefont {Sharma}, \citenamefont {Sokolov},\ and\ \citenamefont
  {Kin-Lic~Chan}}]{Sun26JCP}%
  \BibitemOpen
  \bibfield  {author} {\bibinfo {author} {\bibfnamefont {Q.}~\bibnamefont
  {Sun}}, \bibinfo {author} {\bibfnamefont {M.~R.}\ \bibnamefont {Hermes}},
  \bibinfo {author} {\bibfnamefont {X.}~\bibnamefont {Wu}}, \bibinfo {author}
  {\bibfnamefont {H.}~\bibnamefont {Zhai}}, \bibinfo {author} {\bibfnamefont
  {X.}~\bibnamefont {Zhang}}, \bibinfo {author} {\bibfnamefont {A.~M.}\
  \bibnamefont {Ahmed}}, \bibinfo {author} {\bibfnamefont {J.~J.}\ \bibnamefont
  {Aucar}}, \bibinfo {author} {\bibfnamefont {O.~J.}\ \bibnamefont
  {Backhouse}}, \bibinfo {author} {\bibfnamefont {S.}~\bibnamefont {Banerjee}},
  \bibinfo {author} {\bibfnamefont {P.}~\bibnamefont {Bao}}, \bibinfo {author}
  {\bibfnamefont {N.~A.}\ \bibnamefont {Bogdanov}}, \bibinfo {author}
  {\bibfnamefont {K.}~\bibnamefont {Bystrom}}, \bibinfo {author} {\bibfnamefont
  {F.}~\bibnamefont {Chapoton}}, \bibinfo {author} {\bibfnamefont {N.-Y.}\
  \bibnamefont {Chen}}, \bibinfo {author} {\bibfnamefont {I.~Y.}\ \bibnamefont
  {Chernyshov}}, \bibinfo {author} {\bibfnamefont {H.~S.}\ \bibnamefont
  {Clifford}}, \bibinfo {author} {\bibfnamefont {S.}~\bibnamefont
  {Cohen-Janes}}, \bibinfo {author} {\bibfnamefont {Z.-H.}\ \bibnamefont
  {Cui}}, \bibinfo {author} {\bibfnamefont {Y.~D.}\ \bibnamefont {Damour}},
  \bibinfo {author} {\bibfnamefont {N.}~\bibnamefont {Dattani}}, \bibinfo
  {author} {\bibfnamefont {L.~B.}\ \bibnamefont {Dittmer}}, \bibinfo {author}
  {\bibfnamefont {S.}~\bibnamefont {Ehlert}}, \bibinfo {author} {\bibfnamefont
  {J.~J.}\ \bibnamefont {Eriksen}}, \bibinfo {author} {\bibfnamefont {F.~A.}\
  \bibnamefont {Evangelista}}, \bibinfo {author} {\bibfnamefont {S.~A.}\
  \bibnamefont {Ewing}}, \bibinfo {author} {\bibfnamefont {A.}~\bibnamefont
  {Farahvash}}, \bibinfo {author} {\bibfnamefont {K.}~\bibnamefont {Focke}},
  \bibinfo {author} {\bibfnamefont {Y.}~\bibnamefont {Gao}}, \bibinfo {author}
  {\bibfnamefont {K.~E.}\ \bibnamefont {Gasperich}}, \bibinfo {author}
  {\bibfnamefont {N.}~\bibnamefont {Gillispie}}, \bibinfo {author}
  {\bibfnamefont {J.}~\bibnamefont {Greiner}}, \bibinfo {author} {\bibfnamefont
  {M.~R.}\ \bibnamefont {Hennefarth}}, \bibinfo {author} {\bibfnamefont
  {J.}~\bibnamefont {Hermann}}, \bibinfo {author} {\bibfnamefont
  {C.}~\bibnamefont {Hillenbrand}}, \bibinfo {author} {\bibfnamefont
  {J.}~\bibnamefont {Huhtasalo}}, \bibinfo {author} {\bibfnamefont
  {B.}~\bibnamefont {Ibrahim}}, \bibinfo {author} {\bibfnamefont
  {B.}~\bibnamefont {Jangid}}, \bibinfo {author} {\bibfnamefont {A.~N.}\
  \bibnamefont {Javaremi}}, \bibinfo {author} {\bibfnamefont {A.~J.}\
  \bibnamefont {Jenkins}}, \bibinfo {author} {\bibfnamefont {Y.}~\bibnamefont
  {Jin}}, \bibinfo {author} {\bibfnamefont {D.~S.}\ \bibnamefont {King}},
  \bibinfo {author} {\bibfnamefont {D.~P.}\ \bibnamefont {Kooi}}, \bibinfo
  {author} {\bibfnamefont {J.~S.}\ \bibnamefont {Kurian}}, \bibinfo {author}
  {\bibfnamefont {H.~R.}\ \bibnamefont {Larsson}}, \bibinfo {author}
  {\bibfnamefont {B.~T.~G.}\ \bibnamefont {Lau}}, \bibinfo {author}
  {\bibfnamefont {S.}~\bibnamefont {Lee}}, \bibinfo {author} {\bibfnamefont
  {S.}~\bibnamefont {Lehtola}}, \bibinfo {author} {\bibfnamefont
  {C.}~\bibnamefont {Li}}, \bibinfo {author} {\bibfnamefont {H.}~\bibnamefont
  {Li}}, \bibinfo {author} {\bibfnamefont {J.}~\bibnamefont {Li}}, \bibinfo
  {author} {\bibfnamefont {R.}~\bibnamefont {Li}}, \bibinfo {author}
  {\bibfnamefont {S.}~\bibnamefont {Li}}, \bibinfo {author} {\bibfnamefont
  {A.~O.}\ \bibnamefont {Lykhin}}, \bibinfo {author} {\bibfnamefont
  {A.}~\bibnamefont {Mahajan}}, \bibinfo {author} {\bibfnamefont
  {N.}~\bibnamefont {Mauger}}, \bibinfo {author} {\bibfnamefont {P.~d.}\
  \bibnamefont {Mazo-Sevillano}}, \bibinfo {author} {\bibfnamefont
  {J.}~\bibnamefont {Moussa}}, \bibinfo {author} {\bibfnamefont
  {K.}~\bibnamefont {Nakano}}, \bibinfo {author} {\bibfnamefont {V.~A.}\
  \bibnamefont {Neufeld}}, \bibinfo {author} {\bibfnamefont {L.}~\bibnamefont
  {Peng}}, \bibinfo {author} {\bibfnamefont {H.~Q.}\ \bibnamefont {Pham}},
  \bibinfo {author} {\bibfnamefont {P.}~\bibnamefont {Pinski}}, \bibinfo
  {author} {\bibfnamefont {P.}~\bibnamefont {Pokhilko}}, \bibinfo {author}
  {\bibfnamefont {Z.}~\bibnamefont {Pu}}, \bibinfo {author} {\bibfnamefont
  {Y.}~\bibnamefont {Qian}}, \bibinfo {author} {\bibfnamefont {S.~J.}\
  \bibnamefont {Quiton}}, \bibinfo {author} {\bibfnamefont {W.~T.}\
  \bibnamefont {Schulze}}, \bibinfo {author} {\bibfnamefont {T.~R.}\
  \bibnamefont {Scott}}, \bibinfo {author} {\bibfnamefont {A.}~\bibnamefont
  {Seal}}, \bibinfo {author} {\bibfnamefont {J.~D.}\ \bibnamefont {Serna}},
  \bibinfo {author} {\bibfnamefont {J.~E.~T.}\ \bibnamefont {Smith}}, \bibinfo
  {author} {\bibfnamefont {K.~E.}\ \bibnamefont {Smyser}}, \bibinfo {author}
  {\bibfnamefont {T.}~\bibnamefont {Stahl}}, \bibinfo {author} {\bibfnamefont
  {C.}~\bibnamefont {Sun}}, \bibinfo {author} {\bibfnamefont {K.~J.}\
  \bibnamefont {Sung}}, \bibinfo {author} {\bibfnamefont {E.}~\bibnamefont
  {Trushin}}, \bibinfo {author} {\bibfnamefont {S.}~\bibnamefont {Upadhyay}},
  \bibinfo {author} {\bibfnamefont {E.~A.}\ \bibnamefont {Vo}}, \bibinfo
  {author} {\bibfnamefont {T.}~\bibnamefont {Vogels}}, \bibinfo {author}
  {\bibfnamefont {S.}~\bibnamefont {Wang}}, \bibinfo {author} {\bibfnamefont
  {T.}~\bibnamefont {Wang}}, \bibinfo {author} {\bibfnamefont {X.}~\bibnamefont
  {Wang}}, \bibinfo {author} {\bibfnamefont {X.}~\bibnamefont {Wang}}, \bibinfo
  {author} {\bibfnamefont {Y.}~\bibnamefont {Wang}}, \bibinfo {author}
  {\bibfnamefont {M.}~\bibnamefont {Williamson}}, \bibinfo {author}
  {\bibfnamefont {J.}~\bibnamefont {Yang}}, \bibinfo {author} {\bibfnamefont
  {H.-Z.}\ \bibnamefont {Ye}}, \bibinfo {author} {\bibfnamefont {C.-N.}\
  \bibnamefont {Yeh}}, \bibinfo {author} {\bibfnamefont {H.}~\bibnamefont
  {Yu}}, \bibinfo {author} {\bibfnamefont {J.}~\bibnamefont {Yu}}, \bibinfo
  {author} {\bibfnamefont {V.~W.-z.}\ \bibnamefont {Yu}}, \bibinfo {author}
  {\bibfnamefont {C.}~\bibnamefont {Zhang}}, \bibinfo {author} {\bibfnamefont
  {D.}~\bibnamefont {Zhang}}, \bibinfo {author} {\bibfnamefont
  {Y.}~\bibnamefont {Zhang}}, \bibinfo {author} {\bibfnamefont
  {Z.}~\bibnamefont {Zhao}}, \bibinfo {author} {\bibfnamefont {Z.}~\bibnamefont
  {Zhou}}, \bibinfo {author} {\bibfnamefont {A.~J.}\ \bibnamefont {Zhu}},
  \bibinfo {author} {\bibfnamefont {T.}~\bibnamefont {Zhu}}, \bibinfo {author}
  {\bibfnamefont {T.~C.}\ \bibnamefont {Berkelbach}}, \bibinfo {author}
  {\bibfnamefont {L.}~\bibnamefont {Gagliardi}}, \bibinfo {author}
  {\bibfnamefont {S.}~\bibnamefont {Sharma}}, \bibinfo {author} {\bibfnamefont
  {A.~Y.}\ \bibnamefont {Sokolov}},\ and\ \bibinfo {author} {\bibfnamefont
  {G.}~\bibnamefont {Kin-Lic~Chan}},\ }\bibfield  {title} {\enquote {\bibinfo
  {title} {The python simulations of chemistry framework: 10 years of an
  open-source quantum chemistry project},}\ }\href
  {https://doi.org/10.1063/5.0337441} {\bibfield  {journal} {\bibinfo
  {journal} {J. Chem. Phys.}\ }\textbf {\bibinfo {volume} {165}},\ \bibinfo
  {pages} {102502} (\bibinfo {year} {2026})}\BibitemShut {NoStop}%
\bibitem [{\citenamefont {Stoychev}, \citenamefont {Auer},\ and\ \citenamefont
  {Neese}(2017)}]{Stoychev17JCTC}%
  \BibitemOpen
  \bibfield  {author} {\bibinfo {author} {\bibfnamefont {G.~L.}\ \bibnamefont
  {Stoychev}}, \bibinfo {author} {\bibfnamefont {A.~A.}\ \bibnamefont {Auer}},\
  and\ \bibinfo {author} {\bibfnamefont {F.}~\bibnamefont {Neese}},\ }\bibfield
   {title} {\enquote {\bibinfo {title} {Automatic generation of auxiliary basis
  sets},}\ }\href {https://doi.org/10.1021/acs.jctc.6b01041} {\bibfield
  {journal} {\bibinfo  {journal} {J. Chem. Theory Comput.}\ }\textbf {\bibinfo
  {volume} {13}},\ \bibinfo {pages} {554--562} (\bibinfo {year}
  {2017})}\BibitemShut {NoStop}%
\bibitem [{\citenamefont {Neese}(2012)}]{Neese12WIRCMS}%
  \BibitemOpen
  \bibfield  {author} {\bibinfo {author} {\bibfnamefont {F.}~\bibnamefont
  {Neese}},\ }\bibfield  {title} {\enquote {\bibinfo {title} {The orca program
  system},}\ }\href {https://doi.org/10.1002/wcms.81} {\bibfield  {journal}
  {\bibinfo  {journal} {Wiley Interdiscip. Rev. Comput. Mol. Sci}\ }\textbf
  {\bibinfo {volume} {2}},\ \bibinfo {pages} {73--78} (\bibinfo {year}
  {2012})}\BibitemShut {NoStop}%
\bibitem [{\citenamefont {Knizia}(2013)}]{Knizia13JCTC}%
  \BibitemOpen
  \bibfield  {author} {\bibinfo {author} {\bibfnamefont {G.}~\bibnamefont
  {Knizia}},\ }\bibfield  {title} {\enquote {\bibinfo {title} {Intrinsic atomic
  orbitals: An unbiased bridge between quantum theory and chemical concepts},}\
  }\href {https://doi.org/10.1021/ct400687b} {\bibfield  {journal} {\bibinfo
  {journal} {J. Chem. Theory Comput.}\ }\textbf {\bibinfo {volume} {9}},\
  \bibinfo {pages} {4834--4843} (\bibinfo {year} {2013})}\BibitemShut {NoStop}%
\bibitem [{\citenamefont {{Ye Research Group}}(2026)}]{pygto}%
  \BibitemOpen
  \bibfield  {author} {\bibinfo {author} {\bibnamefont {{Ye Research Group}}},\
  }\href@noop {} {\enquote {\bibinfo {title} {{PyGTO}},}\ }\bibinfo
  {howpublished} {\url{https://github.com/ye-chem-group/pygto}} (\bibinfo
  {year} {2026}),\ \bibinfo {note} {accessed November 9, 2026}\BibitemShut
  {NoStop}%
\bibitem [{\citenamefont {Nocedal}(1980)}]{Nocedal80MC}%
  \BibitemOpen
  \bibfield  {author} {\bibinfo {author} {\bibfnamefont {J.}~\bibnamefont
  {Nocedal}},\ }\bibfield  {title} {\enquote {\bibinfo {title} {Updating
  quasi-newton matrices with limited storage},}\ }\href
  {https://doi.org/10.1090/S0025-5718-1980-0572855-7} {\bibfield  {journal}
  {\bibinfo  {journal} {Math. Comput}\ }\textbf {\bibinfo {volume} {35}},\
  \bibinfo {pages} {773--782} (\bibinfo {year} {1980})}\BibitemShut {NoStop}%
\bibitem [{\citenamefont {Petersson}\ \emph {et~al.}(2003)\citenamefont
  {Petersson}, \citenamefont {Zhong}, \citenamefont {Montgomery},\ and\
  \citenamefont {Frisch}}]{Petersson03JCP}%
  \BibitemOpen
  \bibfield  {author} {\bibinfo {author} {\bibfnamefont {G.~A.}\ \bibnamefont
  {Petersson}}, \bibinfo {author} {\bibfnamefont {S.}~\bibnamefont {Zhong}},
  \bibinfo {author} {\bibfnamefont {J.}~\bibnamefont {Montgomery},
  \bibfnamefont {John~A.}},\ and\ \bibinfo {author} {\bibfnamefont {M.~J.}\
  \bibnamefont {Frisch}},\ }\bibfield  {title} {\enquote {\bibinfo {title} {On
  the optimization of gaussian basis sets},}\ }\href
  {https://doi.org/10.1063/1.1516801} {\bibfield  {journal} {\bibinfo
  {journal} {J. Chem. Phys.}\ }\textbf {\bibinfo {volume} {118}},\ \bibinfo
  {pages} {1101--1109} (\bibinfo {year} {2003})}\BibitemShut {NoStop}%
\bibitem [{\citenamefont {Sun}\ \emph {et~al.}(2017)\citenamefont {Sun},
  \citenamefont {Berkelbach}, \citenamefont {McClain},\ and\ \citenamefont
  {Chan}}]{Sun17JCP}%
  \BibitemOpen
  \bibfield  {author} {\bibinfo {author} {\bibfnamefont {Q.}~\bibnamefont
  {Sun}}, \bibinfo {author} {\bibfnamefont {T.~C.}\ \bibnamefont {Berkelbach}},
  \bibinfo {author} {\bibfnamefont {J.~D.}\ \bibnamefont {McClain}},\ and\
  \bibinfo {author} {\bibfnamefont {G.~K.-L.}\ \bibnamefont {Chan}},\
  }\bibfield  {title} {\enquote {\bibinfo {title} {Gaussian and plane-wave
  mixed density fitting for periodic systems},}\ }\href
  {https://doi.org/10.1063/1.4998644} {\bibfield  {journal} {\bibinfo
  {journal} {J. Chem. Phys.}\ }\textbf {\bibinfo {volume} {147}},\ \bibinfo
  {pages} {164119} (\bibinfo {year} {2017})}\BibitemShut {NoStop}%
\bibitem [{\citenamefont {Giannozzi}\ \emph {et~al.}(2009)\citenamefont
  {Giannozzi}, \citenamefont {Baroni}, \citenamefont {Bonini}, \citenamefont
  {Calandra}, \citenamefont {Car}, \citenamefont {Cavazzoni}, \citenamefont
  {Ceresoli}, \citenamefont {Chiarotti}, \citenamefont {Cococcioni},
  \citenamefont {Dabo}, \citenamefont {Dal~Corso}, \citenamefont
  {de~Gironcoli}, \citenamefont {Fabris}, \citenamefont {Fratesi},
  \citenamefont {Gebauer}, \citenamefont {Gerstmann}, \citenamefont
  {Gougoussis}, \citenamefont {Kokalj}, \citenamefont {Lazzeri}, \citenamefont
  {Martin-Samos}, \citenamefont {Marzari}, \citenamefont {Mauri}, \citenamefont
  {Mazzarello}, \citenamefont {Paolini}, \citenamefont {Pasquarello},
  \citenamefont {Paulatto}, \citenamefont {Sbraccia}, \citenamefont {Scandolo},
  \citenamefont {Sclauzero}, \citenamefont {Seitsonen}, \citenamefont
  {Smogunov}, \citenamefont {Umari},\ and\ \citenamefont
  {Wentzcovitch}}]{Giannozzi09JPCM}%
  \BibitemOpen
  \bibfield  {author} {\bibinfo {author} {\bibfnamefont {P.}~\bibnamefont
  {Giannozzi}}, \bibinfo {author} {\bibfnamefont {S.}~\bibnamefont {Baroni}},
  \bibinfo {author} {\bibfnamefont {N.}~\bibnamefont {Bonini}}, \bibinfo
  {author} {\bibfnamefont {M.}~\bibnamefont {Calandra}}, \bibinfo {author}
  {\bibfnamefont {R.}~\bibnamefont {Car}}, \bibinfo {author} {\bibfnamefont
  {C.}~\bibnamefont {Cavazzoni}}, \bibinfo {author} {\bibfnamefont
  {D.}~\bibnamefont {Ceresoli}}, \bibinfo {author} {\bibfnamefont {G.~L.}\
  \bibnamefont {Chiarotti}}, \bibinfo {author} {\bibfnamefont {M.}~\bibnamefont
  {Cococcioni}}, \bibinfo {author} {\bibfnamefont {I.}~\bibnamefont {Dabo}},
  \bibinfo {author} {\bibfnamefont {A.}~\bibnamefont {Dal~Corso}}, \bibinfo
  {author} {\bibfnamefont {S.}~\bibnamefont {de~Gironcoli}}, \bibinfo {author}
  {\bibfnamefont {S.}~\bibnamefont {Fabris}}, \bibinfo {author} {\bibfnamefont
  {G.}~\bibnamefont {Fratesi}}, \bibinfo {author} {\bibfnamefont
  {R.}~\bibnamefont {Gebauer}}, \bibinfo {author} {\bibfnamefont
  {U.}~\bibnamefont {Gerstmann}}, \bibinfo {author} {\bibfnamefont
  {C.}~\bibnamefont {Gougoussis}}, \bibinfo {author} {\bibfnamefont
  {A.}~\bibnamefont {Kokalj}}, \bibinfo {author} {\bibfnamefont
  {M.}~\bibnamefont {Lazzeri}}, \bibinfo {author} {\bibfnamefont
  {L.}~\bibnamefont {Martin-Samos}}, \bibinfo {author} {\bibfnamefont
  {N.}~\bibnamefont {Marzari}}, \bibinfo {author} {\bibfnamefont
  {F.}~\bibnamefont {Mauri}}, \bibinfo {author} {\bibfnamefont
  {R.}~\bibnamefont {Mazzarello}}, \bibinfo {author} {\bibfnamefont
  {S.}~\bibnamefont {Paolini}}, \bibinfo {author} {\bibfnamefont
  {A.}~\bibnamefont {Pasquarello}}, \bibinfo {author} {\bibfnamefont
  {L.}~\bibnamefont {Paulatto}}, \bibinfo {author} {\bibfnamefont
  {C.}~\bibnamefont {Sbraccia}}, \bibinfo {author} {\bibfnamefont
  {S.}~\bibnamefont {Scandolo}}, \bibinfo {author} {\bibfnamefont
  {G.}~\bibnamefont {Sclauzero}}, \bibinfo {author} {\bibfnamefont {A.~P.}\
  \bibnamefont {Seitsonen}}, \bibinfo {author} {\bibfnamefont {A.}~\bibnamefont
  {Smogunov}}, \bibinfo {author} {\bibfnamefont {P.}~\bibnamefont {Umari}},\
  and\ \bibinfo {author} {\bibfnamefont {R.~M.}\ \bibnamefont {Wentzcovitch}},\
  }\bibfield  {title} {\enquote {\bibinfo {title} {Quantum espresso: a modular
  and open-source software project for quantum simulations of materials},}\
  }\href {https://doi.org/10.1088/0953-8984/21/39/395502} {\bibfield  {journal}
  {\bibinfo  {journal} {J. Phys. Condens. Matter}\ }\textbf {\bibinfo {volume}
  {21}},\ \bibinfo {pages} {395502} (\bibinfo {year} {2009})}\BibitemShut
  {NoStop}%
\bibitem [{\citenamefont {Giannozzi}\ \emph {et~al.}(2017)\citenamefont
  {Giannozzi}, \citenamefont {Andreussi}, \citenamefont {Brumme}, \citenamefont
  {Bunau}, \citenamefont {Buongiorno~Nardelli}, \citenamefont {Calandra},
  \citenamefont {Car}, \citenamefont {Cavazzoni}, \citenamefont {Ceresoli},
  \citenamefont {Cococcioni}, \citenamefont {Colonna}, \citenamefont
  {Carnimeo}, \citenamefont {Dal~Corso}, \citenamefont {de~Gironcoli},
  \citenamefont {Delugas}, \citenamefont {DiStasio}, \citenamefont {Ferretti},
  \citenamefont {Floris}, \citenamefont {Fratesi}, \citenamefont {Fugallo},
  \citenamefont {Gebauer}, \citenamefont {Gerstmann}, \citenamefont {Giustino},
  \citenamefont {Gorni}, \citenamefont {Jia}, \citenamefont {Kawamura},
  \citenamefont {Ko}, \citenamefont {Kokalj}, \citenamefont {Küçükbenli},
  \citenamefont {Lazzeri}, \citenamefont {Marsili}, \citenamefont {Marzari},
  \citenamefont {Mauri}, \citenamefont {Nguyen}, \citenamefont {Nguyen},
  \citenamefont {Otero-de-la Roza}, \citenamefont {Paulatto}, \citenamefont
  {Poncé}, \citenamefont {Rocca}, \citenamefont {Sabatini}, \citenamefont
  {Santra}, \citenamefont {Schlipf}, \citenamefont {Seitsonen}, \citenamefont
  {Smogunov}, \citenamefont {Timrov}, \citenamefont {Thonhauser}, \citenamefont
  {Umari}, \citenamefont {Vast}, \citenamefont {Wu},\ and\ \citenamefont
  {Baroni}}]{Giannozzi17JPCM}%
  \BibitemOpen
  \bibfield  {author} {\bibinfo {author} {\bibfnamefont {P.}~\bibnamefont
  {Giannozzi}}, \bibinfo {author} {\bibfnamefont {O.}~\bibnamefont
  {Andreussi}}, \bibinfo {author} {\bibfnamefont {T.}~\bibnamefont {Brumme}},
  \bibinfo {author} {\bibfnamefont {O.}~\bibnamefont {Bunau}}, \bibinfo
  {author} {\bibfnamefont {M.}~\bibnamefont {Buongiorno~Nardelli}}, \bibinfo
  {author} {\bibfnamefont {M.}~\bibnamefont {Calandra}}, \bibinfo {author}
  {\bibfnamefont {R.}~\bibnamefont {Car}}, \bibinfo {author} {\bibfnamefont
  {C.}~\bibnamefont {Cavazzoni}}, \bibinfo {author} {\bibfnamefont
  {D.}~\bibnamefont {Ceresoli}}, \bibinfo {author} {\bibfnamefont
  {M.}~\bibnamefont {Cococcioni}}, \bibinfo {author} {\bibfnamefont
  {N.}~\bibnamefont {Colonna}}, \bibinfo {author} {\bibfnamefont
  {I.}~\bibnamefont {Carnimeo}}, \bibinfo {author} {\bibfnamefont
  {A.}~\bibnamefont {Dal~Corso}}, \bibinfo {author} {\bibfnamefont
  {S.}~\bibnamefont {de~Gironcoli}}, \bibinfo {author} {\bibfnamefont
  {P.}~\bibnamefont {Delugas}}, \bibinfo {author} {\bibfnamefont {R.~A.}\
  \bibnamefont {DiStasio}}, \bibinfo {author} {\bibfnamefont {A.}~\bibnamefont
  {Ferretti}}, \bibinfo {author} {\bibfnamefont {A.}~\bibnamefont {Floris}},
  \bibinfo {author} {\bibfnamefont {G.}~\bibnamefont {Fratesi}}, \bibinfo
  {author} {\bibfnamefont {G.}~\bibnamefont {Fugallo}}, \bibinfo {author}
  {\bibfnamefont {R.}~\bibnamefont {Gebauer}}, \bibinfo {author} {\bibfnamefont
  {U.}~\bibnamefont {Gerstmann}}, \bibinfo {author} {\bibfnamefont
  {F.}~\bibnamefont {Giustino}}, \bibinfo {author} {\bibfnamefont
  {T.}~\bibnamefont {Gorni}}, \bibinfo {author} {\bibfnamefont
  {J.}~\bibnamefont {Jia}}, \bibinfo {author} {\bibfnamefont {M.}~\bibnamefont
  {Kawamura}}, \bibinfo {author} {\bibfnamefont {H.-Y.}\ \bibnamefont {Ko}},
  \bibinfo {author} {\bibfnamefont {A.}~\bibnamefont {Kokalj}}, \bibinfo
  {author} {\bibfnamefont {E.}~\bibnamefont {Küçükbenli}}, \bibinfo {author}
  {\bibfnamefont {M.}~\bibnamefont {Lazzeri}}, \bibinfo {author} {\bibfnamefont
  {M.}~\bibnamefont {Marsili}}, \bibinfo {author} {\bibfnamefont
  {N.}~\bibnamefont {Marzari}}, \bibinfo {author} {\bibfnamefont
  {F.}~\bibnamefont {Mauri}}, \bibinfo {author} {\bibfnamefont {N.~L.}\
  \bibnamefont {Nguyen}}, \bibinfo {author} {\bibfnamefont {H.-V.}\
  \bibnamefont {Nguyen}}, \bibinfo {author} {\bibfnamefont {A.}~\bibnamefont
  {Otero-de-la Roza}}, \bibinfo {author} {\bibfnamefont {L.}~\bibnamefont
  {Paulatto}}, \bibinfo {author} {\bibfnamefont {S.}~\bibnamefont {Poncé}},
  \bibinfo {author} {\bibfnamefont {D.}~\bibnamefont {Rocca}}, \bibinfo
  {author} {\bibfnamefont {R.}~\bibnamefont {Sabatini}}, \bibinfo {author}
  {\bibfnamefont {B.}~\bibnamefont {Santra}}, \bibinfo {author} {\bibfnamefont
  {M.}~\bibnamefont {Schlipf}}, \bibinfo {author} {\bibfnamefont {A.~P.}\
  \bibnamefont {Seitsonen}}, \bibinfo {author} {\bibfnamefont {A.}~\bibnamefont
  {Smogunov}}, \bibinfo {author} {\bibfnamefont {I.}~\bibnamefont {Timrov}},
  \bibinfo {author} {\bibfnamefont {T.}~\bibnamefont {Thonhauser}}, \bibinfo
  {author} {\bibfnamefont {P.}~\bibnamefont {Umari}}, \bibinfo {author}
  {\bibfnamefont {N.}~\bibnamefont {Vast}}, \bibinfo {author} {\bibfnamefont
  {X.}~\bibnamefont {Wu}},\ and\ \bibinfo {author} {\bibfnamefont
  {S.}~\bibnamefont {Baroni}},\ }\bibfield  {title} {\enquote {\bibinfo {title}
  {Advanced capabilities for materials modelling with quantum espresso},}\
  }\href {https://doi.org/10.1088/1361-648X/aa8f79} {\bibfield  {journal}
  {\bibinfo  {journal} {J. Phys. Condens. Matter}\ }\textbf {\bibinfo {volume}
  {29}},\ \bibinfo {pages} {465901} (\bibinfo {year} {2017})}\BibitemShut
  {NoStop}%
\bibitem [{\citenamefont {Ye}\ and\ \citenamefont
  {Berkelbach}(2021{\natexlab{a}})}]{Ye21JCPa}%
  \BibitemOpen
  \bibfield  {author} {\bibinfo {author} {\bibfnamefont {H.-Z.}\ \bibnamefont
  {Ye}}\ and\ \bibinfo {author} {\bibfnamefont {T.~C.}\ \bibnamefont
  {Berkelbach}},\ }\bibfield  {title} {\enquote {\bibinfo {title} {{Fast
  periodic Gaussian density fitting by range separation}},}\ }\href
  {https://doi.org/10.1063/5.0046617} {\bibfield  {journal} {\bibinfo
  {journal} {J. Chem. Phys.}\ }\textbf {\bibinfo {volume} {154}},\ \bibinfo
  {pages} {131104} (\bibinfo {year} {2021}{\natexlab{a}})}\BibitemShut
  {NoStop}%
\bibitem [{\citenamefont {Ye}\ and\ \citenamefont
  {Berkelbach}(2021{\natexlab{b}})}]{Ye21JCPb}%
  \BibitemOpen
  \bibfield  {author} {\bibinfo {author} {\bibfnamefont {H.-Z.}\ \bibnamefont
  {Ye}}\ and\ \bibinfo {author} {\bibfnamefont {T.~C.}\ \bibnamefont
  {Berkelbach}},\ }\bibfield  {title} {\enquote {\bibinfo {title} {{Tight
  distance-dependent estimators for screening two-center and three-center
  short-range Coulomb integrals over Gaussian basis functions}},}\ }\href
  {https://doi.org/10.1063/5.0064151} {\bibfield  {journal} {\bibinfo
  {journal} {J. Chem. Phys.}\ }\textbf {\bibinfo {volume} {155}},\ \bibinfo
  {pages} {124106} (\bibinfo {year} {2021}{\natexlab{b}})}\BibitemShut
  {NoStop}%
\bibitem [{\citenamefont {Dinh}\ \emph {et~al.}(2026)\citenamefont {Dinh},
  \citenamefont {Rettig}, \citenamefont {Feng},\ and\ \citenamefont
  {Lee}}]{Dinh26JCP}%
  \BibitemOpen
  \bibfield  {author} {\bibinfo {author} {\bibfnamefont {H.~Q.}\ \bibnamefont
  {Dinh}}, \bibinfo {author} {\bibfnamefont {A.}~\bibnamefont {Rettig}},
  \bibinfo {author} {\bibfnamefont {X.}~\bibnamefont {Feng}},\ and\ \bibinfo
  {author} {\bibfnamefont {J.}~\bibnamefont {Lee}},\ }\bibfield  {title}
  {\enquote {\bibinfo {title} {Efficient all-electron periodic
  fourier-transformed coulomb method},}\ }\href
  {https://doi.org/10.1063/5.0303084} {\bibfield  {journal} {\bibinfo
  {journal} {J. Chem. Phys.}\ }\textbf {\bibinfo {volume} {164}},\ \bibinfo
  {pages} {054110} (\bibinfo {year} {2026})}\BibitemShut {NoStop}%
\bibitem [{\citenamefont {Varandas}(2019)}]{Varandas19PCCP}%
  \BibitemOpen
  \bibfield  {author} {\bibinfo {author} {\bibfnamefont {A.~J.~C.}\
  \bibnamefont {Varandas}},\ }\bibfield  {title} {\enquote {\bibinfo {title}
  {Cbs extrapolation of hartree–fock energy: Pople and dunning basis sets
  hand-to-hand on the endeavour},}\ }\href {https://doi.org/10.1039/c8cp07847e}
  {\bibfield  {journal} {\bibinfo  {journal} {Phys. Chem. Chem. Phys.}\
  }\textbf {\bibinfo {volume} {21}},\ \bibinfo {pages} {8022--8034} (\bibinfo
  {year} {2019})}\BibitemShut {NoStop}%
\bibitem [{\citenamefont {Helgaker}\ \emph {et~al.}(1997)\citenamefont
  {Helgaker}, \citenamefont {Klopper}, \citenamefont {Koch},\ and\
  \citenamefont {Noga}}]{Helgaker97JCP}%
  \BibitemOpen
  \bibfield  {author} {\bibinfo {author} {\bibfnamefont {T.}~\bibnamefont
  {Helgaker}}, \bibinfo {author} {\bibfnamefont {W.}~\bibnamefont {Klopper}},
  \bibinfo {author} {\bibfnamefont {H.}~\bibnamefont {Koch}},\ and\ \bibinfo
  {author} {\bibfnamefont {J.}~\bibnamefont {Noga}},\ }\bibfield  {title}
  {\enquote {\bibinfo {title} {Basis-set convergence of correlated calculations
  on water},}\ }\href {https://doi.org/10.1063/1.473863} {\bibfield  {journal}
  {\bibinfo  {journal} {J. Chem. Phys.}\ }\textbf {\bibinfo {volume} {106}},\
  \bibinfo {pages} {9639--9646} (\bibinfo {year} {1997})}\BibitemShut {NoStop}%
\bibitem [{\citenamefont {Halkier}\ \emph {et~al.}(1998)\citenamefont
  {Halkier}, \citenamefont {Helgaker}, \citenamefont {Jørgensen},
  \citenamefont {Klopper}, \citenamefont {Koch}, \citenamefont {Olsen},\ and\
  \citenamefont {Wilson}}]{Halkier98CPL}%
  \BibitemOpen
  \bibfield  {author} {\bibinfo {author} {\bibfnamefont {A.}~\bibnamefont
  {Halkier}}, \bibinfo {author} {\bibfnamefont {T.}~\bibnamefont {Helgaker}},
  \bibinfo {author} {\bibfnamefont {P.}~\bibnamefont {Jørgensen}}, \bibinfo
  {author} {\bibfnamefont {W.}~\bibnamefont {Klopper}}, \bibinfo {author}
  {\bibfnamefont {H.}~\bibnamefont {Koch}}, \bibinfo {author} {\bibfnamefont
  {J.}~\bibnamefont {Olsen}},\ and\ \bibinfo {author} {\bibfnamefont {A.~K.}\
  \bibnamefont {Wilson}},\ }\bibfield  {title} {\enquote {\bibinfo {title}
  {Basis-set convergence in correlated calculations on ne, n2, and h2o},}\
  }\href {https://doi.org/10.1016/S0009-2614(98)00111-0} {\bibfield  {journal}
  {\bibinfo  {journal} {Chem. Phys. Lett.}\ }\textbf {\bibinfo {volume}
  {286}},\ \bibinfo {pages} {243--252} (\bibinfo {year} {1998})}\BibitemShut
  {NoStop}%
\bibitem [{\citenamefont {Schimka}, \citenamefont {Harl},\ and\ \citenamefont
  {Kresse}(2011)}]{Schimka11JCP}%
  \BibitemOpen
  \bibfield  {author} {\bibinfo {author} {\bibfnamefont {L.}~\bibnamefont
  {Schimka}}, \bibinfo {author} {\bibfnamefont {J.}~\bibnamefont {Harl}},\ and\
  \bibinfo {author} {\bibfnamefont {G.}~\bibnamefont {Kresse}},\ }\bibfield
  {title} {\enquote {\bibinfo {title} {Improved hybrid functional for solids:
  The hsesol functional},}\ }\href {https://doi.org/10.1063/1.3524336}
  {\bibfield  {journal} {\bibinfo  {journal} {J. Chem. Phys.}\ }\textbf
  {\bibinfo {volume} {134}},\ \bibinfo {pages} {024116} (\bibinfo {year}
  {2011})}\BibitemShut {NoStop}%
\bibitem [{\citenamefont {Boys}\ and\ \citenamefont
  {Bernardi}(1970)}]{Boys70MP}%
  \BibitemOpen
  \bibfield  {author} {\bibinfo {author} {\bibfnamefont {S.~F.}\ \bibnamefont
  {Boys}}\ and\ \bibinfo {author} {\bibfnamefont {F.}~\bibnamefont
  {Bernardi}},\ }\bibfield  {title} {\enquote {\bibinfo {title} {The
  calculation of small molecular interactions by the differences of separate
  total energies. some procedures with reduced errors},}\ }\href
  {https://doi.org/10.1080/00268977000101561} {\bibfield  {journal} {\bibinfo
  {journal} {Mol. Phys.}\ }\textbf {\bibinfo {volume} {19}},\ \bibinfo {pages}
  {553--566} (\bibinfo {year} {1970})}\BibitemShut {NoStop}%
\bibitem [{\citenamefont {Baldereschi}(1973)}]{Baldereschi73PRB}%
  \BibitemOpen
  \bibfield  {author} {\bibinfo {author} {\bibfnamefont {A.}~\bibnamefont
  {Baldereschi}},\ }\bibfield  {title} {\enquote {\bibinfo {title} {Mean-value
  point in the brillouin zone},}\ }\href
  {https://doi.org/10.1103/PhysRevB.7.5212} {\bibfield  {journal} {\bibinfo
  {journal} {Phys. Rev. B}\ }\textbf {\bibinfo {volume} {7}},\ \bibinfo {pages}
  {5212--5215} (\bibinfo {year} {1973})}\BibitemShut {NoStop}%
\bibitem [{\citenamefont {Drummond}(2026)}]{Drummond26JPCM}%
  \BibitemOpen
  \bibfield  {author} {\bibinfo {author} {\bibfnamefont {N.~D.}\ \bibnamefont
  {Drummond}},\ }\bibfield  {title} {\enquote {\bibinfo {title} {Baldereschi
  mean value points for three-dimensional bravais lattices},}\ }\href
  {https://doi.org/10.1088/1361-648X/ae9c16} {\bibfield  {journal} {\bibinfo
  {journal} {J. Phys. Condens. Matter}\ }\textbf {\bibinfo {volume} {38}},\
  \bibinfo {pages} {345703} (\bibinfo {year} {2026})}\BibitemShut {NoStop}%
\bibitem [{\citenamefont {Grimme}(2003)}]{Grimme03JCP}%
  \BibitemOpen
  \bibfield  {author} {\bibinfo {author} {\bibfnamefont {S.}~\bibnamefont
  {Grimme}},\ }\bibfield  {title} {\enquote {\bibinfo {title} {Improved
  second-order møller–plesset perturbation theory by separate scaling of
  parallel- and antiparallel-spin pair correlation energies},}\ }\href
  {https://doi.org/10.1063/1.1569242} {\bibfield  {journal} {\bibinfo
  {journal} {J. Chem. Phys.}\ }\textbf {\bibinfo {volume} {118}},\ \bibinfo
  {pages} {9095--9102} (\bibinfo {year} {2003})}\BibitemShut {NoStop}%
\bibitem [{\citenamefont {Paier}\ \emph {et~al.}(2005)\citenamefont {Paier},
  \citenamefont {Hirschl}, \citenamefont {Marsman},\ and\ \citenamefont
  {Kresse}}]{Paier05JCP}%
  \BibitemOpen
  \bibfield  {author} {\bibinfo {author} {\bibfnamefont {J.}~\bibnamefont
  {Paier}}, \bibinfo {author} {\bibfnamefont {R.}~\bibnamefont {Hirschl}},
  \bibinfo {author} {\bibfnamefont {M.}~\bibnamefont {Marsman}},\ and\ \bibinfo
  {author} {\bibfnamefont {G.}~\bibnamefont {Kresse}},\ }\bibfield  {title}
  {\enquote {\bibinfo {title} {The perdew--burke--ernzerhof
  exchange-correlation functional applied to the g2-1 test set using a
  plane-wave basis set},}\ }\href {https://doi.org/10.1063/1.1926272}
  {\bibfield  {journal} {\bibinfo  {journal} {J. Chem. Phys.}\ }\textbf
  {\bibinfo {volume} {122}},\ \bibinfo {pages} {234102} (\bibinfo {year}
  {2005})}\BibitemShut {NoStop}%
\bibitem [{\citenamefont {Broqvist}, \citenamefont {Alkauskas},\ and\
  \citenamefont {Pasquarello}(2009)}]{Broqvist09PRB}%
  \BibitemOpen
  \bibfield  {author} {\bibinfo {author} {\bibfnamefont {P.}~\bibnamefont
  {Broqvist}}, \bibinfo {author} {\bibfnamefont {A.}~\bibnamefont
  {Alkauskas}},\ and\ \bibinfo {author} {\bibfnamefont {A.}~\bibnamefont
  {Pasquarello}},\ }\bibfield  {title} {\enquote {\bibinfo {title}
  {Hybrid-functional calculations with plane-wave basis sets: Effect of
  singularity correction on total energies, energy eigenvalues, and defect
  energy levels},}\ }\href {https://doi.org/10.1103/PhysRevB.80.085114}
  {\bibfield  {journal} {\bibinfo  {journal} {Phys. Rev. B}\ }\textbf {\bibinfo
  {volume} {80}},\ \bibinfo {pages} {085114} (\bibinfo {year}
  {2009})}\BibitemShut {NoStop}%
\bibitem [{\citenamefont {Xing}\ and\ \citenamefont {Lin}(2024)}]{Xing24PRX}%
  \BibitemOpen
  \bibfield  {author} {\bibinfo {author} {\bibfnamefont {X.}~\bibnamefont
  {Xing}}\ and\ \bibinfo {author} {\bibfnamefont {L.}~\bibnamefont {Lin}},\
  }\bibfield  {title} {\enquote {\bibinfo {title} {Inverse volume scaling of
  finite-size error in periodic coupled cluster theory},}\ }\href
  {https://doi.org/10.1103/PhysRevX.14.011059} {\bibfield  {journal} {\bibinfo
  {journal} {Phys. Rev. X}\ }\textbf {\bibinfo {volume} {14}},\ \bibinfo
  {pages} {011059} (\bibinfo {year} {2024})}\BibitemShut {NoStop}%
\bibitem [{\citenamefont {Rolik}\ and\ \citenamefont
  {K\'{a}llay}(2011)}]{Rolik11JCP}%
  \BibitemOpen
  \bibfield  {author} {\bibinfo {author} {\bibfnamefont {Z.}~\bibnamefont
  {Rolik}}\ and\ \bibinfo {author} {\bibfnamefont {M.}~\bibnamefont
  {K\'{a}llay}},\ }\bibfield  {title} {\enquote {\bibinfo {title} {{A
  general-order local coupled-cluster method based on the cluster-in-molecule
  approach}},}\ }\href {https://doi.org/10.1063/1.3632085} {\bibfield
  {journal} {\bibinfo  {journal} {J. Chem. Phys.}\ }\textbf {\bibinfo {volume}
  {135}},\ \bibinfo {pages} {104111} (\bibinfo {year} {2011})}\BibitemShut
  {NoStop}%
\bibitem [{\citenamefont {Rolik}\ \emph {et~al.}(2013)\citenamefont {Rolik},
  \citenamefont {Szegedy}, \citenamefont {Ladj\'{a}nszki}, \citenamefont
  {Lad\'{o}czki},\ and\ \citenamefont {K\'{a}llay}}]{Rolik13JCP}%
  \BibitemOpen
  \bibfield  {author} {\bibinfo {author} {\bibfnamefont {Z.}~\bibnamefont
  {Rolik}}, \bibinfo {author} {\bibfnamefont {L.}~\bibnamefont {Szegedy}},
  \bibinfo {author} {\bibfnamefont {I.}~\bibnamefont {Ladj\'{a}nszki}},
  \bibinfo {author} {\bibfnamefont {B.}~\bibnamefont {Lad\'{o}czki}},\ and\
  \bibinfo {author} {\bibfnamefont {M.}~\bibnamefont {K\'{a}llay}},\ }\bibfield
   {title} {\enquote {\bibinfo {title} {{An efficient linear-scaling CCSD(T)
  method based on local natural orbitals}},}\ }\href
  {https://doi.org/10.1063/1.4819401} {\bibfield  {journal} {\bibinfo
  {journal} {J. Chem. Phys.}\ }\textbf {\bibinfo {volume} {139}},\ \bibinfo
  {pages} {094105} (\bibinfo {year} {2013})}\BibitemShut {NoStop}%
\bibitem [{\citenamefont {Sosa}\ \emph {et~al.}(1989)\citenamefont {Sosa},
  \citenamefont {Geertsen}, \citenamefont {Trucks}, \citenamefont {Bartlett},\
  and\ \citenamefont {Franz}}]{Sosa89CPL}%
  \BibitemOpen
  \bibfield  {author} {\bibinfo {author} {\bibfnamefont {C.}~\bibnamefont
  {Sosa}}, \bibinfo {author} {\bibfnamefont {J.}~\bibnamefont {Geertsen}},
  \bibinfo {author} {\bibfnamefont {G.~W.}\ \bibnamefont {Trucks}}, \bibinfo
  {author} {\bibfnamefont {R.~J.}\ \bibnamefont {Bartlett}},\ and\ \bibinfo
  {author} {\bibfnamefont {J.~A.}\ \bibnamefont {Franz}},\ }\bibfield  {title}
  {\enquote {\bibinfo {title} {Selection of the reduced virtual space for
  correlated calculations. an application to the energy and dipole moment of
  h2o},}\ }\href {https://doi.org/10.1016/0009-2614(89)87399-3} {\bibfield
  {journal} {\bibinfo  {journal} {Chem. Phys. Lett.}\ }\textbf {\bibinfo
  {volume} {159}},\ \bibinfo {pages} {148--154} (\bibinfo {year}
  {1989})}\BibitemShut {NoStop}%
\bibitem [{\citenamefont {Taube}\ and\ \citenamefont
  {Bartlett}(2008)}]{Taube08JCP}%
  \BibitemOpen
  \bibfield  {author} {\bibinfo {author} {\bibfnamefont {A.~G.}\ \bibnamefont
  {Taube}}\ and\ \bibinfo {author} {\bibfnamefont {R.~J.}\ \bibnamefont
  {Bartlett}},\ }\bibfield  {title} {\enquote {\bibinfo {title} {Frozen natural
  orbital coupled-cluster theory: Forces and application to decomposition of
  nitroethane},}\ }\href {https://doi.org/10.1063/1.2902285} {\bibfield
  {journal} {\bibinfo  {journal} {J. Chem. Phys.}\ }\textbf {\bibinfo {volume}
  {128}},\ \bibinfo {pages} {164101} (\bibinfo {year} {2008})}\BibitemShut
  {NoStop}%
\bibitem [{\citenamefont {Landau}\ \emph {et~al.}(2010)\citenamefont {Landau},
  \citenamefont {Khistyaev}, \citenamefont {Dolgikh},\ and\ \citenamefont
  {Krylov}}]{Landau10JCP}%
  \BibitemOpen
  \bibfield  {author} {\bibinfo {author} {\bibfnamefont {A.}~\bibnamefont
  {Landau}}, \bibinfo {author} {\bibfnamefont {K.}~\bibnamefont {Khistyaev}},
  \bibinfo {author} {\bibfnamefont {S.}~\bibnamefont {Dolgikh}},\ and\ \bibinfo
  {author} {\bibfnamefont {A.~I.}\ \bibnamefont {Krylov}},\ }\bibfield  {title}
  {\enquote {\bibinfo {title} {Frozen natural orbitals for ionized states
  within equation-of-motion coupled-cluster formalism},}\ }\href
  {https://doi.org/10.1063/1.3276630} {\bibfield  {journal} {\bibinfo
  {journal} {J. Chem. Phys.}\ }\textbf {\bibinfo {volume} {132}},\ \bibinfo
  {pages} {014109} (\bibinfo {year} {2010})}\BibitemShut {NoStop}%
\bibitem [{\citenamefont {Grimsditch}, \citenamefont {Zouboulis},\ and\
  \citenamefont {Polian}(1994)}]{Grimsditch94JAP}%
  \BibitemOpen
  \bibfield  {author} {\bibinfo {author} {\bibfnamefont {M.}~\bibnamefont
  {Grimsditch}}, \bibinfo {author} {\bibfnamefont {E.~S.}\ \bibnamefont
  {Zouboulis}},\ and\ \bibinfo {author} {\bibfnamefont {A.}~\bibnamefont
  {Polian}},\ }\bibfield  {title} {\enquote {\bibinfo {title} {Elastic
  constants of boron nitride},}\ }\href {https://doi.org/10.1063/1.357757}
  {\bibfield  {journal} {\bibinfo  {journal} {J. Appl. Phys.}\ }\textbf
  {\bibinfo {volume} {76}},\ \bibinfo {pages} {832--834} (\bibinfo {year}
  {1994})}\BibitemShut {NoStop}%
\bibitem [{\citenamefont {Manghnani}(1998)}]{Manghnani1998}%
  \BibitemOpen
  \bibfield  {author} {\bibinfo {author} {\bibfnamefont {M.~H.}\ \bibnamefont
  {Manghnani}},\ }\bibfield  {title} {\enquote {\bibinfo {title} {Elastic
  properties of {cBN} as a function of temperature and pressure: A comparison
  with diamond},}\ }in\ \href@noop {} {\emph {\bibinfo {booktitle} {Proceedings
  of the 5th {NIRIM} International Symposium on Advanced Materials ({ISAM}
  '98)}}}\ (\bibinfo  {publisher} {National Institute for Research in Inorganic
  Materials},\ \bibinfo {address} {Tsukuba, Japan},\ \bibinfo {year} {1998})\
  pp.\ \bibinfo {pages} {73--78},\ \bibinfo {note} {march 1--5,
  1998}\BibitemShut {NoStop}%
\bibitem [{\citenamefont {Goncharov}\ \emph
  {et~al.}(2007{\natexlab{a}})\citenamefont {Goncharov}, \citenamefont
  {Crowhurst}, \citenamefont {Dewhurst}, \citenamefont {Sharma}, \citenamefont
  {Sanloup}, \citenamefont {Gregoryanz}, \citenamefont {Guignot},\ and\
  \citenamefont {Mezouar}}]{Goncharov07PRB}%
  \BibitemOpen
  \bibfield  {author} {\bibinfo {author} {\bibfnamefont {A.~F.}\ \bibnamefont
  {Goncharov}}, \bibinfo {author} {\bibfnamefont {J.~C.}\ \bibnamefont
  {Crowhurst}}, \bibinfo {author} {\bibfnamefont {J.~K.}\ \bibnamefont
  {Dewhurst}}, \bibinfo {author} {\bibfnamefont {S.}~\bibnamefont {Sharma}},
  \bibinfo {author} {\bibfnamefont {C.}~\bibnamefont {Sanloup}}, \bibinfo
  {author} {\bibfnamefont {E.}~\bibnamefont {Gregoryanz}}, \bibinfo {author}
  {\bibfnamefont {N.}~\bibnamefont {Guignot}},\ and\ \bibinfo {author}
  {\bibfnamefont {M.}~\bibnamefont {Mezouar}},\ }\bibfield  {title} {\enquote
  {\bibinfo {title} {Thermal equation of state of cubic boron nitride:
  Implications for a high-temperature pressure scale},}\ }\href
  {https://doi.org/10.1103/PhysRevB.75.224114} {\bibfield  {journal} {\bibinfo
  {journal} {Phys. Rev. B}\ }\textbf {\bibinfo {volume} {75}},\ \bibinfo
  {pages} {224114} (\bibinfo {year} {2007}{\natexlab{a}})}\BibitemShut
  {NoStop}%
\bibitem [{\citenamefont {Goncharov}\ \emph
  {et~al.}(2007{\natexlab{b}})\citenamefont {Goncharov}, \citenamefont
  {Sinogeikin}, \citenamefont {Crowhurst}, \citenamefont {Ahart}, \citenamefont
  {Lakshtanov}, \citenamefont {Prakapenka}, \citenamefont {Bass}, \citenamefont
  {Beck}, \citenamefont {Tkachev}, \citenamefont {Zaug},\ and\ \citenamefont
  {Fei}}]{Goncharov07HPR}%
  \BibitemOpen
  \bibfield  {author} {\bibinfo {author} {\bibfnamefont {A.~F.}\ \bibnamefont
  {Goncharov}}, \bibinfo {author} {\bibfnamefont {S.}~\bibnamefont
  {Sinogeikin}}, \bibinfo {author} {\bibfnamefont {J.~C.}\ \bibnamefont
  {Crowhurst}}, \bibinfo {author} {\bibfnamefont {M.}~\bibnamefont {Ahart}},
  \bibinfo {author} {\bibfnamefont {D.}~\bibnamefont {Lakshtanov}}, \bibinfo
  {author} {\bibfnamefont {V.}~\bibnamefont {Prakapenka}}, \bibinfo {author}
  {\bibfnamefont {J.}~\bibnamefont {Bass}}, \bibinfo {author} {\bibfnamefont
  {P.}~\bibnamefont {Beck}}, \bibinfo {author} {\bibfnamefont {S.~N.}\
  \bibnamefont {Tkachev}}, \bibinfo {author} {\bibfnamefont {J.~M.}\
  \bibnamefont {Zaug}},\ and\ \bibinfo {author} {\bibfnamefont
  {Y.}~\bibnamefont {Fei}},\ }\bibfield  {title} {\enquote {\bibinfo {title}
  {Cubic boron nitride as a primary calibrant for a high temperature pressure
  scale},}\ }\href {https://doi.org/10.1080/08957950701659726} {\bibfield
  {journal} {\bibinfo  {journal} {High Press. Res.}\ }\textbf {\bibinfo
  {volume} {27}},\ \bibinfo {pages} {409--417} (\bibinfo {year}
  {2007}{\natexlab{b}})}\BibitemShut {NoStop}%
\bibitem [{\citenamefont {Prascher}\ \emph {et~al.}(2011)\citenamefont
  {Prascher}, \citenamefont {Woon}, \citenamefont {Peterson}, \citenamefont
  {Dunning},\ and\ \citenamefont {Wilson}}]{Prascher11TCA}%
  \BibitemOpen
  \bibfield  {author} {\bibinfo {author} {\bibfnamefont {B.~P.}\ \bibnamefont
  {Prascher}}, \bibinfo {author} {\bibfnamefont {D.~E.}\ \bibnamefont {Woon}},
  \bibinfo {author} {\bibfnamefont {K.~A.}\ \bibnamefont {Peterson}}, \bibinfo
  {author} {\bibfnamefont {T.~H.}\ \bibnamefont {Dunning}},\ and\ \bibinfo
  {author} {\bibfnamefont {A.~K.}\ \bibnamefont {Wilson}},\ }\bibfield  {title}
  {\enquote {\bibinfo {title} {Gaussian basis sets for use in correlated
  molecular calculations. vii. valence, core-valence, and scalar relativistic
  basis sets for li, be, na, and mg},}\ }\href
  {https://doi.org/10.1007/s00214-010-0764-0} {\bibfield  {journal} {\bibinfo
  {journal} {Theor. Chim. Acta}\ }\textbf {\bibinfo {volume} {128}},\ \bibinfo
  {pages} {69--82} (\bibinfo {year} {2011})}\BibitemShut {NoStop}%
\end{thebibliography}%


\begin{thebibliography}{26}%
\makeatletter
\providecommand \@ifxundefined [1]{%
 \@ifx{#1\undefined}
}%
\providecommand \@ifnum [1]{%
 \ifnum #1\expandafter \@firstoftwo
 \else \expandafter \@secondoftwo
 \fi
}%
\providecommand \@ifx [1]{%
 \ifx #1\expandafter \@firstoftwo
 \else \expandafter \@secondoftwo
 \fi
}%
\providecommand \natexlab [1]{#1}%
\providecommand \enquote  [1]{``#1''}%
\providecommand \bibnamefont  [1]{#1}%
\providecommand \bibfnamefont [1]{#1}%
\providecommand \citenamefont [1]{#1}%
\providecommand \href@noop [0]{\@secondoftwo}%
\providecommand \href [0]{\begingroup \@sanitize@url \@href}%
\providecommand \@href[1]{\@@startlink{#1}\@@href}%
\providecommand \@@href[1]{\endgroup#1\@@endlink}%
\providecommand \@sanitize@url [0]{\catcode `\\12\catcode `\$12\catcode
  `\&12\catcode `\#12\catcode `\^12\catcode `\_12\catcode `\%12\relax}%
\providecommand \@@startlink[1]{}%
\providecommand \@@endlink[0]{}%
\providecommand \url  [0]{\begingroup\@sanitize@url \@url }%
\providecommand \@url [1]{\endgroup\@href {#1}{\urlprefix }}%
\providecommand \urlprefix  [0]{URL }%
\providecommand \Eprint [0]{\href }%
\providecommand \doibase [0]{https://doi.org/}%
\providecommand \selectlanguage [0]{\@gobble}%
\providecommand \bibinfo  [0]{\@secondoftwo}%
\providecommand \bibfield  [0]{\@secondoftwo}%
\providecommand \translation [1]{[#1]}%
\providecommand \BibitemOpen [0]{}%
\providecommand \bibitemStop [0]{}%
\providecommand \bibitemNoStop [0]{.\EOS\space}%
\providecommand \EOS [0]{\spacefactor3000\relax}%
\providecommand \BibitemShut  [1]{\csname bibitem#1\endcsname}%
\let\auto@bib@innerbib\@empty
\bibitem [{\citenamefont {Bosoni}\ \emph {et~al.}(2024)\citenamefont {Bosoni},
  \citenamefont {Beal}, \citenamefont {Bercx}, \citenamefont {Blaha},
  \citenamefont {Bl{\"u}gel}, \citenamefont {Br{\"o}der}, \citenamefont
  {Callsen}, \citenamefont {Cottenier}, \citenamefont {Degomme}, \citenamefont
  {Dikan}, \citenamefont {Eimre}, \citenamefont {Flage-Larsen}, \citenamefont
  {Fornari}, \citenamefont {Garcia}, \citenamefont {Genovese}, \citenamefont
  {Giantomassi}, \citenamefont {Huber}, \citenamefont {Janssen}, \citenamefont
  {Kastlunger}, \citenamefont {Krack}, \citenamefont {Kresse}, \citenamefont
  {K{\"u}hne}, \citenamefont {Lejaeghere}, \citenamefont {Madsen},
  \citenamefont {Marsman}, \citenamefont {Marzari}, \citenamefont {Michalicek},
  \citenamefont {Mirhosseini}, \citenamefont {M{\"u}ller}, \citenamefont
  {Petretto}, \citenamefont {Pickard}, \citenamefont {Ponc{\'e}}, \citenamefont
  {Rignanese}, \citenamefont {Rubel}, \citenamefont {Ruh}, \citenamefont
  {Sluydts}, \citenamefont {Vanpoucke}, \citenamefont {Vijay}, \citenamefont
  {Wolloch}, \citenamefont {Wortmann}, \citenamefont {Yakutovich},
  \citenamefont {Yu}, \citenamefont {Zadoks}, \citenamefont {Zhu},\ and\
  \citenamefont {Pizzi}}]{Bosoni24NRP}%
  \BibitemOpen
  \bibfield  {author} {\bibinfo {author} {\bibfnamefont {E.}~\bibnamefont
  {Bosoni}}, \bibinfo {author} {\bibfnamefont {L.}~\bibnamefont {Beal}},
  \bibinfo {author} {\bibfnamefont {M.}~\bibnamefont {Bercx}}, \bibinfo
  {author} {\bibfnamefont {P.}~\bibnamefont {Blaha}}, \bibinfo {author}
  {\bibfnamefont {S.}~\bibnamefont {Bl{\"u}gel}}, \bibinfo {author}
  {\bibfnamefont {J.}~\bibnamefont {Br{\"o}der}}, \bibinfo {author}
  {\bibfnamefont {M.}~\bibnamefont {Callsen}}, \bibinfo {author} {\bibfnamefont
  {S.}~\bibnamefont {Cottenier}}, \bibinfo {author} {\bibfnamefont
  {A.}~\bibnamefont {Degomme}}, \bibinfo {author} {\bibfnamefont
  {V.}~\bibnamefont {Dikan}}, \bibinfo {author} {\bibfnamefont
  {K.}~\bibnamefont {Eimre}}, \bibinfo {author} {\bibfnamefont
  {E.}~\bibnamefont {Flage-Larsen}}, \bibinfo {author} {\bibfnamefont
  {M.}~\bibnamefont {Fornari}}, \bibinfo {author} {\bibfnamefont
  {A.}~\bibnamefont {Garcia}}, \bibinfo {author} {\bibfnamefont
  {L.}~\bibnamefont {Genovese}}, \bibinfo {author} {\bibfnamefont
  {M.}~\bibnamefont {Giantomassi}}, \bibinfo {author} {\bibfnamefont {S.~P.}\
  \bibnamefont {Huber}}, \bibinfo {author} {\bibfnamefont {H.}~\bibnamefont
  {Janssen}}, \bibinfo {author} {\bibfnamefont {G.}~\bibnamefont {Kastlunger}},
  \bibinfo {author} {\bibfnamefont {M.}~\bibnamefont {Krack}}, \bibinfo
  {author} {\bibfnamefont {G.}~\bibnamefont {Kresse}}, \bibinfo {author}
  {\bibfnamefont {T.~D.}\ \bibnamefont {K{\"u}hne}}, \bibinfo {author}
  {\bibfnamefont {K.}~\bibnamefont {Lejaeghere}}, \bibinfo {author}
  {\bibfnamefont {G.~K.~H.}\ \bibnamefont {Madsen}}, \bibinfo {author}
  {\bibfnamefont {M.}~\bibnamefont {Marsman}}, \bibinfo {author} {\bibfnamefont
  {N.}~\bibnamefont {Marzari}}, \bibinfo {author} {\bibfnamefont
  {G.}~\bibnamefont {Michalicek}}, \bibinfo {author} {\bibfnamefont
  {H.}~\bibnamefont {Mirhosseini}}, \bibinfo {author} {\bibfnamefont
  {T.~M.~A.}\ \bibnamefont {M{\"u}ller}}, \bibinfo {author} {\bibfnamefont
  {G.}~\bibnamefont {Petretto}}, \bibinfo {author} {\bibfnamefont {C.~J.}\
  \bibnamefont {Pickard}}, \bibinfo {author} {\bibfnamefont {S.}~\bibnamefont
  {Ponc{\'e}}}, \bibinfo {author} {\bibfnamefont {G.-M.}\ \bibnamefont
  {Rignanese}}, \bibinfo {author} {\bibfnamefont {O.}~\bibnamefont {Rubel}},
  \bibinfo {author} {\bibfnamefont {T.}~\bibnamefont {Ruh}}, \bibinfo {author}
  {\bibfnamefont {M.}~\bibnamefont {Sluydts}}, \bibinfo {author} {\bibfnamefont
  {D.~E.~P.}\ \bibnamefont {Vanpoucke}}, \bibinfo {author} {\bibfnamefont
  {S.}~\bibnamefont {Vijay}}, \bibinfo {author} {\bibfnamefont
  {M.}~\bibnamefont {Wolloch}}, \bibinfo {author} {\bibfnamefont
  {D.}~\bibnamefont {Wortmann}}, \bibinfo {author} {\bibfnamefont {A.~V.}\
  \bibnamefont {Yakutovich}}, \bibinfo {author} {\bibfnamefont
  {J.}~\bibnamefont {Yu}}, \bibinfo {author} {\bibfnamefont {A.}~\bibnamefont
  {Zadoks}}, \bibinfo {author} {\bibfnamefont {B.}~\bibnamefont {Zhu}},\ and\
  \bibinfo {author} {\bibfnamefont {G.}~\bibnamefont {Pizzi}},\ }\bibfield
  {title} {\enquote {\bibinfo {title} {How to verify the precision of
  density-functional-theory implementations via reproducible and universal
  workflows},}\ }\href {https://doi.org/10.1038/s42254-023-00655-3} {\bibfield
  {journal} {\bibinfo  {journal} {Nat. Rev. Phys.}\ }\textbf {\bibinfo {volume}
  {6}},\ \bibinfo {pages} {45--58} (\bibinfo {year} {2024})}\BibitemShut
  {NoStop}%
\bibitem [{\citenamefont {Grimme}\ \emph {et~al.}(2010)\citenamefont {Grimme},
  \citenamefont {Antony}, \citenamefont {Ehrlich},\ and\ \citenamefont
  {Krieg}}]{Grimme10JCP}%
  \BibitemOpen
  \bibfield  {author} {\bibinfo {author} {\bibfnamefont {S.}~\bibnamefont
  {Grimme}}, \bibinfo {author} {\bibfnamefont {J.}~\bibnamefont {Antony}},
  \bibinfo {author} {\bibfnamefont {S.}~\bibnamefont {Ehrlich}},\ and\ \bibinfo
  {author} {\bibfnamefont {H.}~\bibnamefont {Krieg}},\ }\bibfield  {title}
  {\enquote {\bibinfo {title} {A consistent and accurate ab initio
  parametrization of density functional dispersion correction ({DFT-D}) for the
  94 elements {H--Pu}},}\ }\href {https://doi.org/10.1063/1.3382344} {\bibfield
   {journal} {\bibinfo  {journal} {The Journal of Chemical Physics}\ }\textbf
  {\bibinfo {volume} {132}},\ \bibinfo {pages} {154104} (\bibinfo {year}
  {2010})}\BibitemShut {NoStop}%
\bibitem [{\citenamefont {Chai}\ and\ \citenamefont
  {Head-Gordon}(2008)}]{Chai08JCP}%
  \BibitemOpen
  \bibfield  {author} {\bibinfo {author} {\bibfnamefont {J.-D.}\ \bibnamefont
  {Chai}}\ and\ \bibinfo {author} {\bibfnamefont {M.}~\bibnamefont
  {Head-Gordon}},\ }\bibfield  {title} {\enquote {\bibinfo {title} {Systematic
  optimization of long-range corrected hybrid density functionals},}\ }\href
  {https://doi.org/10.1063/1.2834918} {\bibfield  {journal} {\bibinfo
  {journal} {The Journal of Chemical Physics}\ }\textbf {\bibinfo {volume}
  {128}},\ \bibinfo {pages} {084106} (\bibinfo {year} {2008})}\BibitemShut
  {NoStop}%
\bibitem [{\citenamefont {Lin}\ \emph {et~al.}(2013)\citenamefont {Lin},
  \citenamefont {Li}, \citenamefont {Mao},\ and\ \citenamefont
  {Chai}}]{Lin13JCTC}%
  \BibitemOpen
  \bibfield  {author} {\bibinfo {author} {\bibfnamefont {Y.-S.}\ \bibnamefont
  {Lin}}, \bibinfo {author} {\bibfnamefont {G.-D.}\ \bibnamefont {Li}},
  \bibinfo {author} {\bibfnamefont {S.-P.}\ \bibnamefont {Mao}},\ and\ \bibinfo
  {author} {\bibfnamefont {J.-D.}\ \bibnamefont {Chai}},\ }\bibfield  {title}
  {\enquote {\bibinfo {title} {Long-range corrected hybrid density functionals
  with improved dispersion corrections},}\ }\href
  {https://doi.org/10.1021/ct300715s} {\bibfield  {journal} {\bibinfo
  {journal} {Journal of Chemical Theory and Computation}\ }\textbf {\bibinfo
  {volume} {9}},\ \bibinfo {pages} {263--272} (\bibinfo {year}
  {2013})}\BibitemShut {NoStop}%
\bibitem [{\citenamefont {Weigend}\ and\ \citenamefont
  {Ahlrichs}(2005)}]{Weigend05PCCP}%
  \BibitemOpen
  \bibfield  {author} {\bibinfo {author} {\bibfnamefont {F.}~\bibnamefont
  {Weigend}}\ and\ \bibinfo {author} {\bibfnamefont {R.}~\bibnamefont
  {Ahlrichs}},\ }\bibfield  {title} {\enquote {\bibinfo {title} {Balanced basis
  sets of split valence{,} triple zeta valence and quadruple zeta valence
  quality for h to rn: Design and assessment of accuracy},}\ }\href
  {https://doi.org/10.1039/B508541A} {\bibfield  {journal} {\bibinfo  {journal}
  {Phys. Chem. Chem. Phys.}\ }\textbf {\bibinfo {volume} {7}},\ \bibinfo
  {pages} {3297--3305} (\bibinfo {year} {2005})}\BibitemShut {NoStop}%
\bibitem [{\citenamefont {Neese}(2012)}]{Neese12WIRCMS}%
  \BibitemOpen
  \bibfield  {author} {\bibinfo {author} {\bibfnamefont {F.}~\bibnamefont
  {Neese}},\ }\bibfield  {title} {\enquote {\bibinfo {title} {The orca program
  system},}\ }\href {https://doi.org/10.1002/wcms.81} {\bibfield  {journal}
  {\bibinfo  {journal} {Wiley Interdiscip. Rev. Comput. Mol. Sci}\ }\textbf
  {\bibinfo {volume} {2}},\ \bibinfo {pages} {73--78} (\bibinfo {year}
  {2012})}\BibitemShut {NoStop}%
\bibitem [{\citenamefont {Schimka}, \citenamefont {Harl},\ and\ \citenamefont
  {Kresse}(2011)}]{Schimka11JCP}%
  \BibitemOpen
  \bibfield  {author} {\bibinfo {author} {\bibfnamefont {L.}~\bibnamefont
  {Schimka}}, \bibinfo {author} {\bibfnamefont {J.}~\bibnamefont {Harl}},\ and\
  \bibinfo {author} {\bibfnamefont {G.}~\bibnamefont {Kresse}},\ }\bibfield
  {title} {\enquote {\bibinfo {title} {Improved hybrid functional for solids:
  The hsesol functional},}\ }\href {https://doi.org/10.1063/1.3524336}
  {\bibfield  {journal} {\bibinfo  {journal} {J. Chem. Phys.}\ }\textbf
  {\bibinfo {volume} {134}},\ \bibinfo {pages} {024116} (\bibinfo {year}
  {2011})}\BibitemShut {NoStop}%
\bibitem [{\citenamefont {Bennett}\ \emph {et~al.}(2017)\citenamefont
  {Bennett}, \citenamefont {Melton}, \citenamefont {Annaberdiyev},
  \citenamefont {Wang}, \citenamefont {Shulenburger},\ and\ \citenamefont
  {Mitas}}]{Bennett17JCP}%
  \BibitemOpen
  \bibfield  {author} {\bibinfo {author} {\bibfnamefont {M.~C.}\ \bibnamefont
  {Bennett}}, \bibinfo {author} {\bibfnamefont {C.~A.}\ \bibnamefont {Melton}},
  \bibinfo {author} {\bibfnamefont {A.}~\bibnamefont {Annaberdiyev}}, \bibinfo
  {author} {\bibfnamefont {G.}~\bibnamefont {Wang}}, \bibinfo {author}
  {\bibfnamefont {L.}~\bibnamefont {Shulenburger}},\ and\ \bibinfo {author}
  {\bibfnamefont {L.}~\bibnamefont {Mitas}},\ }\bibfield  {title} {\enquote
  {\bibinfo {title} {A new generation of effective core potentials for
  correlated calculations},}\ }\href {https://doi.org/10.1063/1.4995643}
  {\bibfield  {journal} {\bibinfo  {journal} {J. Chem. Phys.}\ }\textbf
  {\bibinfo {volume} {147}},\ \bibinfo {pages} {224106} (\bibinfo {year}
  {2017})}\BibitemShut {NoStop}%
\bibitem [{\citenamefont {Ye}\ and\ \citenamefont
  {Berkelbach}(2022)}]{Ye22JCTC}%
  \BibitemOpen
  \bibfield  {author} {\bibinfo {author} {\bibfnamefont {H.-Z.}\ \bibnamefont
  {Ye}}\ and\ \bibinfo {author} {\bibfnamefont {T.~C.}\ \bibnamefont
  {Berkelbach}},\ }\bibfield  {title} {\enquote {\bibinfo {title}
  {Correlation-consistent gaussian basis sets for solids made simple},}\ }\href
  {https://doi.org/10.1021/acs.jctc.1c01245} {\bibfield  {journal} {\bibinfo
  {journal} {J. Chem. Theory Comput.}\ }\textbf {\bibinfo {volume} {18}},\
  \bibinfo {pages} {1595--1606} (\bibinfo {year} {2022})}\BibitemShut {NoStop}%
\bibitem [{\citenamefont {Sun}\ \emph {et~al.}(2018)\citenamefont {Sun},
  \citenamefont {Berkelbach}, \citenamefont {Blunt}, \citenamefont {Booth},
  \citenamefont {Guo}, \citenamefont {Li}, \citenamefont {Liu}, \citenamefont
  {McClain}, \citenamefont {Sayfutyarova}, \citenamefont {Sharma},
  \citenamefont {Wouters},\ and\ \citenamefont {Chan}}]{Sun18WIRCMS}%
  \BibitemOpen
  \bibfield  {author} {\bibinfo {author} {\bibfnamefont {Q.}~\bibnamefont
  {Sun}}, \bibinfo {author} {\bibfnamefont {T.~C.}\ \bibnamefont {Berkelbach}},
  \bibinfo {author} {\bibfnamefont {N.~S.}\ \bibnamefont {Blunt}}, \bibinfo
  {author} {\bibfnamefont {G.~H.}\ \bibnamefont {Booth}}, \bibinfo {author}
  {\bibfnamefont {S.}~\bibnamefont {Guo}}, \bibinfo {author} {\bibfnamefont
  {Z.}~\bibnamefont {Li}}, \bibinfo {author} {\bibfnamefont {J.}~\bibnamefont
  {Liu}}, \bibinfo {author} {\bibfnamefont {J.~D.}\ \bibnamefont {McClain}},
  \bibinfo {author} {\bibfnamefont {E.~R.}\ \bibnamefont {Sayfutyarova}},
  \bibinfo {author} {\bibfnamefont {S.}~\bibnamefont {Sharma}}, \bibinfo
  {author} {\bibfnamefont {S.}~\bibnamefont {Wouters}},\ and\ \bibinfo {author}
  {\bibfnamefont {G.~K.-L.}\ \bibnamefont {Chan}},\ }\bibfield  {title}
  {\enquote {\bibinfo {title} {Pyscf: the python-based simulations of chemistry
  framework},}\ }\href {https://doi.org/10.1002/wcms.1340} {\bibfield
  {journal} {\bibinfo  {journal} {Wiley Interdiscip. Rev. Comput. Mol. Sci}\
  }\textbf {\bibinfo {volume} {8}},\ \bibinfo {pages} {e1340} (\bibinfo {year}
  {2018})}\BibitemShut {NoStop}%
\bibitem [{\citenamefont {Sun}\ \emph {et~al.}(2020)\citenamefont {Sun},
  \citenamefont {Zhang}, \citenamefont {Banerjee}, \citenamefont {Bao},
  \citenamefont {Barbry}, \citenamefont {Blunt}, \citenamefont {Bogdanov},
  \citenamefont {Booth}, \citenamefont {Chen}, \citenamefont {Cui},
  \citenamefont {Eriksen}, \citenamefont {Gao}, \citenamefont {Guo},
  \citenamefont {Hermann}, \citenamefont {Hermes}, \citenamefont {Koh},
  \citenamefont {Koval}, \citenamefont {Lehtola}, \citenamefont {Li},
  \citenamefont {Liu}, \citenamefont {Mardirossian}, \citenamefont {McClain},
  \citenamefont {Motta}, \citenamefont {Mussard}, \citenamefont {Pham},
  \citenamefont {Pulkin}, \citenamefont {Purwanto}, \citenamefont {Robinson},
  \citenamefont {Ronca}, \citenamefont {Sayfutyarova}, \citenamefont
  {Scheurer}, \citenamefont {Schurkus}, \citenamefont {Smith}, \citenamefont
  {Sun}, \citenamefont {Sun}, \citenamefont {Upadhyay}, \citenamefont {Wagner},
  \citenamefont {Wang}, \citenamefont {White}, \citenamefont {Whitfield},
  \citenamefont {Williamson}, \citenamefont {Wouters}, \citenamefont {Yang},
  \citenamefont {Yu}, \citenamefont {Zhu}, \citenamefont {Berkelbach},
  \citenamefont {Sharma}, \citenamefont {Sokolov},\ and\ \citenamefont
  {Chan}}]{Sun20JCP}%
  \BibitemOpen
  \bibfield  {author} {\bibinfo {author} {\bibfnamefont {Q.}~\bibnamefont
  {Sun}}, \bibinfo {author} {\bibfnamefont {X.}~\bibnamefont {Zhang}}, \bibinfo
  {author} {\bibfnamefont {S.}~\bibnamefont {Banerjee}}, \bibinfo {author}
  {\bibfnamefont {P.}~\bibnamefont {Bao}}, \bibinfo {author} {\bibfnamefont
  {M.}~\bibnamefont {Barbry}}, \bibinfo {author} {\bibfnamefont {N.~S.}\
  \bibnamefont {Blunt}}, \bibinfo {author} {\bibfnamefont {N.~A.}\ \bibnamefont
  {Bogdanov}}, \bibinfo {author} {\bibfnamefont {G.~H.}\ \bibnamefont {Booth}},
  \bibinfo {author} {\bibfnamefont {J.}~\bibnamefont {Chen}}, \bibinfo {author}
  {\bibfnamefont {Z.-H.}\ \bibnamefont {Cui}}, \bibinfo {author} {\bibfnamefont
  {J.~J.}\ \bibnamefont {Eriksen}}, \bibinfo {author} {\bibfnamefont
  {Y.}~\bibnamefont {Gao}}, \bibinfo {author} {\bibfnamefont {S.}~\bibnamefont
  {Guo}}, \bibinfo {author} {\bibfnamefont {J.}~\bibnamefont {Hermann}},
  \bibinfo {author} {\bibfnamefont {M.~R.}\ \bibnamefont {Hermes}}, \bibinfo
  {author} {\bibfnamefont {K.}~\bibnamefont {Koh}}, \bibinfo {author}
  {\bibfnamefont {P.}~\bibnamefont {Koval}}, \bibinfo {author} {\bibfnamefont
  {S.}~\bibnamefont {Lehtola}}, \bibinfo {author} {\bibfnamefont
  {Z.}~\bibnamefont {Li}}, \bibinfo {author} {\bibfnamefont {J.}~\bibnamefont
  {Liu}}, \bibinfo {author} {\bibfnamefont {N.}~\bibnamefont {Mardirossian}},
  \bibinfo {author} {\bibfnamefont {J.~D.}\ \bibnamefont {McClain}}, \bibinfo
  {author} {\bibfnamefont {M.}~\bibnamefont {Motta}}, \bibinfo {author}
  {\bibfnamefont {B.}~\bibnamefont {Mussard}}, \bibinfo {author} {\bibfnamefont
  {H.~Q.}\ \bibnamefont {Pham}}, \bibinfo {author} {\bibfnamefont
  {A.}~\bibnamefont {Pulkin}}, \bibinfo {author} {\bibfnamefont
  {W.}~\bibnamefont {Purwanto}}, \bibinfo {author} {\bibfnamefont {P.~J.}\
  \bibnamefont {Robinson}}, \bibinfo {author} {\bibfnamefont {E.}~\bibnamefont
  {Ronca}}, \bibinfo {author} {\bibfnamefont {E.~R.}\ \bibnamefont
  {Sayfutyarova}}, \bibinfo {author} {\bibfnamefont {M.}~\bibnamefont
  {Scheurer}}, \bibinfo {author} {\bibfnamefont {H.~F.}\ \bibnamefont
  {Schurkus}}, \bibinfo {author} {\bibfnamefont {J.~E.~T.}\ \bibnamefont
  {Smith}}, \bibinfo {author} {\bibfnamefont {C.}~\bibnamefont {Sun}}, \bibinfo
  {author} {\bibfnamefont {S.-N.}\ \bibnamefont {Sun}}, \bibinfo {author}
  {\bibfnamefont {S.}~\bibnamefont {Upadhyay}}, \bibinfo {author}
  {\bibfnamefont {L.~K.}\ \bibnamefont {Wagner}}, \bibinfo {author}
  {\bibfnamefont {X.}~\bibnamefont {Wang}}, \bibinfo {author} {\bibfnamefont
  {A.}~\bibnamefont {White}}, \bibinfo {author} {\bibfnamefont {J.~D.}\
  \bibnamefont {Whitfield}}, \bibinfo {author} {\bibfnamefont {M.~J.}\
  \bibnamefont {Williamson}}, \bibinfo {author} {\bibfnamefont
  {S.}~\bibnamefont {Wouters}}, \bibinfo {author} {\bibfnamefont
  {J.}~\bibnamefont {Yang}}, \bibinfo {author} {\bibfnamefont {J.~M.}\
  \bibnamefont {Yu}}, \bibinfo {author} {\bibfnamefont {T.}~\bibnamefont
  {Zhu}}, \bibinfo {author} {\bibfnamefont {T.~C.}\ \bibnamefont {Berkelbach}},
  \bibinfo {author} {\bibfnamefont {S.}~\bibnamefont {Sharma}}, \bibinfo
  {author} {\bibfnamefont {A.~Y.}\ \bibnamefont {Sokolov}},\ and\ \bibinfo
  {author} {\bibfnamefont {G.~K.-L.}\ \bibnamefont {Chan}},\ }\bibfield
  {title} {\enquote {\bibinfo {title} {{Recent developments in the PySCF
  program package}},}\ }\href {https://doi.org/10.1063/5.0006074} {\bibfield
  {journal} {\bibinfo  {journal} {J. Chem. Phys.}\ }\textbf {\bibinfo {volume}
  {153}},\ \bibinfo {pages} {024109} (\bibinfo {year} {2020})}\BibitemShut
  {NoStop}%
\bibitem [{\citenamefont {Sun}\ \emph {et~al.}(2026)\citenamefont {Sun},
  \citenamefont {Hermes}, \citenamefont {Wu}, \citenamefont {Zhai},
  \citenamefont {Zhang}, \citenamefont {Ahmed}, \citenamefont {Aucar},
  \citenamefont {Backhouse}, \citenamefont {Banerjee}, \citenamefont {Bao},
  \citenamefont {Bogdanov}, \citenamefont {Bystrom}, \citenamefont {Chapoton},
  \citenamefont {Chen}, \citenamefont {Chernyshov}, \citenamefont {Clifford},
  \citenamefont {Cohen-Janes}, \citenamefont {Cui}, \citenamefont {Damour},
  \citenamefont {Dattani}, \citenamefont {Dittmer}, \citenamefont {Ehlert},
  \citenamefont {Eriksen}, \citenamefont {Evangelista}, \citenamefont {Ewing},
  \citenamefont {Farahvash}, \citenamefont {Focke}, \citenamefont {Gao},
  \citenamefont {Gasperich}, \citenamefont {Gillispie}, \citenamefont
  {Greiner}, \citenamefont {Hennefarth}, \citenamefont {Hermann}, \citenamefont
  {Hillenbrand}, \citenamefont {Huhtasalo}, \citenamefont {Ibrahim},
  \citenamefont {Jangid}, \citenamefont {Javaremi}, \citenamefont {Jenkins},
  \citenamefont {Jin}, \citenamefont {King}, \citenamefont {Kooi},
  \citenamefont {Kurian}, \citenamefont {Larsson}, \citenamefont {Lau},
  \citenamefont {Lee}, \citenamefont {Lehtola}, \citenamefont {Li},
  \citenamefont {Li}, \citenamefont {Li}, \citenamefont {Li}, \citenamefont
  {Li}, \citenamefont {Lykhin}, \citenamefont {Mahajan}, \citenamefont
  {Mauger}, \citenamefont {Mazo-Sevillano}, \citenamefont {Moussa},
  \citenamefont {Nakano}, \citenamefont {Neufeld}, \citenamefont {Peng},
  \citenamefont {Pham}, \citenamefont {Pinski}, \citenamefont {Pokhilko},
  \citenamefont {Pu}, \citenamefont {Qian}, \citenamefont {Quiton},
  \citenamefont {Schulze}, \citenamefont {Scott}, \citenamefont {Seal},
  \citenamefont {Serna}, \citenamefont {Smith}, \citenamefont {Smyser},
  \citenamefont {Stahl}, \citenamefont {Sun}, \citenamefont {Sung},
  \citenamefont {Trushin}, \citenamefont {Upadhyay}, \citenamefont {Vo},
  \citenamefont {Vogels}, \citenamefont {Wang}, \citenamefont {Wang},
  \citenamefont {Wang}, \citenamefont {Wang}, \citenamefont {Wang},
  \citenamefont {Williamson}, \citenamefont {Yang}, \citenamefont {Ye},
  \citenamefont {Yeh}, \citenamefont {Yu}, \citenamefont {Yu}, \citenamefont
  {Yu}, \citenamefont {Zhang}, \citenamefont {Zhang}, \citenamefont {Zhang},
  \citenamefont {Zhao}, \citenamefont {Zhou}, \citenamefont {Zhu},
  \citenamefont {Zhu}, \citenamefont {Berkelbach}, \citenamefont {Gagliardi},
  \citenamefont {Sharma}, \citenamefont {Sokolov},\ and\ \citenamefont
  {Kin-Lic~Chan}}]{Sun26JCP}%
  \BibitemOpen
  \bibfield  {author} {\bibinfo {author} {\bibfnamefont {Q.}~\bibnamefont
  {Sun}}, \bibinfo {author} {\bibfnamefont {M.~R.}\ \bibnamefont {Hermes}},
  \bibinfo {author} {\bibfnamefont {X.}~\bibnamefont {Wu}}, \bibinfo {author}
  {\bibfnamefont {H.}~\bibnamefont {Zhai}}, \bibinfo {author} {\bibfnamefont
  {X.}~\bibnamefont {Zhang}}, \bibinfo {author} {\bibfnamefont {A.~M.}\
  \bibnamefont {Ahmed}}, \bibinfo {author} {\bibfnamefont {J.~J.}\ \bibnamefont
  {Aucar}}, \bibinfo {author} {\bibfnamefont {O.~J.}\ \bibnamefont
  {Backhouse}}, \bibinfo {author} {\bibfnamefont {S.}~\bibnamefont {Banerjee}},
  \bibinfo {author} {\bibfnamefont {P.}~\bibnamefont {Bao}}, \bibinfo {author}
  {\bibfnamefont {N.~A.}\ \bibnamefont {Bogdanov}}, \bibinfo {author}
  {\bibfnamefont {K.}~\bibnamefont {Bystrom}}, \bibinfo {author} {\bibfnamefont
  {F.}~\bibnamefont {Chapoton}}, \bibinfo {author} {\bibfnamefont {N.-Y.}\
  \bibnamefont {Chen}}, \bibinfo {author} {\bibfnamefont {I.~Y.}\ \bibnamefont
  {Chernyshov}}, \bibinfo {author} {\bibfnamefont {H.~S.}\ \bibnamefont
  {Clifford}}, \bibinfo {author} {\bibfnamefont {S.}~\bibnamefont
  {Cohen-Janes}}, \bibinfo {author} {\bibfnamefont {Z.-H.}\ \bibnamefont
  {Cui}}, \bibinfo {author} {\bibfnamefont {Y.~D.}\ \bibnamefont {Damour}},
  \bibinfo {author} {\bibfnamefont {N.}~\bibnamefont {Dattani}}, \bibinfo
  {author} {\bibfnamefont {L.~B.}\ \bibnamefont {Dittmer}}, \bibinfo {author}
  {\bibfnamefont {S.}~\bibnamefont {Ehlert}}, \bibinfo {author} {\bibfnamefont
  {J.~J.}\ \bibnamefont {Eriksen}}, \bibinfo {author} {\bibfnamefont {F.~A.}\
  \bibnamefont {Evangelista}}, \bibinfo {author} {\bibfnamefont {S.~A.}\
  \bibnamefont {Ewing}}, \bibinfo {author} {\bibfnamefont {A.}~\bibnamefont
  {Farahvash}}, \bibinfo {author} {\bibfnamefont {K.}~\bibnamefont {Focke}},
  \bibinfo {author} {\bibfnamefont {Y.}~\bibnamefont {Gao}}, \bibinfo {author}
  {\bibfnamefont {K.~E.}\ \bibnamefont {Gasperich}}, \bibinfo {author}
  {\bibfnamefont {N.}~\bibnamefont {Gillispie}}, \bibinfo {author}
  {\bibfnamefont {J.}~\bibnamefont {Greiner}}, \bibinfo {author} {\bibfnamefont
  {M.~R.}\ \bibnamefont {Hennefarth}}, \bibinfo {author} {\bibfnamefont
  {J.}~\bibnamefont {Hermann}}, \bibinfo {author} {\bibfnamefont
  {C.}~\bibnamefont {Hillenbrand}}, \bibinfo {author} {\bibfnamefont
  {J.}~\bibnamefont {Huhtasalo}}, \bibinfo {author} {\bibfnamefont
  {B.}~\bibnamefont {Ibrahim}}, \bibinfo {author} {\bibfnamefont
  {B.}~\bibnamefont {Jangid}}, \bibinfo {author} {\bibfnamefont {A.~N.}\
  \bibnamefont {Javaremi}}, \bibinfo {author} {\bibfnamefont {A.~J.}\
  \bibnamefont {Jenkins}}, \bibinfo {author} {\bibfnamefont {Y.}~\bibnamefont
  {Jin}}, \bibinfo {author} {\bibfnamefont {D.~S.}\ \bibnamefont {King}},
  \bibinfo {author} {\bibfnamefont {D.~P.}\ \bibnamefont {Kooi}}, \bibinfo
  {author} {\bibfnamefont {J.~S.}\ \bibnamefont {Kurian}}, \bibinfo {author}
  {\bibfnamefont {H.~R.}\ \bibnamefont {Larsson}}, \bibinfo {author}
  {\bibfnamefont {B.~T.~G.}\ \bibnamefont {Lau}}, \bibinfo {author}
  {\bibfnamefont {S.}~\bibnamefont {Lee}}, \bibinfo {author} {\bibfnamefont
  {S.}~\bibnamefont {Lehtola}}, \bibinfo {author} {\bibfnamefont
  {C.}~\bibnamefont {Li}}, \bibinfo {author} {\bibfnamefont {H.}~\bibnamefont
  {Li}}, \bibinfo {author} {\bibfnamefont {J.}~\bibnamefont {Li}}, \bibinfo
  {author} {\bibfnamefont {R.}~\bibnamefont {Li}}, \bibinfo {author}
  {\bibfnamefont {S.}~\bibnamefont {Li}}, \bibinfo {author} {\bibfnamefont
  {A.~O.}\ \bibnamefont {Lykhin}}, \bibinfo {author} {\bibfnamefont
  {A.}~\bibnamefont {Mahajan}}, \bibinfo {author} {\bibfnamefont
  {N.}~\bibnamefont {Mauger}}, \bibinfo {author} {\bibfnamefont {P.~d.}\
  \bibnamefont {Mazo-Sevillano}}, \bibinfo {author} {\bibfnamefont
  {J.}~\bibnamefont {Moussa}}, \bibinfo {author} {\bibfnamefont
  {K.}~\bibnamefont {Nakano}}, \bibinfo {author} {\bibfnamefont {V.~A.}\
  \bibnamefont {Neufeld}}, \bibinfo {author} {\bibfnamefont {L.}~\bibnamefont
  {Peng}}, \bibinfo {author} {\bibfnamefont {H.~Q.}\ \bibnamefont {Pham}},
  \bibinfo {author} {\bibfnamefont {P.}~\bibnamefont {Pinski}}, \bibinfo
  {author} {\bibfnamefont {P.}~\bibnamefont {Pokhilko}}, \bibinfo {author}
  {\bibfnamefont {Z.}~\bibnamefont {Pu}}, \bibinfo {author} {\bibfnamefont
  {Y.}~\bibnamefont {Qian}}, \bibinfo {author} {\bibfnamefont {S.~J.}\
  \bibnamefont {Quiton}}, \bibinfo {author} {\bibfnamefont {W.~T.}\
  \bibnamefont {Schulze}}, \bibinfo {author} {\bibfnamefont {T.~R.}\
  \bibnamefont {Scott}}, \bibinfo {author} {\bibfnamefont {A.}~\bibnamefont
  {Seal}}, \bibinfo {author} {\bibfnamefont {J.~D.}\ \bibnamefont {Serna}},
  \bibinfo {author} {\bibfnamefont {J.~E.~T.}\ \bibnamefont {Smith}}, \bibinfo
  {author} {\bibfnamefont {K.~E.}\ \bibnamefont {Smyser}}, \bibinfo {author}
  {\bibfnamefont {T.}~\bibnamefont {Stahl}}, \bibinfo {author} {\bibfnamefont
  {C.}~\bibnamefont {Sun}}, \bibinfo {author} {\bibfnamefont {K.~J.}\
  \bibnamefont {Sung}}, \bibinfo {author} {\bibfnamefont {E.}~\bibnamefont
  {Trushin}}, \bibinfo {author} {\bibfnamefont {S.}~\bibnamefont {Upadhyay}},
  \bibinfo {author} {\bibfnamefont {E.~A.}\ \bibnamefont {Vo}}, \bibinfo
  {author} {\bibfnamefont {T.}~\bibnamefont {Vogels}}, \bibinfo {author}
  {\bibfnamefont {S.}~\bibnamefont {Wang}}, \bibinfo {author} {\bibfnamefont
  {T.}~\bibnamefont {Wang}}, \bibinfo {author} {\bibfnamefont {X.}~\bibnamefont
  {Wang}}, \bibinfo {author} {\bibfnamefont {X.}~\bibnamefont {Wang}}, \bibinfo
  {author} {\bibfnamefont {Y.}~\bibnamefont {Wang}}, \bibinfo {author}
  {\bibfnamefont {M.}~\bibnamefont {Williamson}}, \bibinfo {author}
  {\bibfnamefont {J.}~\bibnamefont {Yang}}, \bibinfo {author} {\bibfnamefont
  {H.-Z.}\ \bibnamefont {Ye}}, \bibinfo {author} {\bibfnamefont {C.-N.}\
  \bibnamefont {Yeh}}, \bibinfo {author} {\bibfnamefont {H.}~\bibnamefont
  {Yu}}, \bibinfo {author} {\bibfnamefont {J.}~\bibnamefont {Yu}}, \bibinfo
  {author} {\bibfnamefont {V.~W.-z.}\ \bibnamefont {Yu}}, \bibinfo {author}
  {\bibfnamefont {C.}~\bibnamefont {Zhang}}, \bibinfo {author} {\bibfnamefont
  {D.}~\bibnamefont {Zhang}}, \bibinfo {author} {\bibfnamefont
  {Y.}~\bibnamefont {Zhang}}, \bibinfo {author} {\bibfnamefont
  {Z.}~\bibnamefont {Zhao}}, \bibinfo {author} {\bibfnamefont {Z.}~\bibnamefont
  {Zhou}}, \bibinfo {author} {\bibfnamefont {A.~J.}\ \bibnamefont {Zhu}},
  \bibinfo {author} {\bibfnamefont {T.}~\bibnamefont {Zhu}}, \bibinfo {author}
  {\bibfnamefont {T.~C.}\ \bibnamefont {Berkelbach}}, \bibinfo {author}
  {\bibfnamefont {L.}~\bibnamefont {Gagliardi}}, \bibinfo {author}
  {\bibfnamefont {S.}~\bibnamefont {Sharma}}, \bibinfo {author} {\bibfnamefont
  {A.~Y.}\ \bibnamefont {Sokolov}},\ and\ \bibinfo {author} {\bibfnamefont
  {G.}~\bibnamefont {Kin-Lic~Chan}},\ }\bibfield  {title} {\enquote {\bibinfo
  {title} {The python simulations of chemistry framework: 10 years of an
  open-source quantum chemistry project},}\ }\href
  {https://doi.org/10.1063/5.0337441} {\bibfield  {journal} {\bibinfo
  {journal} {J. Chem. Phys.}\ }\textbf {\bibinfo {volume} {165}},\ \bibinfo
  {pages} {102502} (\bibinfo {year} {2026})}\BibitemShut {NoStop}%
\bibitem [{\citenamefont {Ye}\ and\ \citenamefont
  {Berkelbach}(2024)}]{Ye24JCTC}%
  \BibitemOpen
  \bibfield  {author} {\bibinfo {author} {\bibfnamefont {H.-Z.}\ \bibnamefont
  {Ye}}\ and\ \bibinfo {author} {\bibfnamefont {T.~C.}\ \bibnamefont
  {Berkelbach}},\ }\bibfield  {title} {\enquote {\bibinfo {title} {Periodic
  local coupled-cluster theory for insulators and metals},}\ }\href
  {https://doi.org/10.1021/acs.jctc.4c00936} {\bibfield  {journal} {\bibinfo
  {journal} {J. Chem. Theory Comput.}\ }\textbf {\bibinfo {volume} {20}},\
  \bibinfo {pages} {8948--8959} (\bibinfo {year} {2024})}\BibitemShut {NoStop}%
\bibitem [{\citenamefont {Knizia}(2013)}]{Knizia13JCTC}%
  \BibitemOpen
  \bibfield  {author} {\bibinfo {author} {\bibfnamefont {G.}~\bibnamefont
  {Knizia}},\ }\bibfield  {title} {\enquote {\bibinfo {title} {Intrinsic atomic
  orbitals: An unbiased bridge between quantum theory and chemical concepts},}\
  }\href {https://doi.org/10.1021/ct400687b} {\bibfield  {journal} {\bibinfo
  {journal} {J. Chem. Theory Comput.}\ }\textbf {\bibinfo {volume} {9}},\
  \bibinfo {pages} {4834--4843} (\bibinfo {year} {2013})}\BibitemShut {NoStop}%
\bibitem [{\citenamefont {Song}\ \emph {et~al.}(2026)\citenamefont {Song},
  \citenamefont {Gong}, \citenamefont {Bakry},\ and\ \citenamefont
  {Ye}}]{Song26JCP}%
  \BibitemOpen
  \bibfield  {author} {\bibinfo {author} {\bibfnamefont {R.}~\bibnamefont
  {Song}}, \bibinfo {author} {\bibfnamefont {X.}~\bibnamefont {Gong}}, \bibinfo
  {author} {\bibfnamefont {A.}~\bibnamefont {Bakry}},\ and\ \bibinfo {author}
  {\bibfnamefont {H.-Z.}\ \bibnamefont {Ye}},\ }\bibfield  {title} {\enquote
  {\bibinfo {title} {Random phase approximation-based local natural orbital
  coupled cluster theory},}\ }\href {https://doi.org/10.1063/5.0320612}
  {\bibfield  {journal} {\bibinfo  {journal} {J. Chem. Phys.}\ }\textbf
  {\bibinfo {volume} {164}},\ \bibinfo {pages} {214114} (\bibinfo {year}
  {2026})}\BibitemShut {NoStop}%
\bibitem [{\citenamefont {Petersson}\ \emph {et~al.}(2003)\citenamefont
  {Petersson}, \citenamefont {Zhong}, \citenamefont {Montgomery},\ and\
  \citenamefont {Frisch}}]{Petersson03JCP}%
  \BibitemOpen
  \bibfield  {author} {\bibinfo {author} {\bibfnamefont {G.~A.}\ \bibnamefont
  {Petersson}}, \bibinfo {author} {\bibfnamefont {S.}~\bibnamefont {Zhong}},
  \bibinfo {author} {\bibfnamefont {J.}~\bibnamefont {Montgomery},
  \bibfnamefont {John~A.}},\ and\ \bibinfo {author} {\bibfnamefont {M.~J.}\
  \bibnamefont {Frisch}},\ }\bibfield  {title} {\enquote {\bibinfo {title} {On
  the optimization of gaussian basis sets},}\ }\href
  {https://doi.org/10.1063/1.1516801} {\bibfield  {journal} {\bibinfo
  {journal} {J. Chem. Phys.}\ }\textbf {\bibinfo {volume} {118}},\ \bibinfo
  {pages} {1101--1109} (\bibinfo {year} {2003})}\BibitemShut {NoStop}%
\bibitem [{\citenamefont {Yu}\ \emph {et~al.}(2026)\citenamefont {Yu},
  \citenamefont {Zhang}, \citenamefont {Zhang}, \citenamefont {Cao},
  \citenamefont {Sun},\ and\ \citenamefont {Ye}}]{Yu26arXiv}%
  \BibitemOpen
  \bibfield  {author} {\bibinfo {author} {\bibfnamefont {J.}~\bibnamefont
  {Yu}}, \bibinfo {author} {\bibfnamefont {X.}~\bibnamefont {Zhang}}, \bibinfo
  {author} {\bibfnamefont {M.-Y.}\ \bibnamefont {Zhang}}, \bibinfo {author}
  {\bibfnamefont {Y.}~\bibnamefont {Cao}}, \bibinfo {author} {\bibfnamefont
  {Q.}~\bibnamefont {Sun}},\ and\ \bibinfo {author} {\bibfnamefont {H.-Z.}\
  \bibnamefont {Ye}},\ }\href {https://doi.org/10.48550/arXiv.2607.11711}
  {\enquote {\bibinfo {title} {Correlation-consistent gaussian basis sets for
  copper solids from material-constrained atomic optimization},}\ } (\bibinfo
  {year} {2026}),\ \Eprint {https://arxiv.org/abs/2607.11711}
  {arXiv:2607.11711} \BibitemShut {NoStop}%
\bibitem [{\citenamefont {Weigend}\ \emph {et~al.}(1998)\citenamefont
  {Weigend}, \citenamefont {H{\"a}ser}, \citenamefont {Patzelt},\ and\
  \citenamefont {Ahlrichs}}]{Weigend98CPL}%
  \BibitemOpen
  \bibfield  {author} {\bibinfo {author} {\bibfnamefont {F.}~\bibnamefont
  {Weigend}}, \bibinfo {author} {\bibfnamefont {M.}~\bibnamefont {H{\"a}ser}},
  \bibinfo {author} {\bibfnamefont {H.}~\bibnamefont {Patzelt}},\ and\ \bibinfo
  {author} {\bibfnamefont {R.}~\bibnamefont {Ahlrichs}},\ }\bibfield  {title}
  {\enquote {\bibinfo {title} {{RI-MP2}: optimized auxiliary basis sets and
  demonstration of efficiency},}\ }\href
  {https://doi.org/10.1016/S0009-2614(98)00862-8} {\bibfield  {journal}
  {\bibinfo  {journal} {Chem. Phys. Lett.}\ }\textbf {\bibinfo {volume}
  {294}},\ \bibinfo {pages} {143--152} (\bibinfo {year} {1998})}\BibitemShut
  {NoStop}%
\bibitem [{\citenamefont {Weigend}, \citenamefont {K{\"o}hn},\ and\
  \citenamefont {H{\"a}ttig}(2002)}]{Weigend02JCP}%
  \BibitemOpen
  \bibfield  {author} {\bibinfo {author} {\bibfnamefont {F.}~\bibnamefont
  {Weigend}}, \bibinfo {author} {\bibfnamefont {A.}~\bibnamefont {K{\"o}hn}},\
  and\ \bibinfo {author} {\bibfnamefont {C.}~\bibnamefont {H{\"a}ttig}},\
  }\bibfield  {title} {\enquote {\bibinfo {title} {Efficient use of the
  correlation consistent basis sets in resolution of the identity {MP2}
  calculations},}\ }\href {https://doi.org/10.1063/1.1445115} {\bibfield
  {journal} {\bibinfo  {journal} {J. Chem. Phys.}\ }\textbf {\bibinfo {volume}
  {116}},\ \bibinfo {pages} {3175--3183} (\bibinfo {year} {2002})}\BibitemShut
  {NoStop}%
\bibitem [{\citenamefont {Weigend}(2002)}]{Weigend02PCCP}%
  \BibitemOpen
  \bibfield  {author} {\bibinfo {author} {\bibfnamefont {F.}~\bibnamefont
  {Weigend}},\ }\bibfield  {title} {\enquote {\bibinfo {title} {A fully direct
  {RI-HF} algorithm: Implementation, optimised auxiliary basis sets,
  demonstration of accuracy and efficiency},}\ }\href
  {https://doi.org/10.1039/b204199p} {\bibfield  {journal} {\bibinfo  {journal}
  {Phys. Chem. Chem. Phys.}\ }\textbf {\bibinfo {volume} {4}},\ \bibinfo
  {pages} {4285--4291} (\bibinfo {year} {2002})}\BibitemShut {NoStop}%
\bibitem [{\citenamefont {Weigend}(2006)}]{Weigend06PCCP}%
  \BibitemOpen
  \bibfield  {author} {\bibinfo {author} {\bibfnamefont {F.}~\bibnamefont
  {Weigend}},\ }\bibfield  {title} {\enquote {\bibinfo {title} {Accurate
  {Coulomb}-fitting basis sets for {H} to {Rn}},}\ }\href
  {https://doi.org/10.1039/b515623h} {\bibfield  {journal} {\bibinfo  {journal}
  {Phys. Chem. Chem. Phys.}\ }\textbf {\bibinfo {volume} {8}},\ \bibinfo
  {pages} {1057--1065} (\bibinfo {year} {2006})}\BibitemShut {NoStop}%
\bibitem [{\citenamefont {Weigend}(2008)}]{Weigend08JCC}%
  \BibitemOpen
  \bibfield  {author} {\bibinfo {author} {\bibfnamefont {F.}~\bibnamefont
  {Weigend}},\ }\bibfield  {title} {\enquote {\bibinfo {title} {{Hartree--Fock}
  exchange fitting basis sets for {H} to {Rn}},}\ }\href
  {https://doi.org/10.1002/jcc.20702} {\bibfield  {journal} {\bibinfo
  {journal} {J. Comput. Chem.}\ }\textbf {\bibinfo {volume} {29}},\ \bibinfo
  {pages} {167--175} (\bibinfo {year} {2008})}\BibitemShut {NoStop}%
\bibitem [{\citenamefont {H{\"a}ttig}(2005)}]{Hattig05PCCP}%
  \BibitemOpen
  \bibfield  {author} {\bibinfo {author} {\bibfnamefont {C.}~\bibnamefont
  {H{\"a}ttig}},\ }\bibfield  {title} {\enquote {\bibinfo {title} {Optimization
  of auxiliary basis sets for {RI-MP2} and {RI-CC2} calculations: Core--valence
  and quintuple-{$\zeta$} basis sets for {H} to {Ar} and {QZVPP} basis sets for
  {Li} to {Kr}},}\ }\href {https://doi.org/10.1039/b415208e} {\bibfield
  {journal} {\bibinfo  {journal} {Phys. Chem. Chem. Phys.}\ }\textbf {\bibinfo
  {volume} {7}},\ \bibinfo {pages} {59--66} (\bibinfo {year}
  {2005})}\BibitemShut {NoStop}%
\bibitem [{\citenamefont {Hellweg}\ \emph {et~al.}(2007)\citenamefont
  {Hellweg}, \citenamefont {H{\"a}ttig}, \citenamefont {H{\"o}fener},\ and\
  \citenamefont {Klopper}}]{Hellweg07TCA}%
  \BibitemOpen
  \bibfield  {author} {\bibinfo {author} {\bibfnamefont {A.}~\bibnamefont
  {Hellweg}}, \bibinfo {author} {\bibfnamefont {C.}~\bibnamefont {H{\"a}ttig}},
  \bibinfo {author} {\bibfnamefont {S.}~\bibnamefont {H{\"o}fener}},\ and\
  \bibinfo {author} {\bibfnamefont {W.}~\bibnamefont {Klopper}},\ }\bibfield
  {title} {\enquote {\bibinfo {title} {Optimized accurate auxiliary basis sets
  for {RI-MP2} and {RI-CC2} calculations for the atoms {Rb} to {Rn}},}\ }\href
  {https://doi.org/10.1007/s00214-007-0250-5} {\bibfield  {journal} {\bibinfo
  {journal} {Theor. Chim. Acta}\ }\textbf {\bibinfo {volume} {117}},\ \bibinfo
  {pages} {587--597} (\bibinfo {year} {2007})}\BibitemShut {NoStop}%
\bibitem [{\citenamefont {Goldzak}\ \emph {et~al.}(2022)\citenamefont
  {Goldzak}, \citenamefont {Wang}, \citenamefont {Ye},\ and\ \citenamefont
  {Berkelbach}}]{Goldzak22JCP}%
  \BibitemOpen
  \bibfield  {author} {\bibinfo {author} {\bibfnamefont {T.}~\bibnamefont
  {Goldzak}}, \bibinfo {author} {\bibfnamefont {X.}~\bibnamefont {Wang}},
  \bibinfo {author} {\bibfnamefont {H.-Z.}\ \bibnamefont {Ye}},\ and\ \bibinfo
  {author} {\bibfnamefont {T.~C.}\ \bibnamefont {Berkelbach}},\ }\bibfield
  {title} {\enquote {\bibinfo {title} {Accurate thermochemistry of covalent and
  ionic solids from spin-component-scaled mp2},}\ }\href
  {https://doi.org/10.1063/5.0119633} {\bibfield  {journal} {\bibinfo
  {journal} {J. Chem. Phys.}\ }\textbf {\bibinfo {volume} {157}},\ \bibinfo
  {pages} {174112} (\bibinfo {year} {2022})}\BibitemShut {NoStop}%
\bibitem [{\citenamefont {Dinh}\ \emph {et~al.}(2026)\citenamefont {Dinh},
  \citenamefont {Rettig}, \citenamefont {Feng},\ and\ \citenamefont
  {Lee}}]{Dinh26JCP}%
  \BibitemOpen
  \bibfield  {author} {\bibinfo {author} {\bibfnamefont {H.~Q.}\ \bibnamefont
  {Dinh}}, \bibinfo {author} {\bibfnamefont {A.}~\bibnamefont {Rettig}},
  \bibinfo {author} {\bibfnamefont {X.}~\bibnamefont {Feng}},\ and\ \bibinfo
  {author} {\bibfnamefont {J.}~\bibnamefont {Lee}},\ }\bibfield  {title}
  {\enquote {\bibinfo {title} {Efficient all-electron periodic
  fourier-transformed coulomb method},}\ }\href
  {https://doi.org/10.1063/5.0303084} {\bibfield  {journal} {\bibinfo
  {journal} {J. Chem. Phys.}\ }\textbf {\bibinfo {volume} {164}},\ \bibinfo
  {pages} {054110} (\bibinfo {year} {2026})}\BibitemShut {NoStop}%
\end{thebibliography}%

\end{document}


\title{Supporting Information for Correlation-Consistent Gaussian Basis Sets for Periodic Solids.~I.~Hydrogen and $p$-block Elements from Boron to Chlorine}

    \author{Jincheng Yu}
    \affiliation{Department of Chemistry and Biochemistry, University of Maryland, College Park, Maryland, 20742, United States}

    \author{Hong-Zhou Ye}
    \email{hzye@umd.edu}
    \affiliation{Department of Chemistry and Biochemistry, University of Maryland, College Park, Maryland, 20742, United States}
    \affiliation{Institute for Physical Science and Technology, University of Maryland, College Park, Maryland, 20742, United States}

    \maketitle

    \tableofcontents

    \clearpage

    \section{Geometry and basis-set data}

    \begin{table}[!h]
        \caption{Lattice constants of the $7$ semiconductors and aluminum.}
        \label{tab:struct}
        \begin{tabular}{lcc}
            \toprule
            Material & Lattice & $a_0$ (\AA) \\
            \midrule
            BN & B3 & 3.592 \\
            BP & B3 & 4.525 \\
            C & A4 & 3.553 \\
            AlN & B3 & 4.368 \\
            AlP & B3 & 5.451 \\
            Si & A4 & 5.421 \\
            SiC & B3 & 4.346 \\
            Al & A1 & 4.018 \\
            \bottomrule
        \end{tabular}
    \end{table}

    Geometry files (in XYZ or VASP POSCAR format) and Gaussian basis-set data files (in NWChem format) are available in the following GitHub repository:
    \begin{center}
        \href{https://github.com/hongzhouye/supporting_data/tree/main/2026/ccpvxz_solids}{https://github.com/hongzhouye/supporting\_data/tree/main/2026/ccpvxz\_solids}
    \end{center}
    The geometry files include
    \begin{itemize}
        \item the Bosoni set, reproduced from ref~\citenum{Bosoni24NRP} using the reference cell volumes $V_0$;
        \item the $19$ molecular oxides, optimized at the $\omega$B97X-D3~\cite{Grimme10JCP,Chai08JCP,Lin13JCTC}/def2-TZVPP~\cite{Weigend05PCCP} level using ORCA~\cite{Neese12WIRCMS};
        \item the $7$ semiconductors and FCC aluminum, reproduced from ref~\citenum{Schimka11JCP} using the experimental lattice constants $a_0$ listed in \cref{tab:struct}.
    \end{itemize}
    The basis-set data files include
    \begin{itemize}
        \item the cc-pV$X$Z-s basis sets and the associated cc-pV$X$Z-s-FIT auxiliary basis sets for all four nuclear potentials;
        \item the FIT auxiliary basis sets optimized for the cc-pV$X$Z, pob-TZVP, pob-D/TZVP, and MOLOPT-DZVP/TZV2P basis sets.
    \end{itemize}

    \section{Raw data}

    The GitHub repository linked above also provides the raw data for the benchmarks presented in the main text, including
    \begin{itemize}
        \item the all-electron HF and MP2 molecular atomization energies calculated using the standard cc-pV$X$Z ($X = $ D, T, Q, and 5) and cc-pV$X$Z-s ($X = $ D, T, and Q) basis sets;
        \item the PBE free energies for the Bosoni EOS calculations with the GTH-PBE pseudopotentials, including results obtained with cc-pV$X$Z-s ($X = $ D, T, and Q), MOLOPT-DZVP, MOLOPT-TZV2P, and a converged PW basis;
        \item the HF, MP2, LNO-CCSD, and LNO-CCSD(T) energies used in EOS calculations for the $7$ semiconductors, including HF and MP2 results with various pseudopotentials and all-electron LNO-CC results.
    \end{itemize}

    \clearpage

    \section{Supplementary figures}

    \begin{figure}[!h]
        \centering
        \includegraphics[width=4.5in]{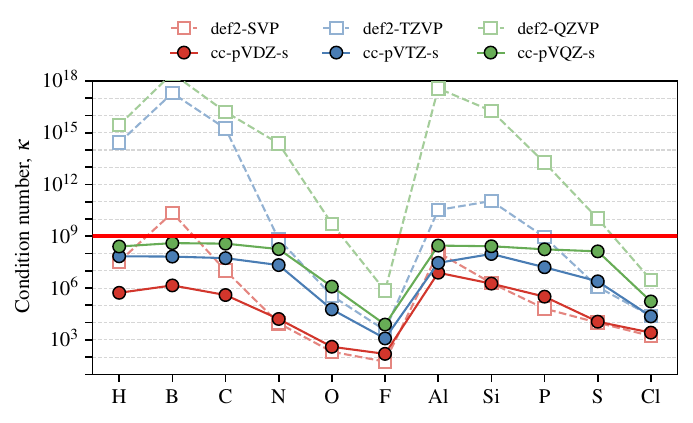}
        \caption{Maximum condition number ($\kappa_{\text{max}}$) over the $10$ reference solids for each element, evaluated with the conventional def2-family basis sets and the all-electron cc-pV$X$Z-s sets developed in this work.
        The horizontal red line indicates $\kappa_{\text{max}} = 10^{9}$.
        }
    \end{figure}

    \begin{figure}[!h]
        \centering
        \includegraphics[width=4.5in]{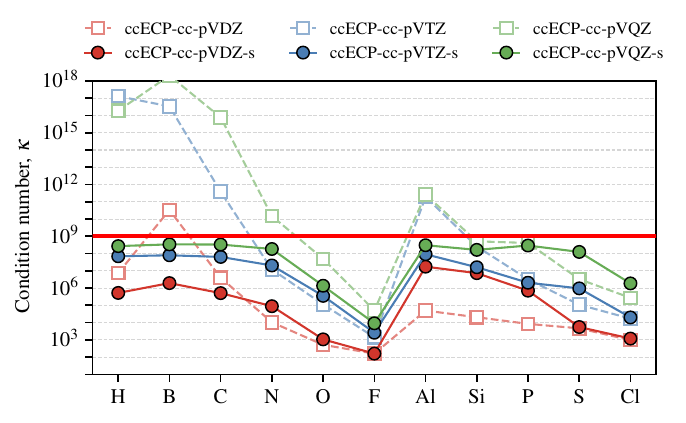}
        \caption{Maximum condition number ($\kappa_{\text{max}}$) over the $10$ reference solids for each element, evaluated with the ccECP-cc-pV$X$Z basis sets provided with ccECP~\cite{Bennett17JCP} and the cc-pV$X$Z-s sets developed for ccECP in this work.
        The horizontal red line indicates $\kappa_{\text{max}} = 10^{9}$.
        }
    \end{figure}

    \begin{figure}[!h]
        \centering
        \includegraphics[width=4.5in]{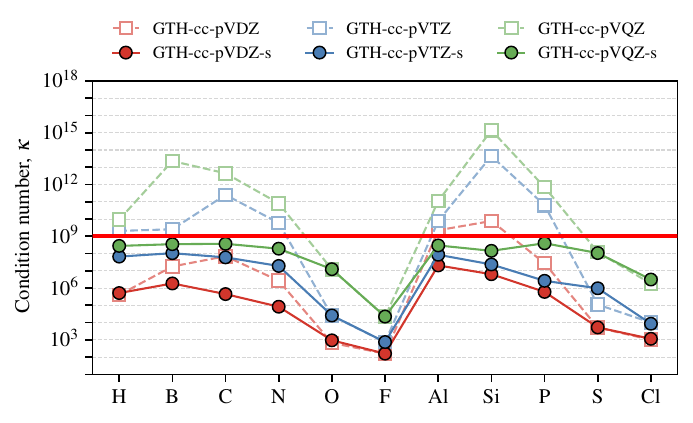}
        \caption{Maximum condition number ($\kappa_{\text{max}}$) over the $10$ reference solids for each element, evaluated with the GTH-cc-pV$X$Z basis sets from ref~\cite{Ye22JCTC} and the cc-pV$X$Z-s sets developed for the GTH-HF pseudopotential in this work.
        The horizontal red line indicates $\kappa_{\text{max}} = 10^{9}$.
        }
    \end{figure}

    \begin{figure}[!h]
        \centering
        \includegraphics[width=4in]{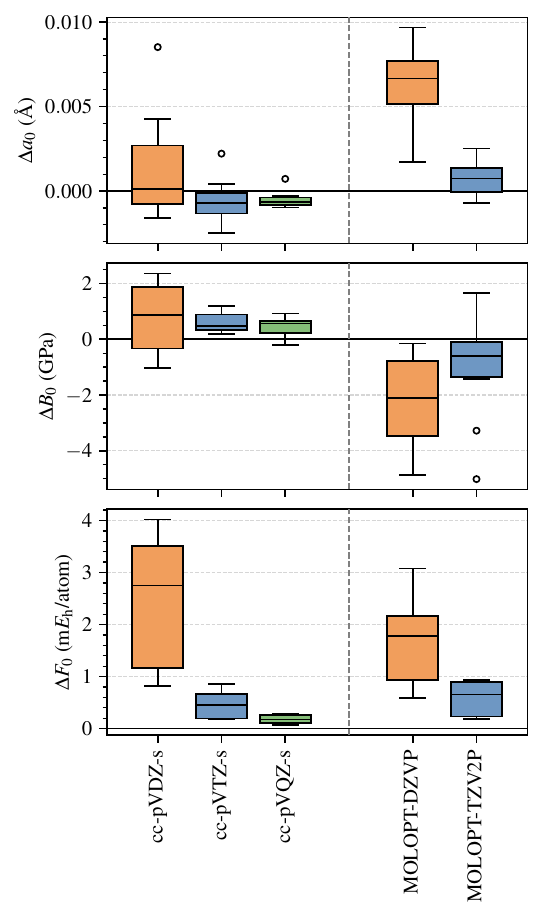}
        \caption{Same as Fig.~8 in the main text, but for the $10$ solids in the H subset of the Bosoni test set.}
    \end{figure}

    \begin{figure}[!h]
        \centering
        \includegraphics[width=4in]{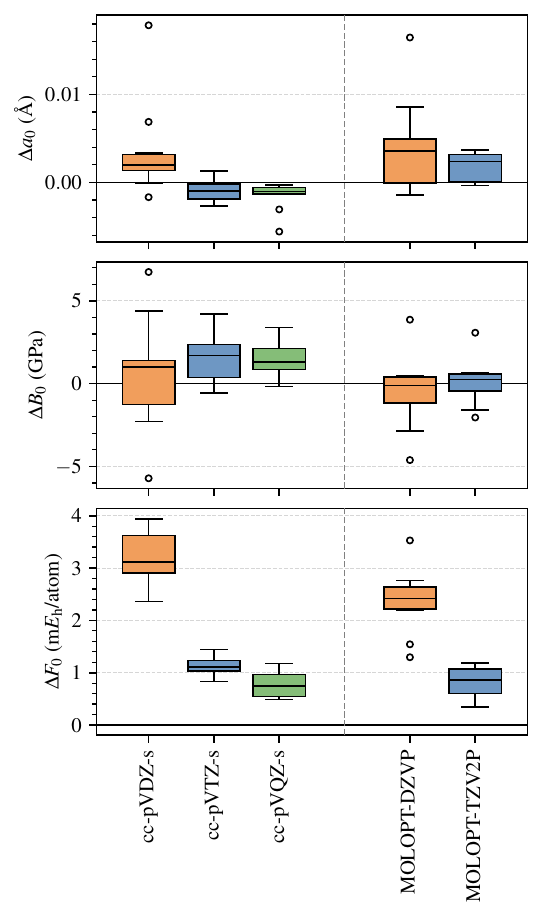}
        \caption{Same as Fig.~8 in the main text, but for the $10$ solids in the B subset of the Bosoni test set.}
    \end{figure}

    \begin{figure}[!h]
        \centering
        \includegraphics[width=4in]{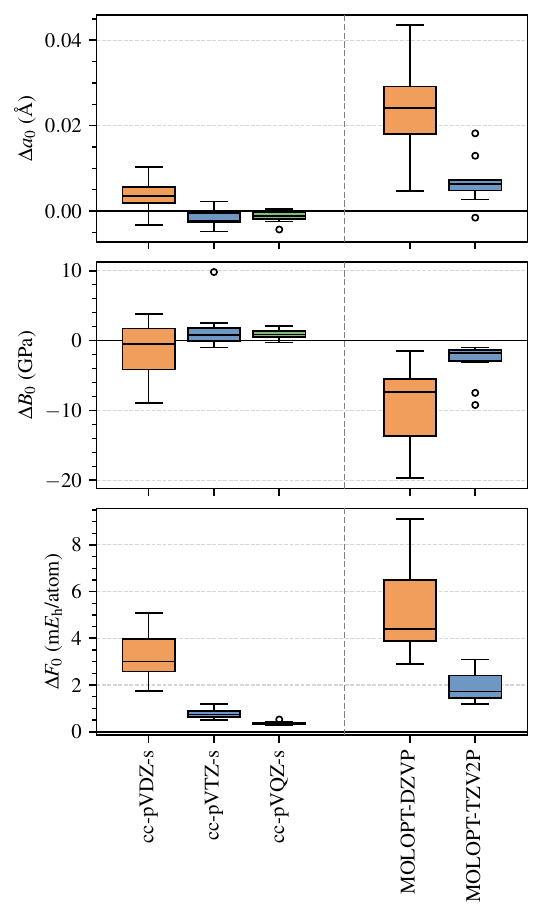}
        \caption{Same as Fig.~8 in the main text, but for the $10$ solids in the C subset of the Bosoni test set.}
    \end{figure}

    \begin{figure}[!h]
        \centering
        \includegraphics[width=4in]{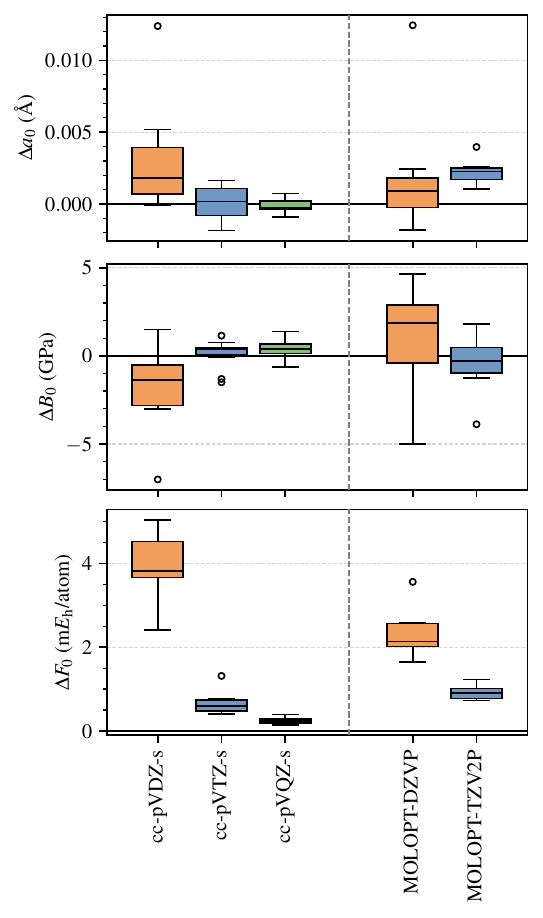}
        \caption{Same as Fig.~8 in the main text, but for the $10$ solids in the N subset of the Bosoni test set.}
    \end{figure}

    \begin{figure}[!h]
        \centering
        \includegraphics[width=4in]{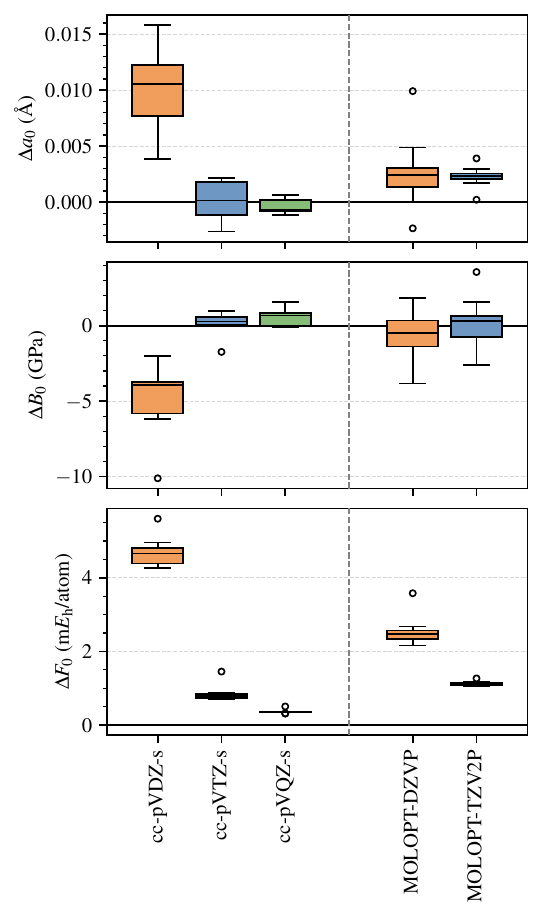}
        \caption{Same as Fig.~8 in the main text, but for the $10$ solids in the O subset of the Bosoni test set.}
    \end{figure}

    \begin{figure}[!h]
        \centering
        \includegraphics[width=4in]{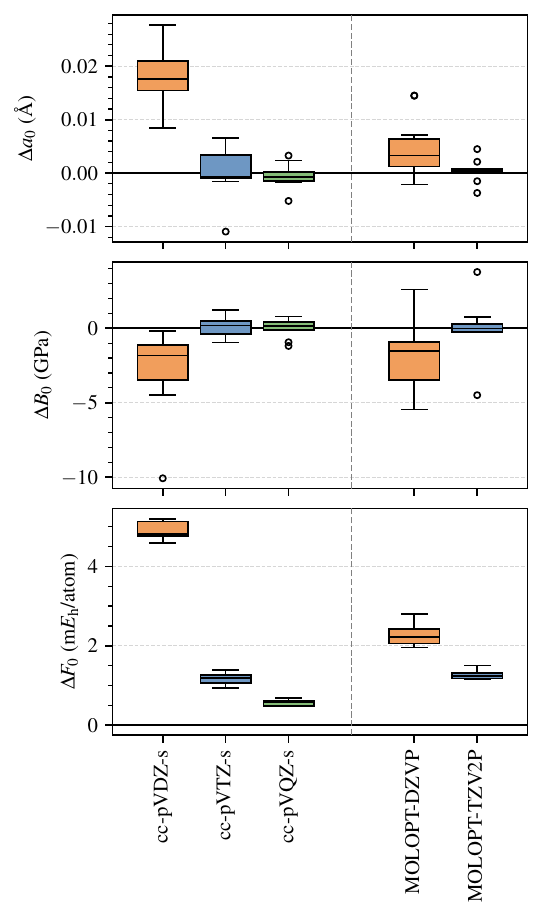}
        \caption{Same as Fig.~8 in the main text, but for the $10$ solids in the F subset of the Bosoni test set.}
    \end{figure}

    \begin{figure}[!h]
        \centering
        \includegraphics[width=4in]{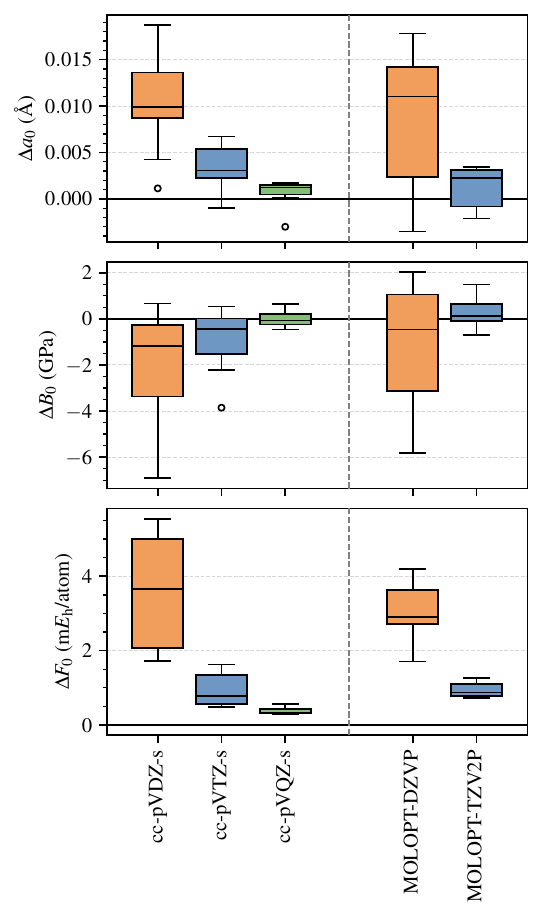}
        \caption{Same as Fig.~8 in the main text, but for the $10$ solids in the Al subset of the Bosoni test set.}
    \end{figure}

    \begin{figure}[!h]
        \centering
        \includegraphics[width=4in]{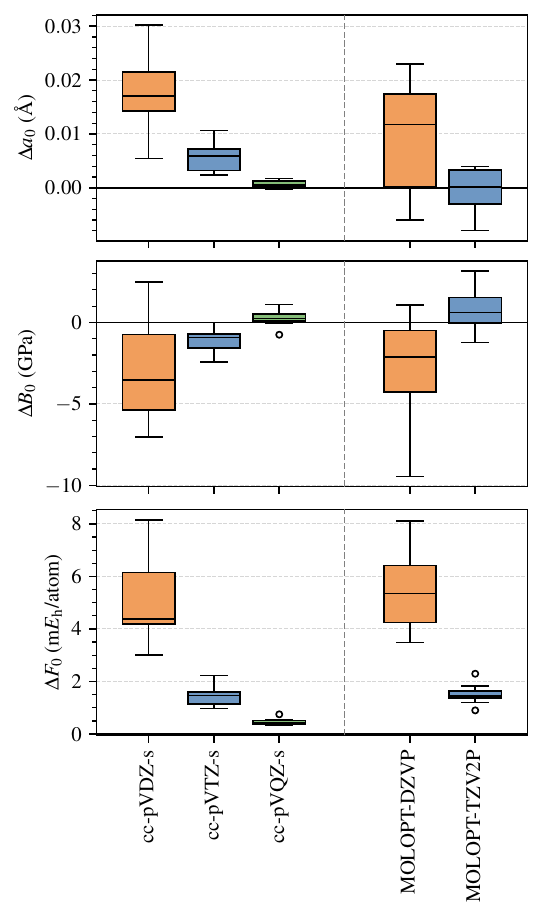}
        \caption{Same as Fig.~8 in the main text, but for the $10$ solids in the Si subset of the Bosoni test set.}
    \end{figure}

    \begin{figure}[!h]
        \centering
        \includegraphics[width=4in]{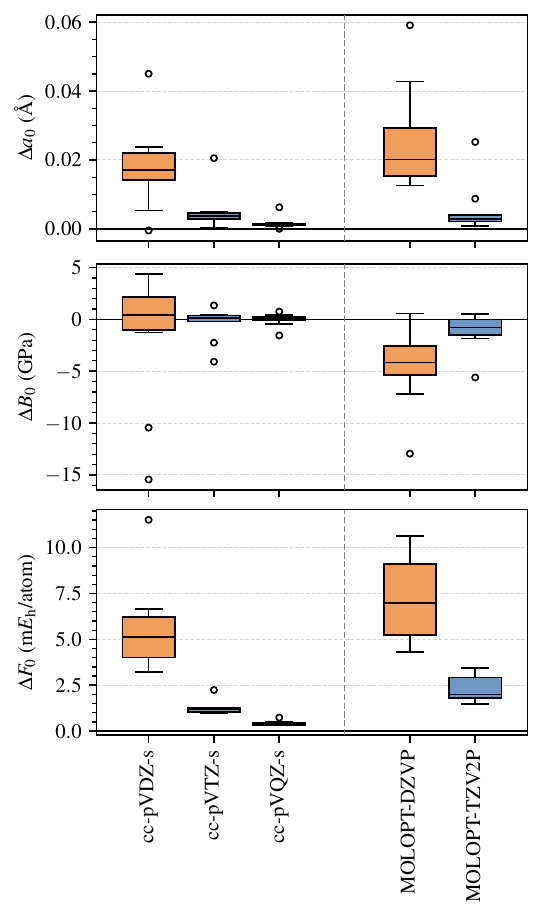}
        \caption{Same as Fig.~8 in the main text, but for the $10$ solids in the P subset of the Bosoni test set.}
    \end{figure}

    \begin{figure}[!h]
        \centering
        \includegraphics[width=4in]{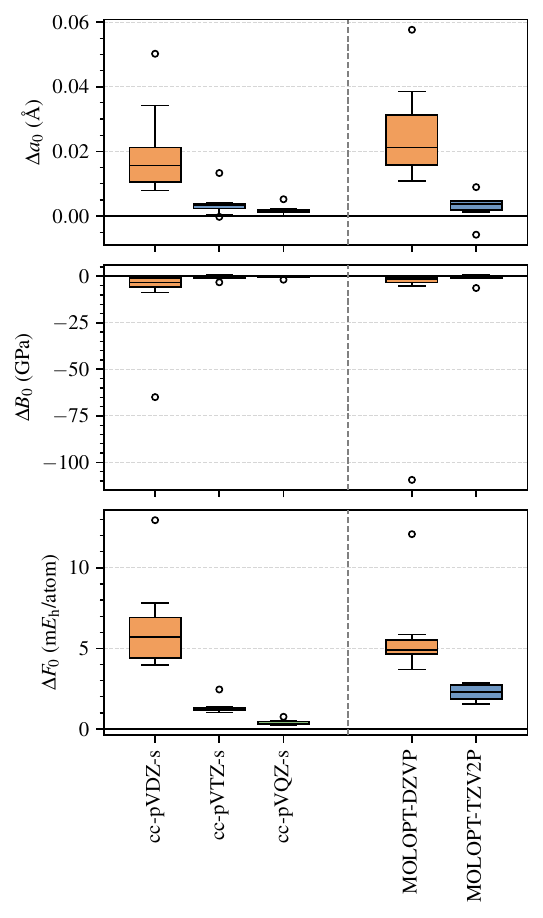}
        \caption{Same as Fig.~8 in the main text, but for the $10$ solids in the S subset of the Bosoni test set.}
    \end{figure}

    \begin{figure}[!h]
        \centering
        \includegraphics[width=4in]{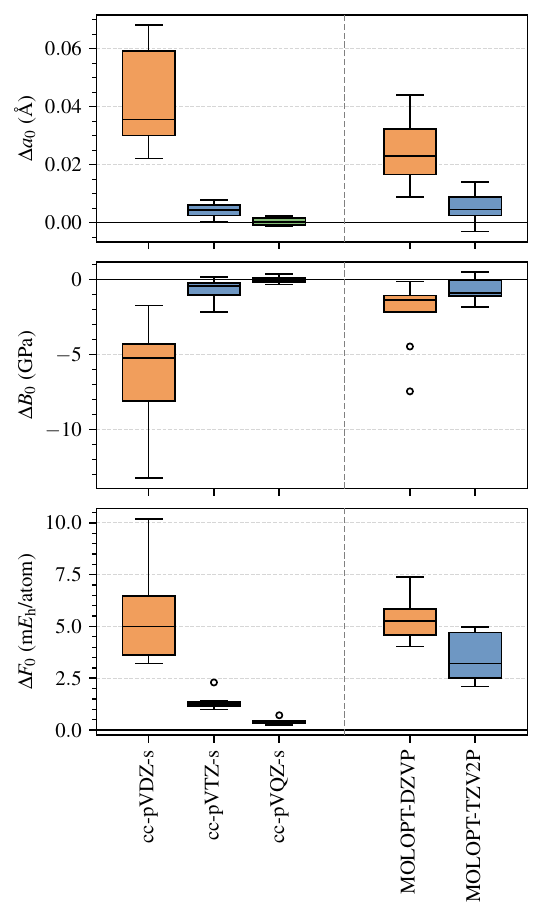}
        \caption{Same as Fig.~8 in the main text, but for the $10$ solids in the Cl subset of the Bosoni test set.}
    \end{figure}

    \begin{figure}[!t]
        \centering
        \includegraphics[width=\linewidth]{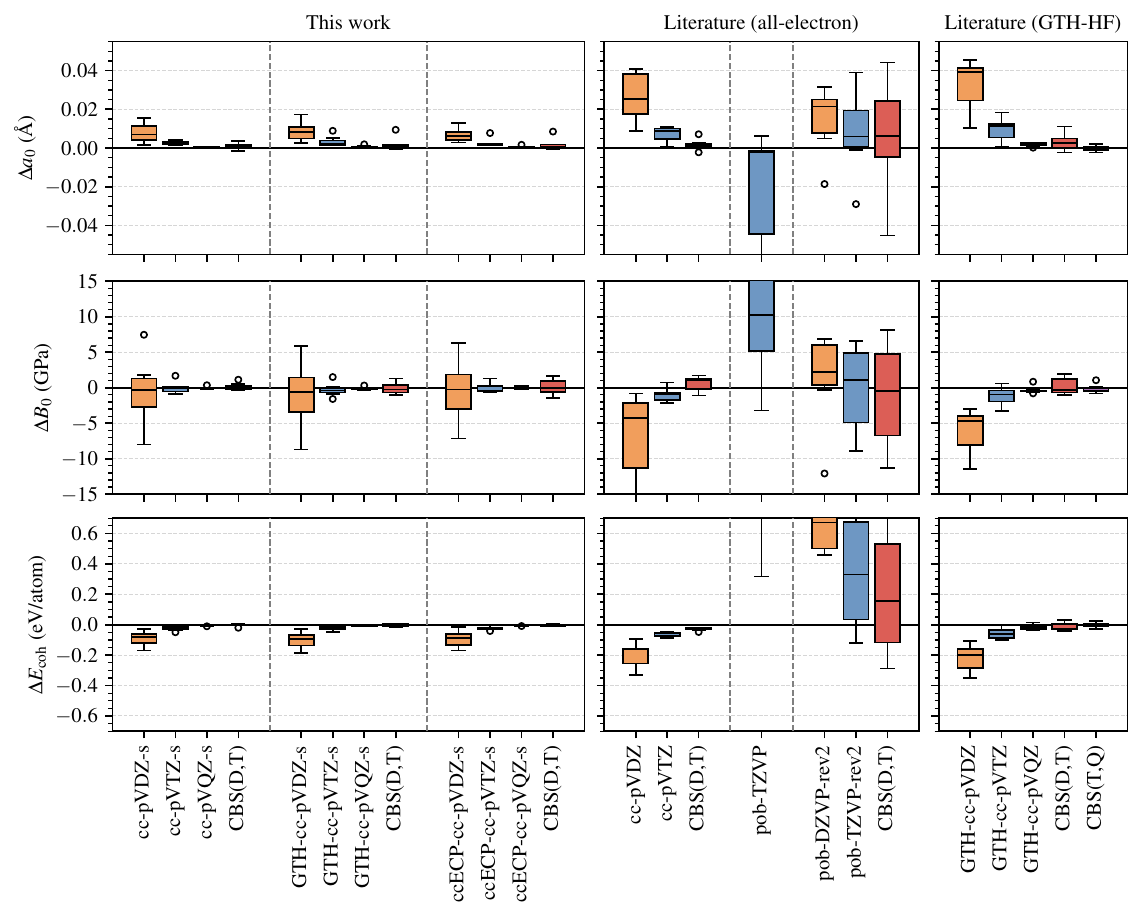}
        \caption{Same as Fig.~10 in the main text, but for errors at the HF level.
        }
    \end{figure}

    \clearpage

    \section{Local natural orbital--coupled cluster}

    In this section, we describe the computational details of the local natural orbital--coupled cluster (LNO-CC) method implemented in PySCF~\cite{Sun18WIRCMS,Sun20JCP,Sun26JCP,Ye24JCTC}.
    We use this method to accurately estimate canonical CCSD and CCSD(T) equations of state (EOS) for the $7$ semiconductors.
    LNO-CC rewrites the canonical CC correlation energy as a sum over fragment contributions,
    \begin{equation}
        E_{\text{LNO-CC}}(\eta)
            = \sum_{i}^{n_{\text{frag}}} E_{\text{CC}}^{(i)}(\eta) + \Delta E_{\text{MP2}}^{\text{ext}}(\eta).
    \end{equation}
    Each fragment correlation energy $E_{\text{CC}}^{(i)}(\eta)$ is calculated within a local active space consisting of (i) a small set of localized orbitals and (ii) their optimized bath orbitals.
    In this work, we use the atom-based fragments defined in ref~\citenum{Ye24JCTC}.
    Each fragment uses all intrinsic atomic orbitals~\cite{Knizia13JCTC} (IAOs) on the corresponding atom as its localized orbitals.
    The bath orbitals are chosen to be the MP2 local natural orbitals with occupations beyond user-selected thresholds, $\eta_{\text{occ}}$ and $\eta_{\text{vir}}$, for the occupied and virtual spaces, respectively.
    The ratio $\eta_{\text{occ}} / \eta_{\text{vir}}$ is fixed at $10$ for optimal convergence in gapped systems, and only $\eta \equiv \eta_{\text{vir}}$ is reported below.
    Orbitals outside the local active space contribute to the LNO-CC correlation energy through the external correction $\Delta E_{\text{MP2}}^{\text{ext}}$, evaluated using MP2.
    As $\eta$ is tightened, the LNO-CCSD and LNO-CCSD(T) correlation energies systematically converge to their canonical counterparts in the same basis set and on the same $k$-point mesh.
    Canonical CC results can be estimated from LNO-CC calculations at finite thresholds by linear extrapolation against $\Delta E_{\text{MP2}}^{\text{ext}}$,
    \begin{equation}    \label{eq:lno_conv}
        E_{\text{LNO-CC}}(\eta)
            = E_{\text{CC}} + A \Delta E_{\text{MP2}}^{\text{ext}}(\eta)
    \end{equation}
    In this work, we use two-point extrapolations following ref~\citenum{Song26JCP}, denoted by $(\eta_1, \eta_2)$, where $\eta_{1,2}$ are the thresholds for the two calculations.

    As discussed in the main text, the final L-CCSD$+\Delta$MP2 and L-CCSD(T)$+\Delta$MP2 total energies are obtained by adding the HF energy and the scaled MP2-based basis-set and finite-size correction to the LNO-CCSD and LNO-CCSD(T) correlation energies evaluated using the TZ basis sets and a $3\times3\times3$ $k$-point mesh.
    Because the HF energy and scaled MP2 correction are independent of $\eta$, the final L-CCSD$+\Delta$MP2 and L-CCSD(T)$+\Delta$MP2 energies can be extrapolated to the canonical CC limit using \cref{eq:lno_conv}.

    \begin{table}[!h]
        \centering
        \caption{L-CCSD$+\Delta$MP2 and L-CCSD(T)$+\Delta$MP2 values of $a_0$, $B_0$, and $E_{\text{coh}}$ for diamond, evaluated at several LNO truncation thresholds $\eta$ and extrapolated using \cref{eq:lno_conv}.
        }
        \label{tab:lno_conv}
        \begin{tabular}{lcccccc}
            \toprule
            \multirow{2}{*}{$\eta$} & \multicolumn{3}{c}{LNO-CCSD} & \multicolumn{3}{c}{LNO-CCSD(T)} \\
            \cmidrule(lr){2-4} \cmidrule(lr){5-7}
             & $a_0$ (\AA) & $B_0$ (GPa) & $E_{\mathrm{coh}}$ (eV/atom) & $a_0$ (\AA) & $B_0$ (GPa) & $E_{\mathrm{coh}}$ (eV/atom) \\
            \midrule
            $10^{-4}$ & 3.551 & 464.6 & 7.59 & 3.558 & 452.6 & 7.66 \\
            $3\times10^{-5}$ & 3.552 & 465.4 & 7.46 & 3.561 & 450.4 & 7.59 \\
            $10^{-5}$ & 3.553 & 466.3 & 7.39 & 3.562 & 449.9 & 7.55 \\
            $(10^{-4}, 3\times 10^{-5})$ & 3.554 & 465.7 & 7.38 & 3.563 & 448.8 & 7.54 \\
            $(3\times 10^{-5}, 10^{-5})$ & 3.554 & 466.1 & 7.35 & 3.563 & 449.3 & 7.53 \\
            \bottomrule
        \end{tabular}
    \end{table}

\begin{figure}[!ht]
        \centering
        \includegraphics[width=6in]{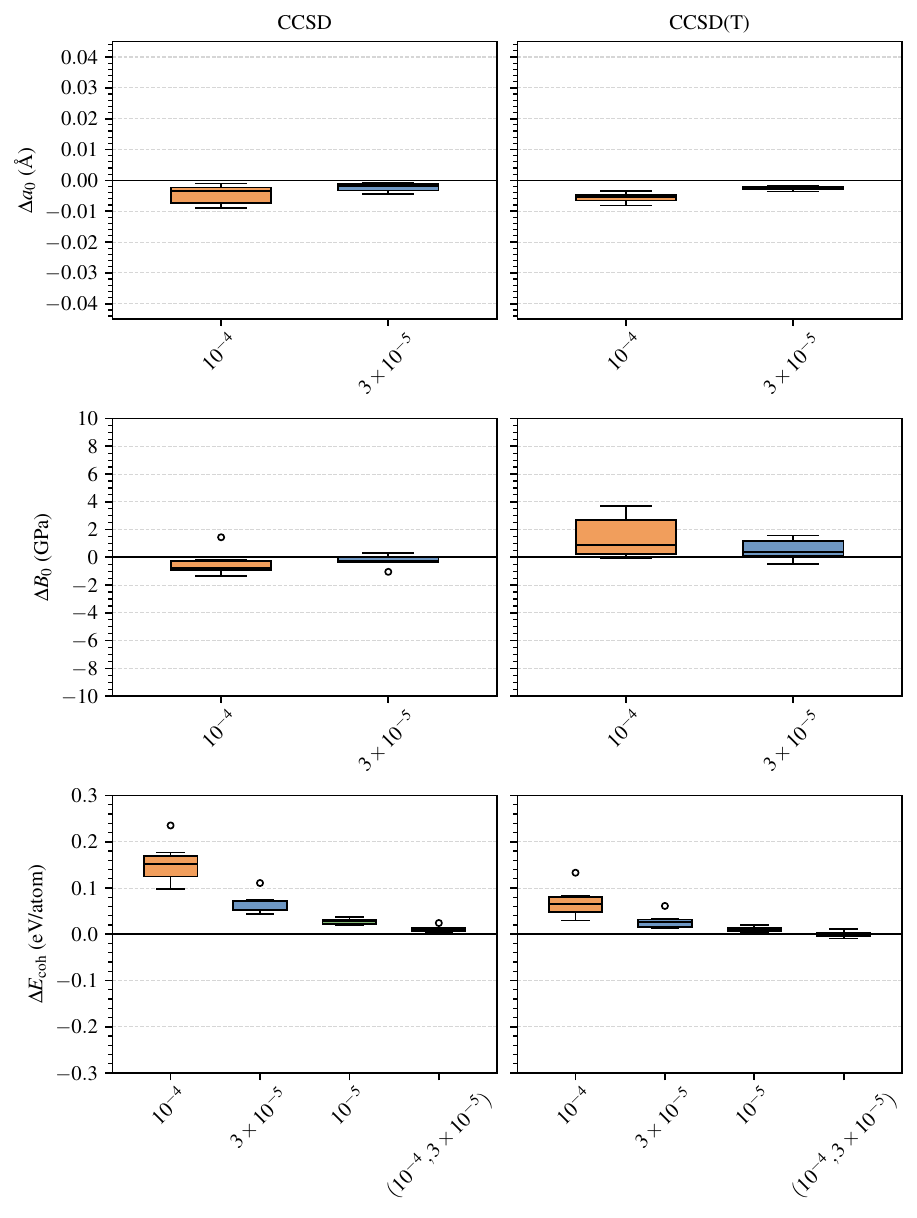}
        \caption{Convergence of $a_0$, $B_0$, and $E_{\text{coh}}$ with the LNO truncation threshold $\eta$ for L-CCSD$+\Delta$MP2 (left) and L-CCSD(T)$+\Delta$MP2 (right).
        The references are the $(10^{-4}, 3\times10^{-5})$ extrapolation for $a_0$ and $B_0$ and the $(3\times10^{-5}, 10^{-5})$ extrapolation for $E_{\text{coh}}$.
        }
        \label{fig:lno_conv}
    \end{figure}

    For diamond, we performed L-CC$+\Delta$MP2 calculations using $\eta = 10^{-4}$, $3\times10^{-5}$, and $10^{-5}$, allowing two extrapolations: a loose extrapolation, $(10^{-4}, 3\times10^{-5})$, and a tight extrapolation, $(3\times10^{-5}, 10^{-5})$.
    As shown in \cref{tab:lno_conv}, the loose and tight extrapolations agree to within \textit{ca.}~$0.001~\text{\AA}$ for $a_0$ and $0.5$~GPa for $B_0$, suggesting that the loose extrapolation is sufficient for both structural properties.
    \Cref{fig:lno_conv} further supports this choice for all $7$ semiconductors: results at $\eta = 3\times 10^{-5}$ differ from the loose extrapolation by \textit{ca.}~$0.005~\text{\AA}$ for $a_0$ and $1$~GPa for $B_0$.
    We estimate the errors of the loose extrapolation as half of these differences, giving \textit{ca.}~$0.003~\text{\AA}$ for $a_0$ and $0.5$~GPa for $B_0$.

    For $E_{\text{coh}}$, \cref{tab:lno_conv} shows that the loose and tight extrapolations differ by $0.03$~eV/atom for CCSD and $0.01$~eV/atom for CCSD(T), supporting the use of the tight extrapolation for $E_{\text{coh}}$.
    \Cref{fig:lno_conv} shows that the loose--tight differences remain well within $0.03$~eV/atom for CCSD and $0.01$~eV/atom for CCSD(T) across all $7$ semiconductors.
    We similarly estimate the error of the tight extrapolation for $E_{\text{coh}}$ as half of the loose--tight difference, giving \textit{ca.}~$0.02$~eV/atom.

    \clearpage

    \section{Optimization of auxiliary basis sets}
    \label{app:aux_basis}

    The cc-pV$X$Z-s-FIT auxiliary basis sets are uncontracted and parameterized using the Legendre polynomial-based approach described in ref~\citenum{Petersson03JCP}, with up to $6$ terms through degree $5$ (i.e.,~$k_{\text{max}} = 5$).
    We found that this parameterization yields more compact optimized auxiliary basis sets than the ETB-based parameterization used in our previous work~\cite{Yu26arXiv}.
    The basis parameters are optimized by minimizing an objective that combines atomic DF errors from HF and MP2 calculations, inspired by previous work on auxiliary basis-set optimization~\cite{Weigend98CPL,Weigend02JCP,Weigend02PCCP,Weigend06PCCP,Weigend08JCC,Hattig05PCCP,Hellweg07TCA},
    \begin{equation}    \label{eq:aux_obj}
    \begin{split}
        L_{\text{aux}}
            = \max \{
                &\gamma_{JK} \Delta J,~
                \gamma_{JK} \Delta K,~
                \Delta E_{J},~
                \Delta E_{K},   \\
                &\Delta E_{\text{MP2}},~
                \gamma_{\text{MP2}} \Delta I_{\text{MP2}}^{\text{SS}},~
                \gamma_{\text{MP2}} \Delta I_{\text{MP2}}^{\text{OS}}
            \} + P_{\text{min-exp}}
    \end{split}
    \end{equation}
    The terms in $L_{\text{aux}}$ are defined below.
    (For H, all MP2-related terms are calculated using \ce{H2} at a bond length of $0.7414~\text{\AA}$.)
    Minimizing $L_{\text{aux}}$ yields the optimized auxiliary basis set for a fixed size.
    Starting from a sufficiently large initial auxiliary basis set, we remove primitives from each $l$ channel, proceeding from high to low $l$, to obtain the smallest auxiliary basis set satisfying
    \begin{equation}
        L_{\text{aux}} \leq 10^{-5}~\text{a.u.}
    \end{equation}
    This criterion yields atomic HF and MP2 energy errors below $10~\mu E_{\text{h}}$, comparable to those of other optimized auxiliary basis sets.~\cite{Weigend98CPL,Weigend02JCP}

    The first four terms in \cref{eq:aux_obj} measure DF errors in HF calculations.
    $\Delta J$ and $\Delta K$ are the maximum absolute errors in the HF Coulomb and exchange matrix elements,
    \begin{subequations}
    \begin{align}
        \Delta J
            &= \max_{\mu\nu} |\tilde{J}_{\mu\nu} - J_{\mu\nu}|, \\
        \Delta K
            &= \max_{\mu\nu} |\tilde{K}_{\mu\nu} - K_{\mu\nu}|,
    \end{align}
    \end{subequations}
    and $\Delta E_{J} = |\tilde{E}_{J} - E_{J}|$ and $\Delta E_{K} = |\tilde{E}_{K} - E_{K}|$ are errors in the HF and exchange energies.
    In all cases, $\tilde{X}$ and $X$ denote quantities evaluated using the density-fitted electron-repulsion integrals (ERIs) $\tilde{V}_{pqrs}$ and the exact ERIs $V_{pqrs}$, respectively.
    All quantities are evaluated using a fixed ROHF density matrix calculated without DF, so the errors arise solely from the DF approximation.

    The last three terms inside the maximum in \cref{eq:aux_obj} measure DF errors in frozen-core MP2 calculations.
    $\Delta E_{\text{MP2}} = |\tilde{E}_{\text{MP2}} - E_{\text{MP2}}|$ is the error in the MP2 correlation energy,
    and $\Delta I_{\text{MP2}}^{\text{SS}}$ and $\Delta I_{\text{MP2}}^{\text{OS}}$ are errors in same-spin and opposite-spin MP2 wavefunction amplitudes adapted from literature~\cite{Weigend98CPL,Weigend02JCP,Hattig05PCCP,Hellweg07TCA},
    \begin{subequations}
    \begin{align}
        \Delta I_{\text{MP2}}^{\text{SS}}
            &= \frac{1}{|E_{\text{MP2}}^{\text{SS}}|}
            \sum_{iajb} \frac{
                |\Delta V_{iajb}^* (\Delta V_{iajb} - \Delta V_{ibja})|
            }{
                \Delta_{iajb}
            },  \\
        \Delta I_{\text{MP2}}^{\text{OS}}
            &= \frac{1}{|E_{\text{MP2}}^{\text{OS}}|}
            \sum_{iajb} \frac{
                |\Delta V_{iajb}|^2
            }{
                \Delta_{iajb}
            }
    \end{align}
    \end{subequations}
    where $\Delta V_{iajb} = \tilde{V}_{iajb} - V_{iajb}$, $\Delta_{iajb} = -\epsilon_i + \epsilon_a - \epsilon_j + \epsilon_b$ is the orbital energy difference, and $E_{\text{MP2}}^{\text{SS}}$ and $E_{\text{MP2}}^{\text{OS}}$ are the same-spin and opposite-spin MP2 correlation energies, respectively.

    The parameters $\gamma_{JK} = 0.1$ and $\gamma_{\text{MP2}} = 10$ impose looser tolerances on errors in the $J$ and $K$ matrices and stricter tolerances on errors in the MP2 amplitudes.
    The former is motivated by the observation that $\Delta J$ and $\Delta K$ are typically larger than $\Delta E_{J}$ and $\Delta E_{K}$ when using standard cc-pV$X$Z-JKFIT basis sets, while the latter follows literature~\cite{Weigend98CPL,Weigend02JCP}, which requires small errors in both the MP2 amplitudes and the MP2 energy.

    The last term in \cref{eq:aux_obj}, $P_{\text{min-exp}}$, penalizes overly diffuse primitives that can cause strong linear dependence in the DF metric matrix and destabilize DF-based SCF calculations~\cite{Goldzak22JCP,Dinh26JCP}.
    Let $\alpha_{0}^{(l)}$ and $\beta_{0}^{(l)}$ be the smallest exponents in angular-momentum channel $l$ of the AO and auxiliary basis sets, respectively.
    Define
    \begin{equation}
    \begin{split}
        &\beta_{\text{min}}^{(l)}
            = \min_{(l_1,l_2)} \alpha_{0}^{(l_1)} + \alpha_{0}^{(l_2)},
    \end{split}
    \end{equation}
    where the minimization is carried out over all $(l_1,l_2)$ pairs such that $|l_1-l_2| \leq l \leq l_1 + l_2$ and $l_1+l_2+l$ is even.
    We require the auxiliary exponents to satisfy
    \begin{equation}    \label{eq:aux_min_exp}
        \beta_{0}^{(l)} / \beta_{\text{min}}^{(l)}
            \gtrapprox r_{\text{min-exp}}
    \end{equation}
    for all $l$ channels in the auxiliary basis set.
    The scaling factor is set to $r_{\text{min-exp}} = 0.45$, obtained by analyzing the exponents of cc-pV$X$Z and cc-pV$X$Z-JKFIT.
    The constraint in \cref{eq:aux_min_exp} is enforced using the exponent-ratio penalty in eq.~(19) of the main text.
    To obtain $P_{\text{min-exp}}$, we apply this penalty to $\beta_{0}^{(l)}/\beta_{\text{min}}^{(l)}$ with threshold $r_{\text{min-exp}}$ and sum over all $l$ channels.
    The minimum-exponent penalty helps regularize the initial stages of optimization but has no effect on the optimized cc-pV$X$Z-s-FIT auxiliary basis sets, whose exponents readily satisfy \cref{eq:aux_min_exp}, with a negligible $P_{\text{min-exp}}$ (below $10^{-30}$).

    \clearpage

    \bibliography{refs_si}